\documentclass[twoside,11pt]{PhDthesisSU} 

\newcommand{\figuremacroW}[4]{
	\begin{figure}[htbp]
		\centering
		\includegraphics[width=#4\textwidth]{#1}
		\caption[#2]{\textbf{#2} - #3}
		\label{#1}
	\end{figure}
}

\title{Debris disks and the search for life in the universe}
\author{\href{mailto:gianni.cataldi@astro.su.se}{Gianni Cataldi}}

\hypersetup{pdfinfo={
Title={Debris disks and the search for life in the universe},
Author={Gianni Cataldi}
}}

\begin{document}

\selectlanguage{english}

\frontmatterSU



\maketitle  


\newpage
\thispagestyle{empty}
The \textbf{cover image} consists of the following parts:
\begin{description}
\item[upper left] Simulation of ALMA observations of C\,I emission from the $\beta$~Pictoris debris disk, shown in paper~I. {\small Reproduced with permission from Astronomy \& Astrophysics, \textcopyright ESO.}
\item[lower left] Emission line from neutral oxygen from the $\beta$~Pictoris debris disk as seen by \textit{Herschel}/PACS, shown in paper~II.
\item[lower right] The Fomalhaut debris belt with respect to the PACS integral field unit, for the observations presented in paper~III.
\item[upper right] The main crater of the Kaali crater field on the Estonian island of Saaremaa. Photo taken during the 2013 summer school `Impacts and their Role in the Evolution of Life' that provided important inspiration for the work presented in paper~IV. 
\item[background] The sky around $\beta$~Pictoris (the bright star in the middle) as seen by the Digitized Sky Survey 2. {\small Acknowledgement: The Digitized Sky Surveys were produced at the Space Telescope Science Institute under U.S.\ Government grant NAG W-2166. The images of these surveys are based on photographic data obtained using the Oschin Schmidt Telescope on Palomar Mountain and the UK Schmidt Telescope. The plates were processed into the present compressed digital form with the permission of these institutions. The present image was generated using the `Aladin sky atlas' developed at CDS, Strasbourg Observatory, France \citep{Bonnarel_etal_2000,Boch_Fernique_2014}.}
\end{description}


\vspace{\stretch{1}}
{\fontfamily{verdana}\selectfont
{\scriptsize
\noindent
\textcopyright Gianni Cataldi, Stockholm University 2016 
 
\vspace{5mm}
\noindent
ISBN 978-91-7649-366-3 

\vspace{5mm}
\noindent
Printed by Holmbergs, Malm\"o 2016 

\noindent
Distributor: Department of Astronomy, Stockholm University 
}
}

\cleardoublepage



\begin{dedication}

\ECFJD
{\Large

Gewidmet meiner Familie: Giovanni, Angelina, Fabiano \& Filippo!

}
\end{dedication} 

\chapter{List of Papers}

\vspace{-5pt} 

The following papers are included in this thesis. They are referred to in the text by their Roman numerals.

\vspace{0pt} 

\begin{enumerate}[P{A}PER I: ]

\setlength{\itemsep}{3.3mm} 


\item\textbf{\textit{Herschel}/HIFI observations of ionised carbon in the $\beta$~Pictoris debris disk}\\
\textbf{Cataldi, G.}, Brandeker, A., Olofsson, G., Larsson, B., Liseau, R., Blommaert, J., Fridlund, M., Ivison, R., Pantin, E., Sibthorpe, B., Vandenbussche, B., \& Wu, Y. 2014, \textit{A\&A}, 563, A66\\
DOI: \href{http://dx.doi.org/10.1051/0004-6361/201323126}{10.1051/0004-6361/201323126}

\item\textbf{\textit{Herschel} detects oxygen in the $\beta$~Pictoris debris disk}\\
Brandeker, A., \textbf{Cataldi, G.}, Olofsson, G. \& 28 colleagues 2016, subm.\ to \textit{A\&A}

\item\textbf{Constraints on the gas content of the Fomalhaut debris belt. Can gas-dust interactions explain the belt's morphology?}\\
\textbf{Cataldi, G.}, Brandeker, A., Olofsson, G., Chen, C.~H., Dent, W.~R.~F., Kamp, I., Roberge, A., \& Vandenbussche, B. 2015, \textit{A\&A}, 574, L1\\
DOI: \href{http://dx.doi.org/10.1051/0004-6361/201425322}{10.1051/0004-6361/201425322}

\item\textbf{Searching for biosignatures in exoplanetary impact ejecta}\\
\textbf{Cataldi, G.}, Brandeker, A., Th\'ebault, P., Ahmed, E., de Vries, B.~L., Neubeck, A., Olofsson, G. \& Singer, K. 2016, subm.\ to \textit{Astrobiology}

\end{enumerate}

\noindent
\rule{\linewidth}{0.5mm}

\vspace{2mm}

\noindent
Papers~I and III reproduced with permission from Astronomy \& Astrophysics, \textcopyright ESO 

\let\cleardoublepage\clearpage
\chapter{Author's contribution}
\begin{itemize}

\item PAPER I\\
I performed all the modelling of the \textit{Herschel}/HIFI data except for the extension of the \texttt{ONTARIO} code. I produced all the figures, except for figure 1. I wrote all the text, except for some parts in sections 2 and 4 and some suggested additions and comments by the co-authors.

\item PAPER II\\
I reduced the \textit{Herschel}/PACS data and measured the line strengths with error estimation. I took part in the modelling work by taking the output of the \texttt{ONTARIO} code and using it to produce synthetic PACS observations that can be compared to the data. I produced all the figures and wrote a section about the data reduction.

\item PAPER III\\
I reduced the \textit{Herschel}/PACS data and carried out their analysis and modelling. I produced all the figures and wrote all the text except for some suggested additions and comments by the co-authors.

\item PAPER IV\\
I did most of the preparing literature search. I wrote the code for all model calculations (impact, collisional evolution of the debris, detectability of biosignatures). I wrote most of the text and produced all figures.

\end{itemize} 


\setcounter{secnumdepth}{2} 
\setcounter{tocdepth}{1}    
\tableofcontents            



\renewcommand{\nomname}{Abbreviations} 





\begin{footnotesize} 

\printnomenclature[2 cm] 
\label{nom} 

\end{footnotesize}









\mainmatterSU


\setcounter{chapter}{-1}




\chapter{Some introducing words} 
\section{Astronomy}
The science of astronomy tries to explore and understand the place we are living in, called \emph{the universe}. In a certain sense, astronomers are analogue to the explorers of the old days that entered unknown areas to fill the empty space on their maps and discover new lands, plants, or animals. Similarly, astronomers try to map the universe and gather information about it, the big difference being of course that astronomy is dependent on remote observations in most of the cases---only the solar system can be explored by means of space travel at the moment. However, astronomers not only try to find out what our universe looks like and what its dimensions and constitutions are. Equally important is it to explain \emph{why} we see what we see. In short, we would like to understand how the universe works. To this end, models based on physical laws are constructed. There are models describing single planets, stars, galaxies or even the whole universe. These models attempt to explain our observations and make predictions that can then be tested with new observations, ultimately leading to a better understanding of the universe. Ideally, we would like to put ourselves into a cosmic context, understand our place in the universe, where we are coming from and where we are going.

\section{This thesis}
This thesis concentrates on a (very) small subfield of astronomy. A high degree of specialisation is very typical for a thesis in the natural sciences today. This is due to the fact that even seemingly small steps forward often require large efforts and a relatively deep understanding of the subject.

The topic of the present thesis is the observational study of so-called debris disks, extrasolar analogues of the solar system's asteroid belt or Kuiper belt. The thesis also touches upon the old question of whether there exist inhabited worlds other than the Earth by looking at the possibility to detect traces of alien life in impact generated debris.





\chapter{Context: star and planet formation}

\section{The standard picture of star formation}
Debris disks can be seen as an end product of the star and planet formation process. Consequently, they carry information about this process and can help us understand how stars and their planets arise. In this section, I will give a brief summary of the current picture of star and planet formation and describe their different phases.
\subsection{Collapse of interstellar cloud cores}
Our Galaxy, the Milky Way, contains roughly $10^{11}$ stars. The space between these stars is not empty, but filled with tenuous matter called the interstellar medium (ISM). By mass, the ISM consists of 99\% gas and 1\% dust \citep{Boulanger_etal_2000}. The gas is composed of hydrogen ($\sim$70\% by mass) and helium ($\sim$28\%), the rest being heavier elements referred to as metals in astronomy. The ISM is far from homogeneous. Temperature and density vary considerably for its different components. Table \ref{table_ISM} gives an overview of temperatures and hydrogen nuclei densities encountered in the ISM. About half of the ISM mass is found in cold interstellar clouds, but these clouds occupy only $\sim$1--2\% of the interstellar volume \citep{Ferriere_2001}. Stars are believed to form inside the cold, molecular component of the ISM, in giant molecular clouds (GMCs). These objects have typical masses of $10^4$--$10^6$\,M$_\odot$ and sizes between 10 and 100\,pc \citep{Natta_2000}. The densest parts of GMCs are called \emph{cores} and are characterised by a typical size of 0.1\,pc, a H$_2$ number density of $10^4$--$10^5$\,cm$^{-3}$, and masses of a few solar masses \citep{Natta_2000}. The number density encountered in the densest parts of a GMC are still very small when compared to terrestrial standards: at sea level, the typical number density is $\sim$$10^{19}$\,cm$^{-3}$. As another example, the Large Hadron Collider at CERN needs an ultra-high vacuum, equivalent to an H$_2$ number density of $10^9$\,cm$^{-3}$, in order for the circulating beam of particles to survive for 100 hours \citep{Jimenez_2009}.
\begin{table}
\centering
\begin{tabular}{lcc}
\hline\hline
ISM component & $T$ [K] & $n_\mathrm{H}$ [cm$^{-3}$] \\ 
\hline
Molecular & 10--20 & $10^2$--$10^6$ \\
Cold atomic & 50--100 & 20--50 \\
Warm atomic & 6000--10'000 & 0.2--0.5 \\
Warm ionized & $\sim$8000 & 0.2--0.5\\
Hot ionized & $\sim$$10^6$ & $\sim$$6.5\times10^{-3}$\\
\hline\hline
\end{tabular}
\caption{Temperature and number density of hydrogen nuclei for different components of the ISM. Adapted from \citet{Ferriere_2001}.}
\label{table_ISM}
\end{table}

The gravitational collapse of a dense core can lead to the formation of one or several stars. The collapse essentially happens on the free-fall timescale $t_\mathrm{ff}$. It can be estimated by considering the equation of motion of a core that freely contracts under its own gravity \citep{Natta_2000}:
\begin{equation}
\frac{\mathrm{d}^2R}{\mathrm{d}t^2}=-\frac{GM}{R^2}
\end{equation}
where $R$ is the radius of the core, $M$ its mass and $G$ the gravitational constant. From this equation, the following approximate relation can be derived:
\begin{equation}
\frac{R}{t_\mathrm{ff}^2}\approx\frac{GM}{R^2}
\end{equation}
leading to $t_\mathrm{ff}\approx\sqrt{R^3/(GM)}\approx\sqrt{1/(G\rho)}$ with $\rho$ the density. Plugging in the typical parameters of a dense core, one finds a $t_\mathrm{ff}$ of the order of $10^5$ years. This is a quite short timescale compared to the time for the overall star formation process.

\subsection{Young stellar objects}\label{YSO}
As the core collapses, the gravitational energy of the molecules is converted into kinetic energy, i.e.\ the temperature of the gas rises. The collapse continues until the increase in temperature leads to thermal pressure high enough to prevent further collapse. At this stage there exists a central object, known as \emph{protostar}, which is still deeply embedded in the collapsing core. The protostar continues to accrete gas via an accretion disk. The formation of a circumstellar disk is a direct consequence of the conservation of angular momentum. The total angular momentum of the initial dense core is given by
\begin{equation}\label{L_tot}
\mathbf{L}_\mathrm{tot}=\sum_i(\mathbf{r}_i\times m_i\dot{\mathbf{r}}_i)
\end{equation}
where $\mathbf{r}_i$ is the position vector\footnote{Note that the value of the angular momentum depends on the choice of the coordinate system.} of particle $i$ and the sum goes over all particles.
\begin{WrapText}{The Universality of Physical Laws}
A fundamental, though implicit assumption in any astronomy study is that physical laws are universal. In other words, we assume that the physical laws we discover in our laboratories are valid throughout the universe, even in the most distant galaxies, at any time. Actually, most people would argue that physical laws\footnote{For simplicity, I do not distinguish here between physical laws (a.k.a.\ scientific laws) and laws of nature, although these are quite distinct concepts \citep[e.g.][]{Swartz}. In metaphysics, physical laws are often seen as scientists' attempts to approximate or model the `true' or `fundamental' laws of nature, which are universal by definition.} are universal by definition \citep[e.g.][]{Swartz}. While the assumption of the universality of the laws of physics might seem reasonable to most of us, we should not forget that it remains an assumption. This universality assumption ultimately allows us to invoke the conservation of angular momentum in accretion disk formation and to interpret astronomical observations in general. For example, when analysing stellar spectra, we assume that atoms behave the same in our laboratories as in distant stars, giving rise to the same line emission.
\end{WrapText}
For an isolated system, $\mathbf{L}_\mathrm{tot}$ is constant. However, it is very unlikely that $\mathbf{L}_\mathrm{tot}=0$ initially. Thus, there exists a preferred direction. During collapse, the particles have to increase their velocity perpendicular to $\mathbf{r}_i$ in order to conserve angular momentum. Through mutual collisions, the angular momentum of individual particles approaches the direction of the total angular momentum. While contraction in the direction perpendicular to $\mathbf{L}_\mathrm{tot}$ is hampered by a corresponding increase in rotational velocity (equation \ref{L_tot}), collapse in the direction parallel to $\mathbf{L}_\mathrm{tot}$ is possible. This leads to the formation of a circumstellar disk.

The protostellar phase last typically $10^5$--$10^6$\,yr \citep[e.g.][]{Maeder_2009_19,Hartmann_1998}. Once the accretion of gas onto the protostar has decreased significantly, the forming star enters the pre-main-sequence phase\footnote{The pre-main-sequence star continues to accrete material from its disk, but the accreted mass is small and does not change the mass of the star significantly anymore.}. It contracts and slowly evolves towards the main-sequence. Low-mass pre-main-sequence stars ($M_*\lesssim M_\odot$) are called T~Tauri stars (after the prototype pre-main-sequence star T~Tauri), while pre-main-sequence stars with $2M_\odot\lesssim M_*\lesssim 8M_\odot$ are called Herbig Ae/Be stars. For a protostar, the luminosity comes from the accretion of gas. In contrast, pre-main-sequence stars get their luminosity from contraction \citep{Natta_2000}.

Once the star starts to fuse hydrogen, it has arrived on the main-sequence. The time it takes to reach the main-sequence depends strongly on the stellar mass. For a star of five solar masses, the pre-main-sequence lifetime is only 1.2\,Myr, while for a star of 0.2\,M$_\odot$ it is 200\,Myr, and 40\,Myr for a solar type star \citep{Maeder_2009_20}.

\section{Protoplanetary disks and planet formation}\label{protoplanetary_disks}
Planet formation is thought to occur in the aforementioned disk surrounding the young star. The circumstellar disk is thus also called a \emph{protoplanetary disk} (figure \ref{ALMA_HL_Tau}), which typically consist of 99\% gas and 1\% dust. The small dust grains gradually grow and eventually form \emph{planetesimals}\footnote{The word \emph{planetesimal} is a combination of \emph{planet} and \emph{infinitesimal}.} that are thought to be the building blocks of planets. By definition, planetesimals are bodies massive enough that their orbital evolution is determined by mutual gravitational interactions, in contrast to smaller dust particles for which aerodynamic interactions with the gas are more important \citep{Armitage_2009}. Thus, planetesimals typically have a radius of 10\,km or larger. How growth over several orders of magnitude from dust grains to planetesimals happens is not exactly understood and an active research area. For example, the relative velocities of meter-sized objects are high enough for collisions to become destructive. Also, meter-sized objects are expected to rapidly drift towards the star before they can grow further. The problem of growing objects larger than a metre is known as the \emph{metre-size barrier} \citep[e.g.][]{Apai_Lauretta_2010}.
\figuremacroW{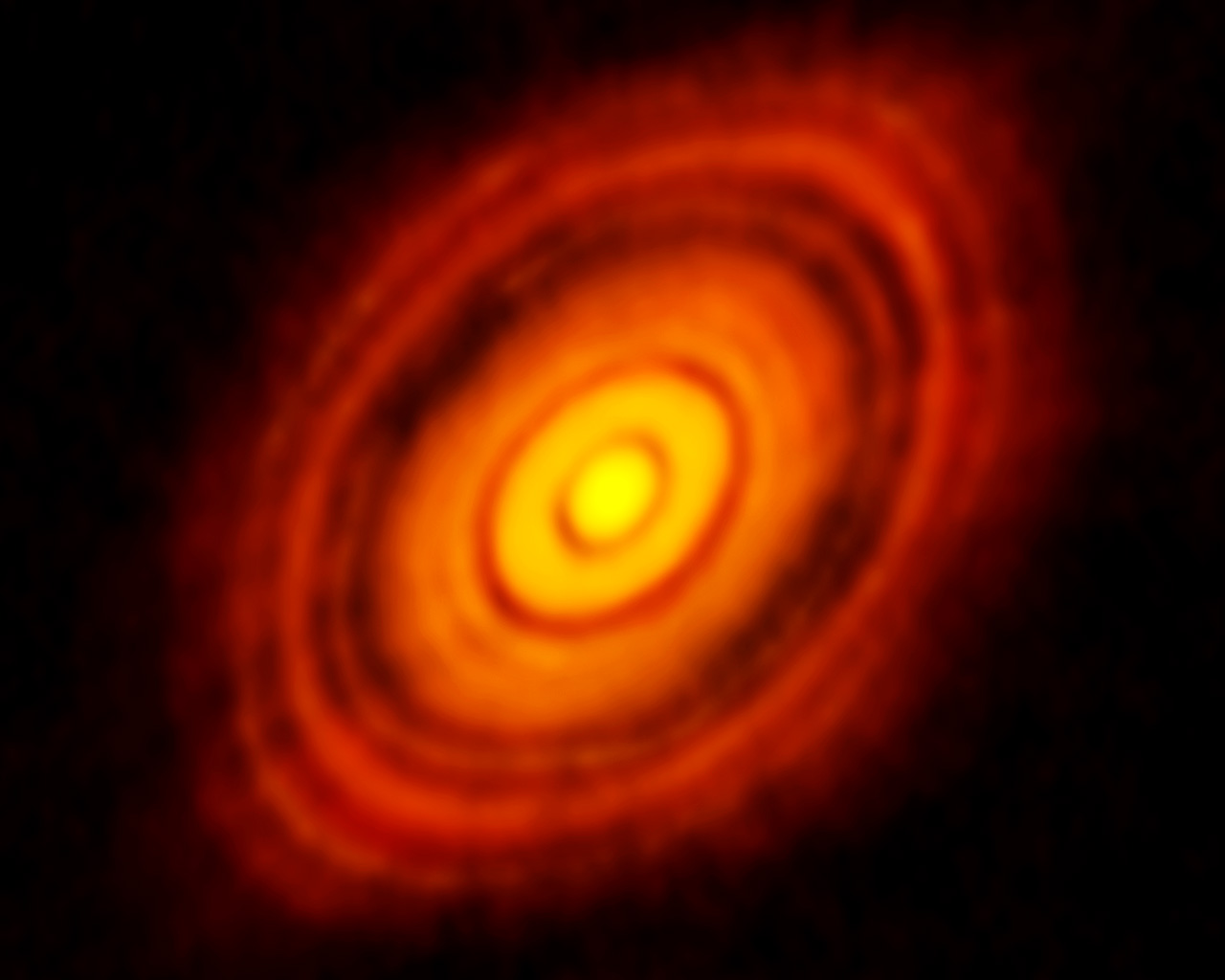}{ALMA protoplanetary disk image}{This spectacular image of the protoplanetary disk around HL~Tauri was taken by ALMA at a wavelength of 1.3\,mm and shows a series of concentric rings and gaps, thought to be due to forming planets \citep{ALMA_etal_2015}. Given that HL~Tau is only 1--2\,Myr old, these observations seem to suggest that planet formation occurs faster than previously thought. Image credit: ALMA (ESO/NAOJ/NRAO).}{0.7}

Once planetesimals have formed, a small fraction of them starts a phase of \emph{runaway growth} \citep{Armitage_2009}. This is due to two effects. On the one hand, a massive body can deflect trajectories of other planetesimals towards it by its larger gravity, thus increasing its collisional cross-section. This effect is called gravitational focussing. On the other hand, in a population consisting of smaller and larger bodies, gravitational interactions between bodies of different sizes lead to a velocity distribution where the relative velocity between two small planetesimals is larger than the relative velocity between a small and a large planetesimal. This effect is called dynamical friction and a consequence of the equipartition of energy between the two populations of bodies. It helps to further increase the growth rate of the largest planetesimals.

The next stage of the planet formation process is called \emph{oligarch growth}. During this stage, a number of larger bodies called oligarchs grow at approximately the same rate by accreting planetesimals from their local environment. Once the oligarch has cleared its local region, it has reached its isolation mass, which marks the end of the oligarch growth phase.

The phases of planet formation described above are completed quite rapidly, within 0.01--1\,Myr. The net result is a population of $10^2$--$10^3$ protoplanets in the terrestrial zone \citep{Armitage_2009}. From N-body simulations, we know that these bodies start to strongly interact dynamically, leading to a chaotic phase of collisions, scattering and merging, ultimately resulting in the formation of terrestrial planets. This final stage lasts 10--100\,Myr \citep{Kenyon_Bromley_2006,Armitage_2009}.

Concerning the formation of giant planets such as the gas giants Jupiter and Saturn or the ice giants Uranus and Neptune in the solar system, there exist two different formation scenarios \citep[e.g.][]{Armitage_2009,DAngelo_etal_2010}. The first is called \emph{core accretion}. In this scenario, the first step is the formation of a solid core, analogous to terrestrial planet formation. The core can accrete a gaseous envelope from the disk. Once the envelope mass is of the same order as the core mass, a critical mass is reached (typically on the order of 10\,M$_\oplus$). Runaway accretion of gas can then occur, allowing the planet to rapidly (within $\sim$$10^5$\,yr) accrete the bulk of its final mass. Gas accretion continues until no supplies are left, either because the protoplanetary disk has dissipated or because the accreting planet has opened a gap in the disk. The overall timescale for the formation of a giant planet via core accretion is about one to a few million years \citep[e.g.][]{DAngelo_etal_2010}. This is one of the difficulties of the model. Indeed, the lifetime of the gas-supplying disk is itself limited to a few million years. Also, the formation of a solid core takes longer at larger orbital distances, making formation by core accretion difficult in the outer disk.

The second model that has been proposed for the formation of giant planets is called \emph{disk instability}. In this scenario, the protoplanetary disk is massive enough for gravitational instabilities to occur \citep[e.g.][]{Boss_2000}. The net result is a fragmentation of the disk into massive clumps, the contractions of which can form giant planets. This can happen on short timescales, thus circumventing one of the problems of the core accretion model. However, disk instability can only occur under specific conditions. For example, the disk needs to be able to cool efficiently. These specific conditions are not readily realised in the inner disk. Thus, the current picture is that core accretion is the dominant formation process in the inner disk (inside of $\sim$100\,AU), while disk instability can be at work in the outer regions of extended and massive disks \citep{Boley_2009,DAngelo_etal_2010}.

As mentioned before, the overall lifetime of the gas-rich protoplanetary disk places a fundamental limit on the time available for giant planet formation. A common approach to determine disk lifetimes is to measure the fraction of stars surrounded by a disk for a given star cluster of known age. In practice, this is done by looking for excess emission in the infrared or sub-millimetre, indicative of circumstellar dust. These studies generally indicate a disk lifetime on the order of 2--6\,Myr \citep[e.g.][]{Ribas_etal_2015}. However, one should bear in mind that protoplanetary disks have a typical gas-to-dust ratio of 100 and that the gas might, in principle, evolve on a different timescale than the dust excess \citep{Gorti_etal_2015}.

Besides the incorporation into giant planets, there are two main processes that clear the gas from the protoplanetary disk. The first is viscous accretion onto the star. In order to accrete, the gas has to loose angular momentum. Understanding how angular momentum can be lost is a central problem in the modelling of accretion disks. Viscosity can allow a parcel of gas to loose angular momentum, however, it can easily be estimated that molecular viscosity, i.e.\ viscosity due to collisions among molecules, is not sufficient since it operates on timescales much longer than the observed evolutionary timescales of protoplanetary disks \citep{Armitage_2009}. Instead, it is believed that turbulence can induce a kind of `effective viscosity'\footnote{Note, however, that this effective viscosity arises from an entirely different physical process. Rather than collisions, it relies on turbulent mixing of gas at neighbouring radii.}, and one writes the strength of this viscosity based on dimensional arguments as
\begin{equation}\label{alpha_prescription}
\nu=\alpha c_\mathrm{s}h
\end{equation}
where $c_\mathrm{s}$ is the sound speed in the disk, $h$ is the disk scale height and $\alpha$ is the Shakura-Sunyaev parameter that measures how efficient turbulent viscosity is transporting angular momentum. Accretion disks modelled with this kind of viscosity are called $\alpha$-disks. The turbulence postulated to write down equation \ref{alpha_prescription} is believed to be due to the presence of magnetic fields that in combination with the differential rotation of the disk lead to an instability known as the \emph{magneto-rotational instability} \citep[MRI, e.g.][]{Armitage_2009}. 
An excellent explanation of the effect is given by \citet{Balbus_2009}. Intuitively, it can be understood in the following way. Consider a gas disk in Keplerian rotation. In general, the gas is well approximated by a perfectly conducting fluid. As a consequence, magnetic field lines are frozen into the fluid, i.e.\ displaced fluid parcels result in a displacement of the magnetic field. It can be shown that the displacement in the magnetic field results in a restoring force, analogue to a spring, due to magnetic tension. It is like neighbouring fluid parcels were connected by a spring (figure \ref{MRI_illustration}). Now imagine that two neighbouring fluid parcels are slightly displaced in the radial direction. Because of the differential rotation of the disk, the inner parcel moves faster than the outer parcel. The restoring force slows down the inner parcel and accelerates the outer parcel. Thus, the inner parcel looses angular momentum and moves inward, while the opposite is true for the outer parcel. This further increases the spring force\footnote{This mechanism only works if the spring (i.e.\ the magnetic field) is weak.}, resulting in an instability (figure \ref{MRI_illustration}). The MRI can thus provide the turbulence needed to allow viscous accretion of gas.
\figuremacroW{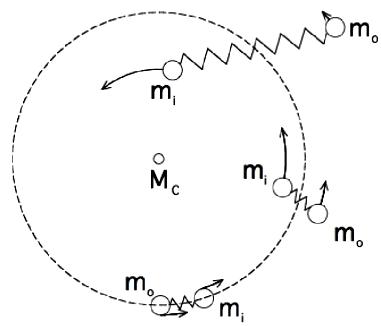}{Illustration of the MRI}{In a perfectly conducting fluid, a magnetic field has the effect to connect fluid elements by `springs' (magnetic tension). Consider two fluid elements slightly displaced in the radial direction. The disk rotates differentially, i.e.\ the inner fluid parcel moves faster than the outer parcel. The restoring force slows down the inner parcel and accelerates the outer parcel. Thus, the inner parcel loses angular momentum while the outer parcel gains angular momentum. This further increases the distance between the parcels, resulting in an instability. Figure from \citet{Balbus_2009}, courtesy of H.\ Ji.}{0.5}

The second mechanism that helps dispersing the gas of protoplanetary disks is called photoevaporation, which occurs if radiation from the central star (or from surrounding massive stars in a cluster environment) has heated the gas to temperatures where the thermal velocity exceeds the escape velocity. Photoevaporation is thought to be particularly important at the late stages of disk evolution \citep{Gorti_etal_2015}.

The disk dispersion process is believed to proceed from inside-out, leading to a class of objects known as \emph{transition disks} with inner dust holes \citep[e.g.][]{Gorti_etal_2015}. These objects seem to mark the transition between optically thick protoplanetary disks and optically thin debris disks, which are described more in detail in the next chapter.




\hyphenation{Fomalhaut}

\chapter{Debris disks}

Once the gas of the protoplanetary disk has been dispersed, the remaining dust disk is called a \emph{debris disk}. The formation of giant planets ceased, but terrestrial planet formation may continue for up to $\sim$100\,Myr \citep{Kenyon_Bromley_2006}. Debris disks are dusty disks that basically consist of leftover planetesimals and comets. Debris disks are much more long-lived than protoplanetary disks. They are indeed seen around main-sequence stars of all ages, although they are more often detected around young stars \citep[e.g.][]{Wyatt_2008}. There are also white dwarfs surrounded by debris disks \citep[e.g.][]{Rocchetto_etal_2015}. The solar system has its own debris disk in the form of the asteroid belt and the Kuiper belt (as well as the zodiacal dust).

Since the lifetime of the dust is generally much shorter than the age of the system, the dust in debris disks is thought to be continuously produced in a collisional cascade among the planetesimals and cometary bodies: collisions produce smaller fragments, which in turn collide to produce even smaller fragments \citep{Backman_Paresce_1993}. Thus, by studying the dust (for example its composition), we can learn more about the building blocks of exoplanets.

\section{Collisional cascade and radiation forces}\label{collcasc_radforces}
A convenient way to describe the fragment sizes in a debris disk is by means of a power law:
\begin{equation}\label{powerlaw_sizedist}
N(D)\propto D^{\alpha}
\end{equation}
where $N(D)$ is the number of fragments within an infinitesimal size interval around the diameter $D$ and $\alpha$ is the power law exponent. In the idealised case of an infinite, steady-state collisional cascade, it can be shown that $\alpha=-7/2$ \citep{Dohnanyi_1969,Tanaka_etal_1996}. An important property of such a steady-state size distribution is that most of the cross-section is in the small particles, but most of the mass is in the large boulders.

In reality, the collisional cascade does not extend down to arbitrary small particles. There is a lower limit $D_\mathrm{min}$. Particles with $D<D_\mathrm{min}$ are quickly removed by radiation forces. We shall now have a closer look at two radiation forces: radiation pressure and Poynting-Robertson (PR) drag.

Radiation pressure arises because photons carry momentum: $p_\gamma=h\nu/c$ with $\nu$ the frequency. Since momentum is conserved, a body absorbing or scattering photons has to gain momentum. Assume a dust particle is located at a distance $r$ from a star with specific luminosity $L_{*,\nu}$. By considering the number of photons absorbed per unit time, one easily derives the force on the particle due to radiation pressure:
\begin{equation}
F_\mathrm{rad}=\frac{\mathrm{d}p}{\mathrm{d}t}=\int\frac{L_{*,\nu}}{4\pi r^2c}\cdot\left(\frac{D}{2}\right)^2\pi Q_\mathrm{pr}(\nu,D)\mathrm{d}\nu
\end{equation}
where $Q_\mathrm{pr}(\nu,D)$ is the frequency-dependent radiation pressure efficiency. It is related to the absorption efficiency $Q_\mathrm{abs}$ and the scattering efficiency\footnote{$Q_\mathrm{abs}$ and $Q_\mathrm{sca}$ are defined as the ratio of the absorption cross-section and the scattering cross-section to the geometrical cross-section respectively. The extinction efficiency is then $Q_\mathrm{ext}=Q_\mathrm{abs}+Q_\mathrm{sca}$.} $Q_\mathrm{sca}$ by $Q_\mathrm{pr}=Q_\mathrm{abs}+Q_\mathrm{sca}(1-g)$ where the asymmetry parameter $g=\left<\cos\theta\right>$ is the mean of the cosine of the scattering angle $\theta$. A common way to parametrise the radiation pressure force is to define the ratio
\begin{equation}
\beta=\frac{F_\mathrm{rad}}{F_\mathrm{G}}
\end{equation}
with $F_\mathrm{G}=GM_*m/r^2$ the gravitational force ($M_*$ and $m$ denote the mass of the star and the particle respectively). Note that $\beta$ is independent of $r$. By writing down the kinetic energy needed to escape the gravity of the star, one can show that $\beta\geq0.5$ is enough to expel a particle from the system if the particle is created from a parent body in Keplerian orbit. For large grains, the radiation pressure efficiency is approximately unity. In this limit, one derives the following expression for the blowout size (i.e.\ the grain size where $\beta=0.5$):
\begin{equation}\label{eq:D_blowout_largegrain}
D_\mathrm{blowout}=\frac{3L_*}{4cG\pi M_*\rho}
\end{equation}
with $L_*$ the stellar luminosity and $\rho$ the density of the grain. For $\rho=2500$\,kg\,m$^{-3}$ and a solar-type star, this evaluates to about a micrometre. However, for small grains, $Q_\mathrm{pr}(\nu)\neq1$ in general. The value of  $Q_\mathrm{pr}(\nu)$ is dependent on the grain composition, size, temperature and shape. For homogeneous spherical grains, an analytical solution to Maxwell's equations called Mie solution (a.k.a.\ Mie theory) exists. This allows us to compute  $Q_\mathrm{pr}$ by calculating absorption and scattering efficiencies as well as the asymmetry parameter for given optical constants (i.e.\ complex refractive index). Figure \ref{beta} shows $\beta$ calculated using Mie theory for three different materials: astrosilicates, water ice and graphite. I used the \texttt{BHMIE} code \citep{Bohren_Huffman_1998_AppendixA} and assumed a solar-type host star. Optical constants are from \citet{Laor_Draine_1993} for astrosilicates and graphite and from \citet{Warren_Brandt_2008} for water ice. As can be seen, $\beta$ does not rise indefinitely for decreasing grain size, as would be the case if $Q_\mathrm{pr}(\nu)=1$ (equation \ref{eq:D_blowout_largegrain}). Rather, after a maximum is reached, $\beta$ starts to decrease for smaller grain sizes. This means that for example water ice spheres are only blown out in a relatively narrow size range. On the other hand, graphite spheres are removed from the system even for small grain sizes. In a more realistic model, one would for example consider grains consisting of a mixture of materials or with complicated geometrical shapes (e.g.\ fluffy grains). The spectral type of the host star is also important. For example, around M dwarfs it can happen that $\beta<0.5$ for any grain size, i.e.\ grains are never blown away.
\figuremacroW{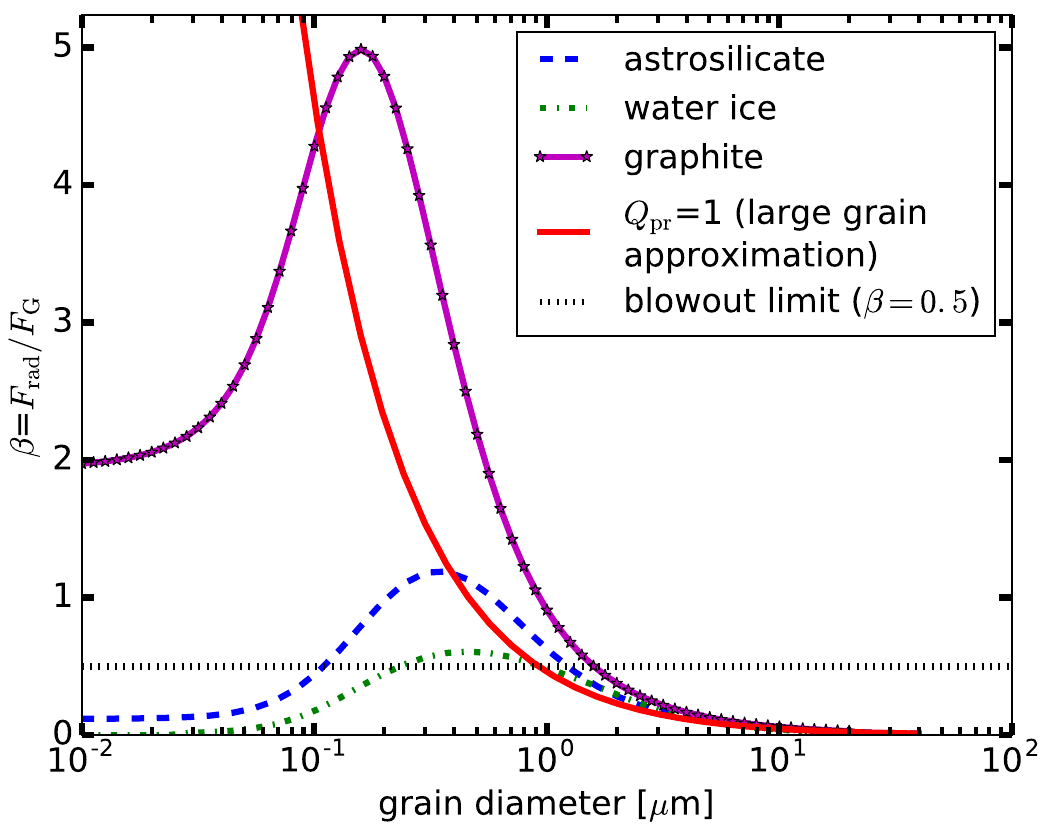}{$\beta$ as a function of grain size}{Using Mie theory, I calculated $\beta$ for three different materials, assuming a solar-type host star. Water ice spheres are only blown out in a narrow size range, while for graphite spheres one finds $\beta>0.5$ for any grain smaller than the blowout size.}{0.8}

Another important radiation force acting on dust grains is PR drag, which in contrast to radiation pressure leads to orbital decay and lets dust grains spiral into the star \citep[e.g.][]{Burns_etal_1979}. PR drag is caused by the re-radiation of absorbed stellar photons, which is isotropic in the reference frame of the particle. However, in the frame of the star, more momentum is carried away by the photons emitted in the direction of motion of the grain because of the Doppler effect. This is equivalent to a drag force. The PR drag force is given by the following expression that depends on the velocity of the grain \citep{Burns_etal_1979}:
\begin{equation}
\mathbf{F}_\mathrm{PR}=-\frac{L_*}{4\pi r^2c}\cdot\left(\frac{D}{2}\right)^2\pi Q_\mathrm{pr}\cdot\left(\frac{2\dot{r}}{c}\mathbf{\hat{r}}+\frac{r\dot{\theta}}{c}\mathbf{\hat{\theta}}\right)
\end{equation}
where $\mathbf{\hat{r}}$ and $\mathbf{\hat{\theta}}$ are the usual unit vectors in a cylindrical coordinate system. Here $Q_\mathrm{pr}$ is an average over the stellar spectrum. The timescale for a dust grain to fall onto the star then reads \citep{Burns_etal_1979,vanLieshout_etal_2014}
\begin{equation}
t_\mathrm{PR}=\frac{cr^2}{4GM_*\beta}=400\left(\frac{M_\odot}{M_*}\right)\left(\frac{r}{r_\oplus}\right)^2\left(\frac{1}{\beta}\right)\quad\mathrm{years}
\end{equation}

The combination of collisions and radiation forces essentially determines how debris disks evolve. We consider the evolution of debris disks in more detail in the next section.

\section{Evolution of debris disks}
The dust in debris disks is thought to originate from collisions among larger bodies such as leftover planetesimals or comets, although other dust sources exist as well. For example, comet sublimation is thought to be the main source for the zodiacal cloud in the solar system \citep{Nesvorny_etal_2010}.

In order for collisions to be frequent enough and destructive (i.e.\ collisions do not lead to net accretion), the colliding bodies need to have acquired a certain eccentricity, typically $10^{-3}$ to $10^{-2}$ \citep{Wyatt_2008}. A debris disk fulfilling this requirement is called \emph{stirred}. A first question we need to answer is thus how a debris disk can be stirred. One obvious possibility is stirring by giant planets that gravitationally perturb the planetesimals in the disk. Another possibility is stirring due to the formation of large ($\sim$2000\,km) planetesimals that again gravitationally perturb the disk. This later stirring mechanism is called \emph{self-stirring}.

Once sufficiently stirred, a collisional cascade is ignited in the debris disk. An important parameter describing the cascade is the collisional lifetime of a dust grain. For grains in a debris belt at a distance $r$ from the star and with width $\Delta r$, it is given by \citep{Wyatt_Dent_2002}
\begin{equation}\label{t_coll}
t_\mathrm{coll}=\frac{2It_\mathrm{pr}r\Delta r}{\sigma_\mathrm{c}(D)f(e,I)}
\end{equation}
with $I$ the mean inclination of the grains, $t_\mathrm{per}$ the orbital period, $\sigma_\mathrm{c}(D)$ the catastrophic cross-section and $f(e,I)$ the ratio between the relative velocity between the fragments $v_\mathrm{rel}$ and the Keplerian velocity $v_\mathrm{Kep}$ at r. $f(e,I)$ depends on the eccentricity $e$ and the inclination. The catastrophic cross-section $\sigma_\mathrm{c}(D)$ is the total cross-section of all particles that could potentially destroy\footnote{Usually, one defines a collision as catastrophic if the most massive remnant of the collision has less than half the mass of the initial body.} a grain of size $D$. For a given relative velocity, the minimum size needed for destruction is
\begin{equation}
D_\mathrm{c}(D)=\left(\frac{2Q}{v_\mathrm{rel}^2}\right)^{1/3}D
\end{equation}
where $Q$, the specific energy needed for destruction, is material-dependent and poorly known in general \citep[e.g.][]{Benz_Asphaug_1999}. The catastrophic cross-section is then given by
\begin{equation}
\sigma_\mathrm{c}(D)=\int_{D_\mathrm{x}(D)}^{D_\mathrm{max}}\frac{(D+D')^2}{4}\pi N(D')\mathrm{d}D'
\end{equation}
where $D_\mathrm{x}(D)=D_\mathrm{c}(D)$ if $D_\mathrm{c}(D)>D_\mathrm{min}$ and $D_\mathrm{x}(D)=D_\mathrm{min}$ otherwise. $D_\mathrm{min}$ and $D_\mathrm{max}$ are the minimum and maximum grain sizes present in the cascade. $D_\mathrm{min}$ is usually set by radiation pressure and equal to the blowout size. However, in tenuous disks with low collision frequency, $D_\mathrm{min}$ might instead be determined by PR drag.

If only collisions are removing mass from the disk, one can write the evolution of the total disk mass as
\begin{equation}
\frac{\mathrm{d}M_\mathrm{tot}}{\mathrm{d}t}=-\frac{M_\mathrm{tot}}{t_\mathrm{coll}(D_\mathrm{max},t)}
\end{equation}
since most of the mass is in the largest fragments for a steady-state collisional cascade. The solution to this equation reads \citep[e.g.][]{Wyatt_etal_2007}
\begin{equation}
M_\mathrm{tot}(t)=\frac{M_\mathrm{tot}(0)}{1+\frac{t}{t_\mathrm{coll}(D_\mathrm{max},0)}}
\end{equation}
This simple model is valid for the case of a steady-state collisional cascade, i.e.\ the size distribution always follows the form of equation \ref{powerlaw_sizedist} with $\alpha=-7/2$. Figure \ref{SteadyState_CollEvolution} shows the steady-state evolution of the fractional luminosity\footnote{The fractional luminosity $f$ of a debris disk is the ratio between the disk luminosity and the stellar luminosity: $f=L_\mathrm{disk}/L_*$.}, which is proportional to the disk mass, for a number of disk models with different initial masses and orbital radii. The same mass put closer to the star results in a brighter disk, which however also fades away faster.
\figuremacroW{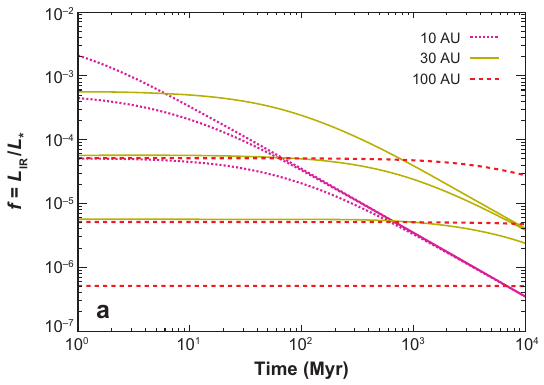}{Steady-state collisional evolution}{This figure by \citet{Wyatt_2008} shows the steady-state evolution of the fractional luminosity for debris belts at different orbital radii. For each radius, three curves corresponding to initial masses of 0.1, 1 and 10\,M$_\oplus$ are shown (from bottom to top). For the same initial mass, belts closer to the host star are brighter, but also fade away quicker because of the higher collisional frequency (equation \ref{t_coll}). Note that the fractional luminosity at late times is independent of the initial mass. The figure also demonstrates that the fractional luminosity of a debris disks at 100\,AU can in principle remain constant for billions of years. Reproduced with permission of Annual Reviews.}{0.8}

Although the steady-state model is arguably the simplest possible, it can still reproduce observations of debris disk evolution reasonably well \citep{Wyatt_2008}. However, it is well known that in reality, the size distribution of the dust grains deviates from the steady-state form. For example, the lower cutoff causes the development of wavy patterns in the size distribution just above the cutoff \citep{Thebault_etal_2003}. Numerical codes can be used to compute the evolution of disk masses and size distributions more realistically without relying on the steady-state assumption \citep[e.g.][]{Kral_etal_2013,Nesvold_etal_2013}. Numerical codes also allow to study the evolution of the spatial distribution of the dust or the interaction with embedded planets. These codes are especially valuable when modelling and interpreting observations of individual systems.

An important property of the steady-state model is the fact that the total mass at late times is independent of the initial disk mass (figure \ref{SteadyState_CollEvolution}), because $t_\mathrm{coll}(D_\mathrm{max},0)\propto 1/M_\mathrm{tot}(0)$. In other words, more massive disks are collisionally more active and therefore remove their mass faster. There exists a maximum mass (or, equivalently, a maximum fractional luminosity $f_\mathrm{max}$) a disk can have at late times if it evolves in steady-state. This fact can be used to observationally test whether the fractional luminosity of a system of known age is consistent with steady-state evolution \citep[e.g.][]{Fujiwara_etal_2013}. For systems that show higher fractional luminosity than what can be explained by the steady-state model, one needs to invoke stochastic events. For example, giant collisions between planetary bodies can add a stochastic element to debris disk evolution. Such collisions, akin to the Moon-forming event in the solar system, produce large amounts of dust and result in a spike in the fractional luminosity of the system \citep[e.g.][]{Jackson_Wyatt_2012,Johnson_etal_2012}. Dynamical instabilities, for example caused by migrating planets, can also cause a transitional spike in dust production. It is thought that such an event occurred in the solar system some 700\,Myr after its formation, thus called the late heavy bombardment (LHB). Evidence for an LHB comes from the dating of lunar craters. \citet{Gomes_etal_2005} proposed that the LHB was caused by the migration of the giant planets that destabilised the orbits of a large number of planetesimals. This resulted in an intense bombardment of the inner solar system and naturally led to an increased dust production, resulting in a zodiacal dust cloud $10^4$ times brighter than today \citep{Nesvorny_etal_2010}. Similar events could occur in other planetary systems, although observations suggest that LHB-like events are rare \citep{Booth_etal_2009}. An example of a system where the high fractional luminosity is thought to be due to an LHB-like event is the 1.4\,Gyr old star $\eta$~Corvi \citep{Lisse_etal_2012}.

\section{Detection and observation of debris disks}
In this section, I briefly describe how the presence of debris disks is inferred and how we can characterise debris disks with observations in various wavelength ranges.

\subsection{Thermal emission}
The dust grains in a debris disk are heated by constantly absorbing stellar photons. The absorbed energy is re-radiated as thermal radiation, typically in the infrared (IR). In steady-state, the absorbed energy equals the emitted energy and the temperature of the grain remains constants. This can be expressed with the following equation:
\begin{equation}\label{dust_grain_temperature}
\int\frac{L_{*,\nu}}{4\pi r^2}\left(\frac{D}{2}\right)^2\pi Q_\mathrm{abs}(\nu,D)\mathrm{d}\nu=\int\pi B_\nu(T)4\pi\left(\frac{D}{2}\right)^2Q_\mathrm{em}(\nu,D)\mathrm{d}\nu
\end{equation}
where $B_\nu(T)$ is the Planck function and $Q_\mathrm{em}(\nu,D)$ the emission efficiency. From Kirchhoff's law of thermal radiation we know that $Q_\mathrm{abs}=Q_\mathrm{em}$. Thus, if the absorption efficiency is known (for example from Mie theory, see section \ref{collcasc_radforces}), one can derive the temperature of the grains at a given distance from the star. Inversely, it is also possible to use equation \ref{dust_grain_temperature} to infer disk radii from observations of the spectral energy distribution (SED). Emission features of minerals are also encoded in the frequency dependence of $Q_\mathrm{abs}$, which allows, to a certain degree, the study of the mineralogy of debris disks using spectroscopic observations.

Debris disks can be detected from photometric observations in the IR. One simply measures the flux coming from a star (of known spectral type) in an infrared band and compares to the flux that would be expected from a stellar atmosphere model. An observed flux in excess of the expected flux is indicative of additional thermal emission from circumstellar dust grains. This technique allows the detection of disks without resolving them. Approximate disk radii can also be determined by fitting one or more black body\footnote{Sometimes also so-called modified black bodies are used. A modified black body's emissivity is reduced by a factor $(\lambda/\lambda_0)^{-\beta}$ for $\lambda>\lambda_0$ where $\lambda_0$ is comparable to the grain's size and $\beta$ is a power law index. This models the fact that grains do not emit efficiently at wavelengths longer than their size.} functions to the excess emission to determine the dust temperature. Observations of excess emission are particularly useful for statistical studies, for example to determine the fraction of stars surrounded by a disk.

In general, the more densely the SED of a system is sampled, the more can be said about the disk properties such as radius, grain properties or the presence of multiple belts. Still, to get a more complete picture of a system, it is necessary to conduct spatially resolved imaging. For example, there is a degeneracy between the radius of the belt and sizes and optical properties of the dust grain. This can be seen from equation \ref{dust_grain_temperature}. If one infers a certain dust temperature from the data, it is not clear whether this temperature is associated with grains that efficiently re-emit the absorbed energy (black body grains) and are close to the star, or with grains that inefficiently emit in the IR\footnote{For example small grains. As mentioned before, the emission efficiency of a grain is small at wavelengths larger than the grain itself.} and that are further away from the star. Resolved imaging can break this degeneracy and provide accurate disk radii as well as other parameters such as the disk width or inclination. For example, \citet{Booth_etal_2013} find that a number of debris disks around A-type stars resolved by \textit{Herschel} have radii up to 2.5 times larger than inferred from black body SED fitting. Depending on the resolution of the observations, imaging is also able to discover features such as gaps, clumps, warps or disk eccentricities that often hint to the presence of planets.

Observations at different wavelengths probe dust populations at different distances from their host star. Table \ref{table_dust_populations} shows the typical observation wavelengths for dust a different radii, following \citet{Su_Rieke_2014}. Different wavelengths also probe grains of different sizes: small grains are observed at shorter wavelengths than large grains\footnote{As a rule of thumb, the observation wavelength roughly corresponds to the grain size that is probed.}.
\begin{table}
\centering
\begin{tabular}{lccc}
\hline\hline
 dust class& $r$ [AU] & $T$ [K] & wavelength region \\ 
\hline
very hot  & $\ll1$& $\sim$1500& near-IR\\

hot & $\sim$1 &$\sim$300& mid-IR\\
& (terrestrial region) & & \\

warm & a few & $\sim$150& mid-IR\\
& (asteroid belt analogue) & & \\

cold & tens to hundreds & 20--100 &far-IR, (sub-)millimetre  \\
& (Kuiper belt analogue) & &  \\
\hline\hline
\end{tabular}
\caption{Different circumstellar dust populations with their temperature and associated observation wavelength \citep{Su_Rieke_2014}.}
\label{table_dust_populations}
\end{table}

In recent years, the \textit{Herschel Space Observatory} \citep{Pilbratt_etal_2010} has resolved various debris disks in the far-IR. Figure \ref{Herschel_PACS_Fomalhaut} shows an image of the Fomalhaut debris belt by \citet{Acke_etal_2012}, taken with the Photoconductor Array Camera and Spectrometer (PACS) aboard \textit{Herschel}. Fomalhaut is an A-type star with an age of $440\pm40$\,Myr \citep{Mamajek_2012}. It is exceptionally nearby, which allows a detailed study of the belt. More often, \textit{Herschel} just marginally resolves disks, i.e.\ the disk appears just slightly extended compared to the point-spread function (PSF) of the observations. By fitting a model image convolved with the PSF to the data, parameters such as disk radius or inclination can still be derived \citep[e.g.][]{Booth_etal_2013}.
\figuremacroW{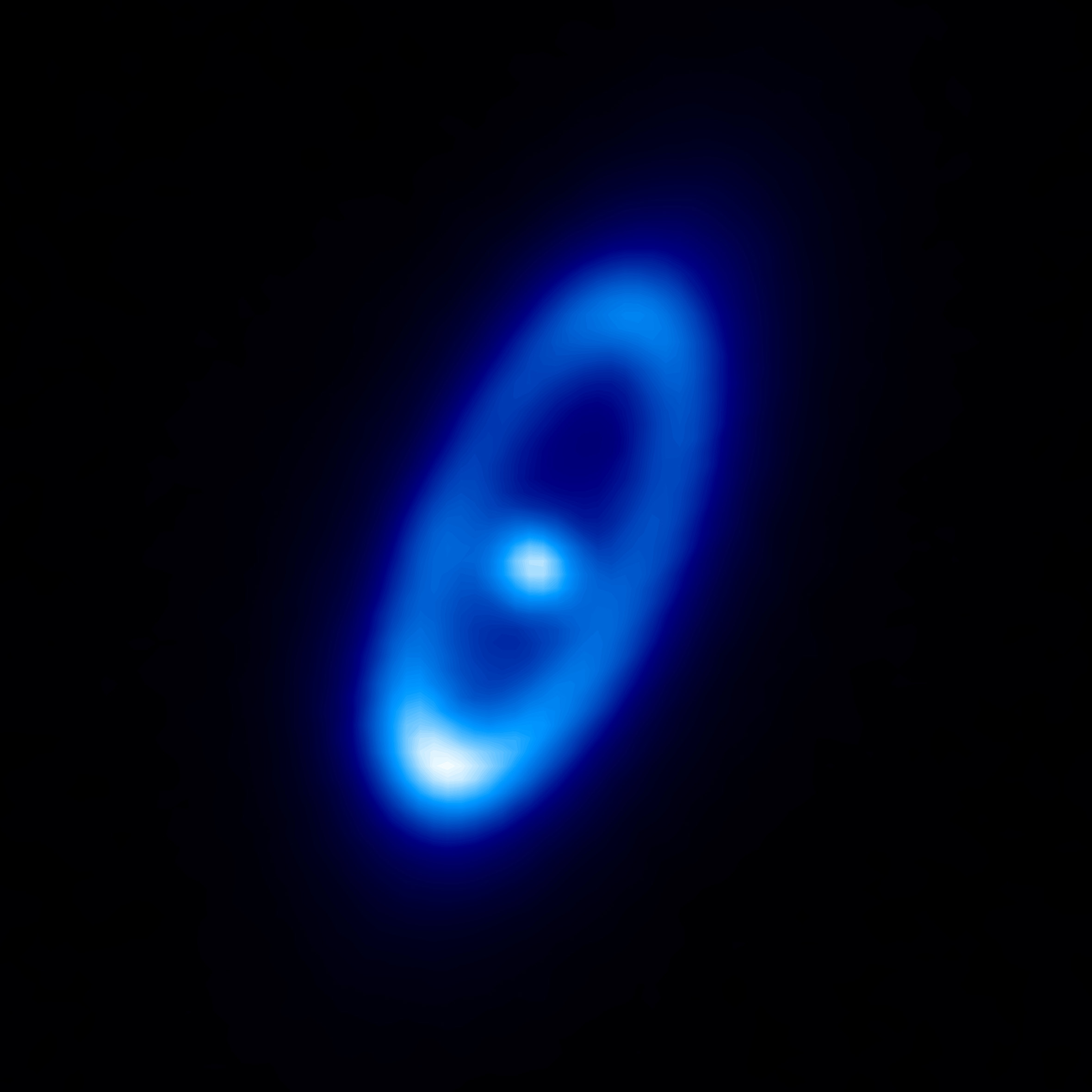}{\textit{Herschel} image of the Fomalhaut belt}{This image of the Fomalhaut debris belt by \citet{Acke_etal_2012} was obtained using \textit{Herschel}/PACS at 70\,$\mu$m, and is shown with a linear black-blue-white colour scale. The belt is eccentric, with the southern ansa being closer to the star and therefore warmer and brighter. Excess emission is also seen at the position of the star, possibly due to hot dust close to the star. Modelling of these \textit{Herschel} data indicates cometary dust grains and a high collisional activity. Image credit: ESA/Herschel/PACS/Bram Acke, KU Leuven, Belgium.}{0.6}

Another instrument that has delivered spectacular images of cold circumstellar dust is the \textit{Atacama Large Millimeter/submillimeter Array} (ALMA), an array of 12\,m telescopes\footnote{There are additional 7\,m telescopes arranged in a compact configuration to image extended structures. This is called the Atacama Compact Array (ACA).} located in the Atacama desert in Chile at 5000\,m altitude. The array functions as an interferometer, giving it an effective resolution corresponding to a telescope with a diameter equal to the longest baseline\footnote{The line connecting two telescopes of the array is called baseline.} used in an observation. The maximum baseline available is 16\,km, corresponding to a resolution of  6\,mas at 675\,GHz to 37\,mas at 110 GHz. In addition to high angular resolution, ALMA also provides exceptional high sensitivity and spectral resolution, making it one of the most powerful instruments available today. An advantage of ALMA observations at (sub-)millimetre wavelengths is that relatively large (millimetre-sized) grains are probed that are not strongly affected by radiation pressure (figure \ref{beta}). Therefore, these grains accurately trace the population of parent planetesimals.

\subsection{Scattered light}
The dust in debris disks not only absorbs and re-emits stellar light, it also scatters stellar photons. The specific luminosity of light scattered into a solid angle $\mathrm{d}\Omega$ about the direction $\Omega$ by a single dust grain of size $D$ can be written
\begin{equation}
\mathrm{d}L_\mathrm{sca,\nu}=\frac{L_{*,\nu}}{4\pi r^2}\cdot\left(\frac{D}{2}\right)^2\pi Q_\mathrm{sca}(\nu,D)\cdot\phi_\nu(\Omega)\mathrm{d}\Omega
\end{equation}
where $\phi_\nu$ is the phase function describing the directional dependance of the scattering process. For isotropic scattering, $\phi_\nu=(4\pi)^{-1}$. However, depending on the dust properties, scattering can be highly anisotropic. A common approach is to use the Henyey-Greenstein phase function which, depending on the value of the asymmetry parameter $g$, can describe backscattering, isotropic scattering and forward scattering.

Observing debris disks in scattered light is intrinsically difficult since one observes at wavelengths where the star is bright and outshines the disk. Usually, one employs a telescopic attachment, called coronagraph, to block direct stellar light and allow the faint scattered light to be observed. On the other hand, the shorter observing wavelength translates into better angular resolution. Figure \ref{Fomalhaut_HST} shows again the Fomalhaut debris belt, this time in scattered light as observed by \citet{Kalas_etal_2005} using the \textit{Hubble Space Telescope} (HST) with a coronagraph. This image revealed that the belt is eccentric (i.e.\ there is an offset between the centre of the belt and the stellar position), suggesting the presence of a perturbing planet, although other mechanisms have been proposed to explain the observed eccentricity (see paper~III). A planetary candidate, named Dagon\footnote{A.k.a.\ Fomalhaut~b.}, has subsequently been detected \citep{Kalas_etal_2008}, but constraints on its orbital parameters show that it cannot be the cause of the belt's eccentricity \citep{Kalas_etal_2013,Beust_etal_2014,Tamayo_2014}. Thus, a yet unseen planet (Fomalhaut~c) might be needed, especially because in paper~III we showed that gas-dust interactions are unlikely to be at the origin of the observed eccentricity.
\figuremacroW{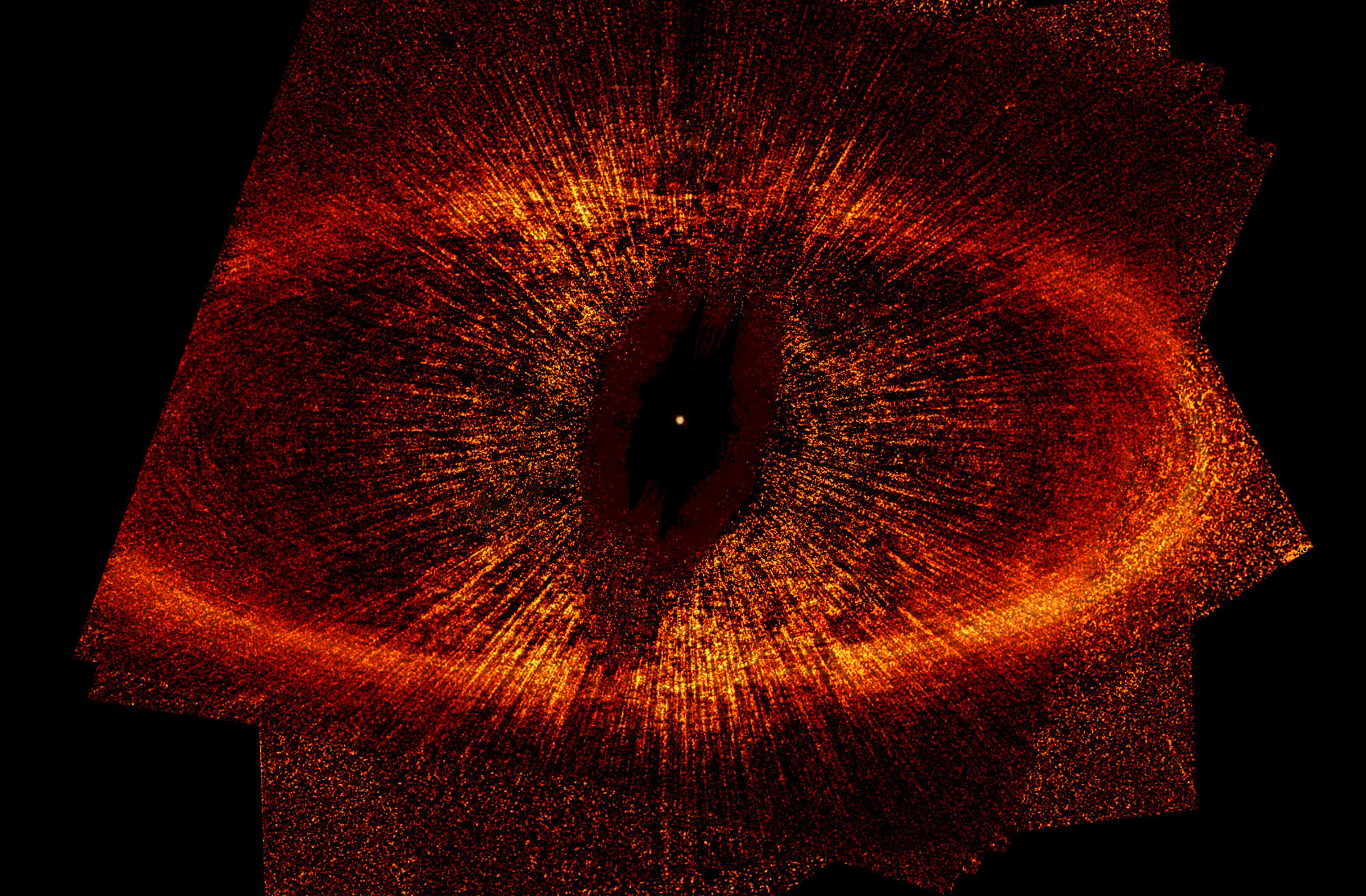}{HST image of the Fomalhaut belt}{This image of the Fomalhaut debris belt in scattered light by \citet{Kalas_etal_2005} was obtained using the HST at optical wavelengths with a coronagraph. The image revealed an offset between the stellar position and the centre of the belt, possibly due to a perturbing planet. Note that the `rays' visible in the image are instrumental artefacts. Image credit:  NASA, ESA, P.\ Kalas and J.\ Graham (University of California, Berkeley) and M.\ Clampin (NASA/GSFC).}{0.8}

\subsection{Microlensing}
It has been suggested to detect debris disks using microlensing, akin to microlensing detections of exoplanets \citep{Zheng_Menard_2005,Heng_Keeton_2009,Hundertmark_etal_2009,Sajadian_Rahvar_2015}. Debris disks surrounding the source or the lens star could in principle be detected. Microlensing would allow the examination of debris disks at kilo-parsec distances, for example in environments of different metallicities. To my knowledge, no debris disk has yet been detected by microlensing. It might become possible in the future, for example with the upcoming James Webb Space Telescope (JWST) \citep{Zheng_Menard_2005}.

\section{Interaction with planets}
As previously mentioned, planets can gravitationally perturb debris disks and cause a variety of features. For example, a planet can clear the region around its orbit from debris and cause a gap in the disk. It can also create different kinds of asymmetries such as clumps or warps. By observing such features, the presence of planets can be inferred that would otherwise be difficult to detect. From modelling, it is also possible to predict the perturbing planet's parameters such as its orbit or mass.

The debris disk around $\beta$~Pictoris is a good example of a system where the existence of a planet was predicted from disk features and subsequently confirmed. In the case of $\beta$~Pic, the inner disk appears to be warped by 4--5 degrees with respect to the main disk \citep{Heap_etal_2000}. The warp can be reproduced by models that include a massive planet with an inclined orbit \citep[e.g.][]{Augereau_etal_2001}. The predicted planet, $\beta$~Pic~b, was finally detected by \citet{Lagrange_etal_2010} by direct imaging. The planet orbits $\beta$~Pic at a distance of approximately 9\,AU \citep{Millar-Blanchaer_etal_2015} and has an estimated mass of roughly ten Jupiter masses \citep{Currie_etal_2013,Morzinski_etal_2015}.

Recently, \citet{Dent_etal_2014} imaged the $\beta$~Pic disk with ALMA and discovered a CO clump at a radial distance of $\sim$85\,AU. Since the CO lifetime is much shorter than the age of the system (due to photodissociation), the CO needs to be currently produced from collisions of cometary bodies. The clump corresponds to a region of enhanced collision rate, possibly due to a mean motion resonance with a yet unseen giant planet.

\section{The debris disks around $\beta$~Pictoris and Fomalhaut}\label{betaPic_Fomalhaut}
One of the best-studied debris disks is found around the already mentioned young \citep[$23\pm3$\,Myr,][]{Mamajek_Bell_2014} main-sequence A6 \citep{Gray_etal_2006} star $\beta$~Pictoris. Infrared excess suggesting circumstellar dust was discovered with the Infrared Astronomical Satellite (IRAS) in 1983. The disk was imaged for the first time shortly afterwards by \citet{Smith_Terrile_1984}. The image showed a nearly edge-on disk extending more than 400\,AU from the star. Since then, the $\beta$~Pic disk has been extensively studied. $\beta$~Pic is located at a distance of only 19.4\,pc \citep{vanLeeuwen_2007}. Thus, detailed observations of the disk structure are possible. Because of its young age, the system is generally regarded an analogue of the young solar system where the early stages of evolution in a planetary system can be studied.

As mentioned earlier, $\beta$~Pic harbours a giant planet ($\beta$~Pic~b) at an orbital distance of $\sim$9\,AU. Recently, \citet{Snellen_etal_2014} measured the spin velocity of $\beta$~Pic~b. They found the planet to exhibit a fast spin of 25\,km\,s$^{-1}$, consistent with expectations given its high (though uncertain) mass of $\sim$10 Jupiter masses. The presence of additional planets in the system is quite possible \citep{Dent_etal_2014}. In addition to the dust, the $\beta$~Pic disk also harbours circumstellar gas. This gaseous component is discussed in more detail in section \ref{beta_Pic_gas}. Papers~I and II present observations of gas emission from the $\beta$~Pic disk.

Another famous debris disk is found around the main-sequence A4 \citep{Gray_etal_2006} star Fomalhaut, shown in figures \ref{Herschel_PACS_Fomalhaut} and \ref{Fomalhaut_HST}. It has an age of $440\pm40$\,Myr \citep{Mamajek_2012} and is thus substantially older than $\beta$~Pic. At a distance of only 7.7\,pc \citep{vanLeeuwen_2007}, Fomalhaut is even closer than $\beta$~Pic. Fomalhaut harbours a prominent, cold dust belt (Kuiper belt analogue) at an orbital distance of $\sim$140\,AU \citep[e.g.][]{Boley_etal_2012}. This belt is remarkably active with a very high rate of dust production by collisions \citep{Acke_etal_2012}. As was described before, the belt is eccentric, as was first noted by \citet{Kalas_etal_2005}. This suggests the presence of a planet at the inner edge of the belt that would force the eccentricity. \citet{Kalas_etal_2008} observed a planetary candidate, Dagon, at optical wavelengths at an orbital distance of 120\,AU, approximately where the perturbing planet was expected. However, the nature of Dagon remains unclear. \citet{Kalas_etal_2008} detected Dagon at 0.6 and 0.8\,$\mu$m, but not at longer wavelengths. This is inconsistent with emission from a young, giant planet that is expected to be bright in the near-IR. \citet{Janson_etal_2012} presented a deep non-detection of Dagon at 4.5\,$\mu$m by the \textit{Spitzer Space Telescope}, rejecting the possibility that the observed flux originates from a planetary surface. They argued that Dagon might instead be a star light scattering, transient dust cloud, for example from a recent planetesimal collision. \citet{Lawler_etal_2015} recently also investigated this possibility. By analogy with the (young) Kuiper belt, these authors argued that collision probabilities could be high enough to make the dust cloud scenario viable. Other possibilities are a circumplanetary disk \citep{Kalas_etal_2008} or a dust producing swarm of planetesimals surrounding a Super-Earth \citep{Kennedy_Wyatt_2011}. It has even been suggested that Dagon is in fact a background neutron star \citep{Neuhauser_etal_2015}.

To make things more complicated, more recent observations showed that Dagon is on a highly eccentric ($e=0.8\pm0.1$) orbit, actually crossing the dust belt in projection. Thus, it is unlikely that Dagon is at the origin of the observed belt eccentricity \citep{Kalas_etal_2013,Beust_etal_2014,Tamayo_2014}. Therefore, there might be a second, yet unseen planet sculpting the belt \citep[e.g.][]{Faramaz_etal_2015}. Alternatively, gas-dust interactions have been proposed to drive the belt's eccentricity \citep{Lyra_Kuchner_2013}, but in paper~III we show that there is not enough gas in the Fomalhaut belt to make this mechanism work. Stellar encounters could also sculpt the belt---Fomalhaut is part of a wide triple system. \citet{Shannon_etal_2014} showed that secular interactions or close encounters could be responsible for the observed belt eccentricity.

In addition to the cold dust, Fomalhaut also harbours dust closer to the star---very hot, hot and warm dust has been inferred from interferometric observations and SED modelling. This inner dust might be connected to an asteroid belt analogue at $\sim$8--15\,AU that delivers dust to the inner regions by PR-drag \citep{Su_etal_2016}.

\section{Setting the solar system into context}
It seems prudent to briefly set the solar system into context with respect to debris disks observed around other stars. The solar system has its own debris disk in the form of the zodiacal dust in the terrestrial region, the asteroid belt at $\sim$3\,AU and the Kuiper belt hosting cold dust in the region between 30 and 50\,AU. A first thing to note is that until a few years ago, the dust levels present in the solar system were not possible to detect around other stars \citep{Wyatt_2008}. The sensitivity of \textit{Herschel} was at least approaching the fractional luminosity of the Kuiper belt \citep{Matthews_etal_2014}. ALMA should also be able to detect debris disks similar to the Kuiper belt \citep{Holland_etal_2009}. Concerning warm dust in the terrestrial region, the ground-based Large Binocular Telescope Interferometer (LBTI) has recently started operating and is expected to have a sensitivity equivalent to a few times the level of the zodiacal dust \citep{Roberge_etal_2012,Weinberger_etal_2015}.

It is also important to remember that the Kuiper belt in the young solar system was presumably some two orders of magnitude more massive than today \citep[e.g.][and references therein]{Chiang_etal_2007}. A possible explanation for the depletion of the Kuiper belt is the occurrence of the LHB. As discussed earlier, it is suggested that migrating giant planets destabilised the orbits of a large number of planetesimals and comets in the primordial Kuiper belt \citep{Gomes_etal_2005}. According to the model by \citet{Booth_etal_2009}, the fractional luminosity of the Kuiper belt was four orders of magnitude higher before the LHB than today.

Since it is at the moment difficult to detect dust levels as observed in the solar system, it is also difficult to say whether the solar system debris disk is typical or not. In the near future, the Hunt for Observable Signatures of Terrestrial planetary Systems (HOSTS) program on the LBTI will constrain the luminosity function of exozodiacal dust down to levels a few times the solar system's zodiacal cloud \citep{Weinberger_etal_2015}. In general, previous surveys have reported detection rates of debris disks of $\sim$10--30\% \citep{Matthews_etal_2014}.

Another interesting question is how the dust grain properties (e.g.\ shape, composition) in the solar system compare to other debris disks. For example, \citet{Donaldson_etal_2013} find that the grains in the outer disk around HD~32297 are similar to cometary grains found in the solar system: highly porous and consisting of silicates, carbonaceous material and water ice. \citet{deVries_etal_2012} observed olivine in the $\beta$~Pic debris disk. The abundance of the olivine compared to the total dust mass and the fact that it is magnesium-rich are strongly reminiscent of dust from primitive solar system comets. On the other hand, \citet{Beichman_etal_2011} find the dust around HD~69830 to be similar in composition to main-belt asteroids in the solar system.



\chapter{Gas in debris disks}
Debris disks are, almost by definition, gas-poor. However, some debris disks show observable amounts of gas beside the dust. In some cases, the gas might be remnant from the protoplanetary phase, while in other cases the gas is most probably of secondary origin. By giving a brief overview of the topic and explaining why this gas is interesting to study, this chapter will set papers~I, II and III into context.

\section{Why study gas in debris disks?}
Only a small fraction of the debris disk population shows observable amounts of gas. In addition, for gaseous debris disks, the dust mass is usually larger than the gas mass. Still, studying the gas is interesting for a number of reasons. First of all, gas of secondary origin is somehow derived from the dust. For example, grain-grain collisions \citep{Czechowski_Mann_2007} or photodesorption \citep{Chen_etal_2007,Grigorieva_etal_2007} could act as gas sources. Volatile-rich colliding comets can also produce gas \citep{Zuckerman_Song_2012}. In these cases, studying the gas composition can give us information about the dust composition. Since the dust is derived from leftover planetesimals, there is a link to the composition of the building blocks of exoplanets. Gas derived from sublimating or colliding comets can allow us to study some aspects of cometary bodies.

In general, understanding the gas producing mechanism will help us to understand the processes occurring in a debris disk and what the connection between the gas, dust, planets and the host star exactly is. If the gas distribution can be spatially resolved, asymmetries can indicate the presence of exoplanets. For example, in the $\beta$~Pic system, the clumpy structure of the CO gas \citep{Dent_etal_2014} hints to the presence of a hitherto unseen giant planet. With high enough spectral resolution, gas observations also allow to study the disk dynamics (unless the disk is face-on). We can also learn about the physical state of the disk by measuring quantities such as the gas temperature.

Gas can also dynamically influence the dust and cause the formation of features such as narrow and eccentric dust belts even in the absence of planets \citep{Lyra_Kuchner_2013}. This is yet another example of the interlocking between gas and dust in debris disks and another motivation to study the gaseous component.

\section{Physics of debris disk gas}
In this section, I discuss some aspects of the physics of debris disk gas. A proper understanding of these concepts is important when interpreting observations of gaseous debris disks.
\subsection{Line emission}
When observing gas emission from debris disks, we measure the amount as well as the spectral and angular distribution of photons emitted or scattered by the gas. From such a measurement, we would typically like to infer the mass, temperature and spatial distribution of the observed species. As we will see shortly, this is not a trivial task. To understand this, we consider how light travels through the gas and how the photons arise in the first place.

\subsubsection{Radiative transfer}
We follow here the formalism described by \citet{Rybicki_Lightman_2007}. The specific intensity $I_\nu$ is defined by 
\begin{equation}
\mathrm{d}E=I_\nu\mathrm{d}A\mathrm{d}\Omega\mathrm{d}t\mathrm{d}\nu
\end{equation}
Here, $\mathrm{d}E$ is the amount of energy in a frequency interval $\mathrm{d}\nu$ passing through an area $\mathrm{d}A$ in a time interval $\mathrm{d}t$ into the solid angle $\mathrm{d}\Omega$. We also define the monochromatic emission coefficient $j_\nu$ by
\begin{equation}
\mathrm{d}E=j_\nu\mathrm{d}V\mathrm{d}\Omega\mathrm{d}t\mathrm{d}\nu
\end{equation}
Thus, $j_\nu$ tells out how much energy $\mathrm{d}E$ is emitted per unit time $\mathrm{d}t$ and volume $\mathrm{d}V$ into an element of solid angle $\mathrm{d}\Omega$ within a frequency interval $\mathrm{d}\nu$. Let us now consider a beam into a direction parametrised by the variable $s$. Radiation is not only emitted, but also absorbed. This can be described by the absorption coefficient $\alpha_\nu$, defined by
\begin{equation}
\mathrm{d}I_\nu^\mathrm{abs}=-\alpha_\nu I_\nu\mathrm{d}s
\end{equation}
where $\mathrm{d}I_\nu^\mathrm{abs}$ is the amount of radiation removed by absorption. From this, it follows that the equation of radiative transfer is given by
\begin{equation}\label{eq:radiative_transfer}
\frac{\mathrm{d}I_\nu}{ds}=-\alpha_\nu I_\nu+j_\nu
\end{equation}
Thus, for each length interval, radiation is removed by absorption and added by emission. This equation governs the transport of radiation within the gas. In general, equation \ref{eq:radiative_transfer} needs to be solved numerically. For simple situations, analytical solutions are possible. For example, if only absorption is present (i.e.\ $j_\nu=0$), one finds that
\begin{equation}
I_\nu(s)=I_\nu(0)e^{-\tau_\nu(s)}
\end{equation}
where $\tau_\nu(s)=\int_0^s\alpha_\nu(s')\mathrm{d}s'$ is the optical depth. The next step is now to connect this description of the radiative transfer to microscopic properties of the gas.

\subsubsection{Level population}
An atom\footnote{The same arguments apply for ions or molecules.} emits a photon when changing from a higher to a lower energy level. The photon carries away an energy equal to the energy difference between the two levels. It is instructive to consider a two-level-system. The generalisation to a multi-level-system is straightforward. We then write $j_\nu$ as
\begin{equation}\label{j_nu_Einstein}
j_\nu=\frac{h\nu}{4\pi}n_2A_{21}\phi(\nu)
\end{equation}
where $A_{21}$ is the Einstein coefficient for spontaneous emission and gives the transition probability per unit time to go from level 2 to level 1. $n_2$ is the number density of atoms in the upper level and $\phi(\nu)$ describes the line shape and is normalised such that $\int\phi(\nu)\mathrm{d}\nu=1$. The absorption coefficient can also be expressed in terms of microscopic properties:
\begin{equation}\label{alpha_nu_Einstein}
\alpha_\nu=\frac{h\nu}{4\pi}\phi(\nu)(n_1B_{12}-n_2B_{21})
\end{equation}
Here $B_{12}$ is the Einstein coefficient for a transition from the lower to the upper level via absorption of a photon: $B_{12}\bar{J}$ is the transition probability per unit time, where $\bar{J}=\int J_\nu\phi(\nu)\mathrm{d}\nu$ and $J_\nu=\frac{1}{4\pi}\int I_\nu\mathrm{d}\Omega$ is the mean intensity. Similarly, $B_{21}$ is the Einstein coefficient for a transition form the upper to the lower level via emission of a photon stimulated by the radiation field, i.e.\ not spontaneously. $B_{21}\bar{J}$ is the transition probability per unit time for stimulated emission, and is zero if there is no ambient radiation field. Stimulated emission is conveniently treated as `negative absorption'.

From equations \ref{j_nu_Einstein} and \ref{alpha_nu_Einstein}, we see that we need to know the fractions of atoms in the upper and lower level respectively. Assuming that these fractions do not change with time, we can write the equations of statistical equilibrium where excitation processes populating a level balance de-excitation processes that depopulate the same level. We consider excitation by collisions and by absorption of a photon and de-excitation by spontaneous and stimulated emission as well as collisions:
\begin{align}
\frac{\mathrm{d}n_1}{\mathrm{d}t}&=-n_1(B_{12}\bar{J}+K_{12}n_\mathrm{col})+n_2(A_{21}+B_{21}\bar{J}+K_{21}n_\mathrm{col})=0 \label{SE_1}\\
\frac{\mathrm{d}n_2}{\mathrm{d}t}&=n_1(B_{12}\bar{J}+K_{12}n_\mathrm{col})-n_2(A_{21}+B_{21}\bar{J}+K_{21}n_\mathrm{col})=0\label{SE_2}
\end{align}
Here, $n_\mathrm{i}n_\mathrm{col}K_\mathrm{ij}$ is the rate of a collisionally induced transitions from level $i$ to level $j$, and $n_\mathrm{col}$ is the number density of the collisional partner (for example electrons). Note that in general, the $K_\mathrm{ij}$ are functions of the kinetic temperature. For an atom with $n$ levels, one writes $n$ equations of this type. We see that the statistical equilibrium equations depend themselves on the radiation field\footnote{The overall radiation field includes external contributions such as stellar radiation, thermal dust radiation or the cosmic microwave background.}, i.e.\ equations \ref{eq:radiative_transfer}, \ref{SE_1} and \ref{SE_2} are coupled. This is the essence of the problem when interpreting observations of gas emission. Various approximations and strategies exist to solve the problem.

First, in the optically thin case where $\tau_\nu\approx0$, one can assume that all emitted photons can leave the gas. In this case, no radiative transfer is necessary and one only needs to solve the equations of statistical equilibrium. This is the original approach of the \texttt{ONTARIO} code by \citet{Zagorovsky_etal_2010}, which was developed to model gas emission from debris disks. In papers~I and II, an updated version of the code including an approximate treating of radiative transfer is used.

Another simplification occurs if local thermodynamic equilibrium (LTE) applies to the gas. In this case, the energy levels are populated according to the Boltzmann distribution and one can forget about the statistical equilibrium equations:
\begin{equation}
\frac{n_1}{n_2}=\frac{g_1}{g_2}\exp\left(\frac{h\nu}{kT}\right)
\end{equation}
where $g_\mathrm{i}$ is the statistical weight of level $i$ and $T$ is the temperature. LTE occurs if the atom's energy levels are efficiently `coupled' to the kinetic energy of the gas. In other words, collisions should occur frequently enough---LTE applies in dense media. This can be quantified by defining a critical density of the colliding bodies where the rate of collisional de-excitation equals spontaneous decay:
\begin{equation}
A_{21}=n_\mathrm{col}^\mathrm{crit}K_{21}
\end{equation}
and thus $n_\mathrm{col}^\mathrm{crit}=A_{21}/K_{21}$. For LTE, the density of colliding bodies should be larger than the critical density. While the LTE assumption provides significant simplification, care must be taken when employing it to the generally low-density environments of debris disks. For example, \citet{Dent_etal_2014} assume LTE to derive a CO mass from their ALMA observations of the CO (3--2) transition. However, \citet{Matra_etal_2015} subsequently showed by means of non-LTE calculations that the mass derived by \citet{Dent_etal_2014} could be too small by four orders of magnitude.

For cases where neither $\tau_\nu\approx0$ nor LTE holds, other approaches are necessary. For example, the \texttt{RADEX} code \citep{vanderTak_etal_2007} is a simple non-LTE radiative transfer code that was applied in paper~III. \texttt{RADEX} employs a so-called escape probability formalism to decouple the statistical equilibrium calculation from the radiative transfer.

\subsection{Thermal balance}
The gas temperature is determined by various heating and cooling processes as outlined in \citet{Zagorovsky_etal_2010}. For example, \texttt{ONTARIO} considers photoelectric heating by dust grains, photoionisation of gas by stellar radiation and gas-grain collisions. Cooling can occur through line emission, free-free radiation and radiative recombination. In order to find the thermal structure, one can assume that heating and cooling processes balance locally. Note that the thermal balance and the level populations are coupled: emission line cooling depends on the level populations, which in turn depend on the temperature through the rate coefficient for collisional (de-)excitation.

\subsection{Ionisation structure}
Energetic photons from the host star (in particular if the star is of early type) or the interstellar radiation field and cosmic rays can ionise atoms. To determine the ionisation structure, one equates the ionisation rate (which depends on the ionising radiation field at the considered point) and the recombination rate (which depends on the electron density and the temperature):
\begin{equation}
\int\frac{L_{*,\nu}}{4\pi r^2}\frac{1}{h\nu}\sigma_\nu^\mathrm{X}\mathrm{d}\nu=n_{\mathrm{X}^+}n_\mathrm{e}\alpha(T)
\end{equation}
where $\sigma_\nu^\mathrm{X}$ is the ionisation cross-section of a specie $\mathrm{X}$, $n_{\mathrm{X}^+}$ and $n_\mathrm{e}$ are the number densities of ions and electrons respectively and $\alpha(T)$ is the recombination coefficient. In papers~I and II, the \texttt{ONTARIO} code is used to calculate the ionisation fraction of carbon in order to estimate the total carbon mass from observations of C$^+$. In paper~III, the ionisation is calculated with a simple, customised routine.

\subsection{Radiation pressure}
We have already discussed how radiation pressure acts on dust grains. Gas is also subject to radiation pressure. Every time an atom absorbs or scatters a photon, there is a net momentum transfer. One can write the radiation pressure force as
\begin{equation}\label{f_rad_gas}
F_\mathrm{rad}=\frac{\mathrm{d}p}{\mathrm{d}t}=\int\frac{L_{*,\nu}}{4\pi r^2c}\sigma(\nu)\mathrm{d}\nu
\end{equation}
with $\sigma(\nu)$ the frequency-dependent cross-section of the atom. It can be written
\begin{equation}
\sigma(\nu)=\sum_{i<j}\sigma_{ij}(\nu)
\end{equation}
with $\sigma_{ij}(\nu)$ the cross-section for a transition from the lower level $i$ to the higher
\begin{WrapText}{A closer look at scattering by atoms}
Because of the conservation of energy, and since the atom gains kinetic energy in the scattering process, the wavelength of the scattered photon is different from the wavelength of the incoming photon. Applying the formula familiar from Compton scattering, the change in wavelength is $\Delta\lambda=\frac{h}{mc}(1-\cos\theta)$ with $m$ the mass of the atom and $\theta$ the scattering angle. For a sodium atom, $\Delta\lambda\approx6\times10^{-8}$\,nm, which is clearly negligible. 
\end{WrapText}
level $j$. This means that different species can experience very different radiation pressure, depending on whether or not they have strong transitions in a wavelength range where the star emits a lot of photons. For example, in the $\beta$~Pic disk, radiation pressure on neutral sodium is very strong, $\beta(\mathrm{Na})\approx360$, while for singly ionised sodium $\beta(\mathrm{Na}^+)=0$ \citep{Fernandez_etal_2006}.

\section{The gas in the $\beta$~Pictoris system}\label{beta_Pic_gas}
Arguably the best-studied gaseous debris disk is found around $\beta$~Pictoris. Interestingly, the gas around $\beta$~Pic was noticed before the discovery of the circumstellar dust. \citet{Slettebak_1975} reported the observation of ``peculiar Ca\,II H and K'' absorption lines that ``are evidently of interstellar or circumstellar origin''. \citet{Slettebak_Carpenter_1983} classified $\beta$~Pic as a `shell star': they imagined the absorption lines to arise in a circumstellar gas shell. Today we know that the gas lies in a circumstellar disk instead.

\subsection{Origin, composition and spatial distribution of the $\beta$~Pic circumstellar gas}
The gas around $\beta$~Pic has been a subject of intense studies over the last decades. The edge-on orientation of the disk allows the study of the gas in absorption against the star. Absorption lines revealed the existence of two distinct components of the gas \citep[e.g.][]{Roberge_2014}: a stable component producing narrow, unvarying absorption features at the velocity of the star, and a fluctuating component producing velocity-shifted absorption features varying on timescales of hours or days. It is believed that the stable component corresponds to the bulk of the circumstellar gas. On the other hand, the varying component is due to so-called falling evaporating bodies (FEBs). These are essentially star-grazing planetesimals or comets (`exocomets'). Close to the star, they undergo sublimation and produce various metallic ions (observed as e.g.\ Ca\,II, Mg\,II, Al\,III) that are seen when the FEB crosses the line of sight \citep[e.g.][]{Beust_Valiron_2007,Beust_2014}. Recently, \citet{Kiefer_etal_2014} showed that the observed
\begin{WrapText}{C$^+$ vs.\ C\,II}
Although many astronomers seem to use the two ways of denoting an ion and a spectrum interchangeably (e.g.\ C$^+$ and C\,II), they have actually different meanings and care should be taken when employing them. Following J.\ Ferland\footnote{\url{http://www.ferland.org/cloudy/cii_vs_cp.htm}}, C$^+$ denotes singly ionised carbon. C\,II on the other hand denotes a set of photons. It stands for the spectrum produced by carbon with one electron removed, for example from collisional excitation of C$^+$, but also from recombination of C$^{2+}$. Thus, C\,II can be produced from both C$^+$ and C$^{2+}$.
\end{WrapText}
FEBs can be divided into two different families of exocomets. One of the families is suggested to be trapped in a mean motion resonance with a massive planet, possibly $\beta$~Pic~b. While FEBs certainly produce gas close to the star, in paper~I we argue that FEBs are likely not the source of the bulk of the gas.

Regarding the composition of the gas, various species (e.g.\ C, O, Na, Al, Ca, Fe,\dots) have been observed. CO is also known to be present in the disk. An inventory of the gas is given in \citet{Roberge_etal_2006}.

Refractory elements are observed to have solar abundances relative to each other \citep{Lagrange_etal_1995}. However, the gas differs from a solar composition in a number of ways. First, no hydrogen is detected, neither in atomic nor molecular form \citep{Freudling_etal_1995,LecavelierdesEtangs_etal_2001}. Second, relative to solar abundances, carbon (and oxygen) are observed to be strongly overabundant with respect to other metals such as Fe: \citet{Roberge_etal_2006} find a C/Fe ratio 16 times the solar value, and the results from papers~I and II as well as \citet{Brandeker_2011} suggest an even stronger overabundance. This has important dynamical consequences. Indeed, the stability of the circumstellar gas was a long-standing puzzle. Observations show that the gas\footnote{Excluding the velocity-shifted gas from FEBs.} is at rest relative to the star, however, some species experience very strong radiation pressure and are thus expected to be blown out. For example, neutral sodium is seen in Keplerian rotation \citep{Olofsson_etal_2001} although radiation pressure exceeds gravity by more than a factor 300. However, \citet{Fernandez_etal_2006} noticed that all species strongly affected by radiation pressure are largely ionised. The result is the coupling of the species into a single fluid by Coulomb interactions. Thus, one can assign an effective radiation pressure to the fluid given by
\begin{equation}
\beta_\mathrm{eff}=\frac{\sum_i\beta_i\rho_i}{\sum_i\rho_i}
\end{equation}
with $\rho_i$ the density of specie $i$. Depending on the composition of the fluid, self-braking occurs. \citet{Fernandez_etal_2006} find that for solar abundances, the mechanism does not work. However, a gas disk enhanced in carbon is self-braking, since carbon is appreciably ionised (thus allowing Coulomb interactions) and is not affected by radiation pressure ($\beta(C)\approx0$). Thus, carbon acts as an efficient braking agent in the $\beta$~Pic disk, as was also confirmed by \citet{Brandeker_2011}.

The carbon overabundance explains the stability of the gaseous disk, however, one wonders how the overabundance arises in the first place. This leads us to discuss the origin of the circumstellar gas. First of all, it is generally believed that the gas is of secondary origin (i.e.\ currently produced) rather than primordial (i.e.\ leftover from the protoplanetary phase). Indeed, the dynamical lifetime of the gas is short compared to the age of the system \citep{Fernandez_etal_2006}. The presence of CO \citep{Dent_etal_2014} also argues for a secondary origin, since CO is quickly photodissociated, on timescales of the order of $\sim$120 years. Thus, it needs to be constantly replenished. The correlation between the spatial distribution of the gas and dust \citep{Brandeker_etal_2004,Nilsson_etal_2012} is suggestive of gas originating from the dust. The gas might thus be produced through photodesorption by UV radiation. \citet{Chen_etal_2007} modelled the production of sodium through photodesorption. Their model predicts production rates consistent with the observed amount of sodium. The spatial distribution of sodium does not fit observations, however, this problem might be mitigated by transport of sodium via radiation pressure. Whether other species could also be produced by photodesorption is difficult to assess because of the lack of laboratory data. In another study, \citet{Grigorieva_etal_2007} considered UV photodesorption from icy grains. CO photodesorption from icy grains \citep{Oberg_etal_2009} with subsequent photodissociation could be a source for the observed C and O gas in the disk. Thus, if photodesorption is the main process producing the gas in the $\beta$~Pic disk, one might expect an enrichment in C and O, as is indeed observed.

Another gas production mechanism is collisional evaporation of dust grains \citep{Czechowski_Mann_2007}. For evaporation to occur, high collisional velocities are needed. For bound grains, this happens only very close to the star. However, collisions among bound grains can produce smaller grains with $\beta>0.5$. These unbound grains are accelerated outwards. On their way, they may collide with grains further out in the disk and produce even more unbound grains, resulting in an avalanche of so-called $\beta$-meteoroids. Sufficiently accelerated $\beta$-meteoroids can induce evaporation of impacted dust grains. Radiation pressure accelerates unbound grains to an asymptotic velocity, the value of which depends on $r_0$, the radius at which the $\beta$-meteoroid was released. Thus, a prediction of the model is the existence of a maximum radius within which $\beta$-meteoroids need to be produced in order to be sufficiently accelerated to induce evaporation. For disks with large gaps, gas production by grain-grain collisions is thus not viable. \citet{Czechowski_Mann_2007} conclude that collisional vaporisation may be an important source of gas in the $\beta$~Pic disk. In this case, the gas composition should be similar to the dust composition. \citet{Czechowski_Mann_2007} also briefly discuss two other possible gas production mechanisms related to the dust: stellar wind-dust interactions and sputtering from dust surfaces.

\citet{Xie_etal_2013} presented a model aimed to explain the C and O overabundance. They conclude that two different scenarios can produce a gas enriched in C and O. Either the gas is produced at solar abundances (by collisional evaporation) with subsequent preferential depletion of elements other than C and O by radiation pressure\footnote{Radiation pressure on C and O is negligible.}. Or C and O are preferentially produced, for example via photodesorption from C/O-rich icy grains. In the latter case, \citet{Xie_etal_2013} predict the circumstellar gas to viscously accrete onto the star under the influence of the MRI (see section \ref{protoplanetary_disks}). In paper~I this prediction was tested by fitting a simple accretion disk model to the observed line profile of C\,II. Since the accretion model did not produce a good fit, the data presented in paper~I seem to suggest a preferential depletion explanation for the C/O overabundance. However, more observation and modelling efforts are needed to determine the gas producing mechanism. The analysis presented in paper~I also showed that most of the C gas is located at $\sim$100\,AU or beyond, further strengthening the case that the bulk of the gas is produced from the dust rather than being supplied by FEBs.

The recent ALMA observations by \citet{Dent_etal_2014} detected and spatially resolved CO in emission. The spatial distribution of CO is highly asymmetric, with a clump at 85\,AU in the southwest side of the disk. Interestingly, a similar asymmetry was tentatively detected in C in paper~I. Since the lifetime of CO is very short compared to the age of the $\beta$~Pic system, the CO needs to be produced currently, potentially from colliding cometary bodies. Photodissociation of CO might then provide a natural explanation for the observed C and O enrichment of the gas. In paper~II, we analyse observations of C\,II 158\,$\mu$m and O\,I 63\,$\mu$m emission from \textit{Herschel}/PACS. The measured oxygen emission was much stronger than expected and turns out to be optically thick. In fact, it is challenging to explain the O\,I emission without postulating that a region of high density exists where the O atoms are efficiently excited. This high density region might be in the form of a clump, possibly corresponding to the CO clump.

\subsection{Outlook}
Although our understanding of the $\beta$~Pic circumstellar gas has improved considerably over the last years and decades, a lot of aspects remain unclear. In particular, the production mechanism remains to be determined. An upcoming analysis of our ALMA observations of C\,I emission will be directly comparable to the aforementioned CO observations and tell us if indeed all the carbon (and oxygen) is produced from photodissociation of CO. These data will also help to understand the strong oxygen emission reported in paper~II.

If the gas is accreting, the detection of accretion signatures in the spectrum of $\beta$~Pic would be another useful piece of information, allowing to determine the gas accretion rate and chemical abundances \citep{Xie_etal_2013}. New, more detailed models of the production and evolution of the gas \citep[e.g.][]{Kral_etal_2015} will also be essential to progress.

\section{No gas in the Fomalhaut system?}
The debris disk around Fomalhaut was introduced in section \ref{betaPic_Fomalhaut}. Here I discuss what we know about the gas content of the Fomalhaut disk and what conclusions we are able to draw from this knowledge.
\subsection{Gas-dust interactions---a way to form a narrow and eccentric dust belt}
The Fomalhaut belt is remarkably narrow. \citet{Boley_etal_2012} thus suggested that the edges of the belt are shaped by shepherding planets. The eccentricity of the belt also suggests a planet at the inner edge of the disk. The planetary candidate Dagon was imaged at an orbital distance where such a perturbing planet would be expected, but we now know from its orbit that it cannot force the belt's eccentricity.

An alternative way to organise dust into narrow belts is via gas-dust interactions, more precisely a clumping instability \citep{Klahr_Lin_2005,Besla_Wu_2007}. Basically, the instability occurs as follows: starting with some enhancement in the dust density (for example from a major collision), photoelectric heating increases the temperature, and thus the pressure of the gas locally. This means that gas orbiting just inside the dust enhancement will orbit faster. Dust grains coupled to the gas thus move outwards (towards the heated region, due to a tailwind), or at least their headwind is reduced\footnote{In general, gas in a disk is expected to orbit with sub-Keplerian speed because of the negative pressure gradient in the radial direction.}, thus slowing down their inward migration. Similarly, gas just outwards of the heated region is slowed down. Dust grains in this region therefore move inward (headwind), towards the heated region. The additional dust further increases the heating, attracting even more dust, i.e.\ there is a positive feedback, or instability, concentrating the dust into a narrow belt. \citet{Lyra_Kuchner_2013} presented the first 2D simulations of this instability. They find that some dust belts develop eccentricities, reminiscent of the Fomalhaut belt. They also find that a necessary condition for the instability to develop is a dust-to-gas ratio $\epsilon \lesssim1$. Thus, it is necessary to determine $\epsilon$ for the Fomalhaut system in order to judge whether the photoelectric instability could cause the observed morphology of the belt.

\subsection{Upper limits on the gas content of the Fomalhaut dust belt}
As we have seen, the presence of gas in the Fomalhaut belt could have important consequences and might make it needless to invoke unseen planets to explain the dust belt's morphology. However, gas has not (yet) been detected in the Fomalhaut debris disk. \citet{Liseau_1999} presented upper limits on the CO (1--0) and (2--1) flux obtained from the \textit{Swedish ESO Submillimetre Telescope} (SEST). \citet{Matra_etal_2015} used an ALMA non-detection of the CO (3--2) transition and non-LTE modelling to place an upper limit of $4.9\times10^{-4}$\,M$_\oplus$ on the CO mass co-spatial with the dust belt. In this section, I will present an independent, previously unpublished analysis of the same ALMA data.

The ALMA data under consideration where first presented by \citet{Boley_etal_2012}. The data are publicly available from the ALMA archive. I used two different methods to estimate an upper limit on the CO (3--2) flux: a Monte Carlo (MC) method working directly with the visibilities\footnote{The visibility is the basic quantity measured by an interferometer like ALMA. It essentially corresponds to the Fourier transform of the sky surface brightness.} and an estimation based on a CLEANed\footnote{In other words, applying the CLEAN algorithm to image the visibilities.} image.

The data were transformed to the barycentric reference frame and the continuum subtracted. This was done using the \textit{Common Astronomy Software Applications} \citep[\texttt{CASA,}][]{McMullin_etal_2007}.
\subsubsection{Monte Carlo approach}
The basic idea of the MC approach is to generate new realisations of the data by characterising the noise in the data, generating random noise with the same properties and add this noise to the original data. Then, a model is fitted to this new data realisation. This is done many times in order to get a distribution of fitted parameters from which the error on the parameter can be estimated. This procedure can be interpreted as repeating the original measurement many times.

A complication arises from the fact that adjacent frequency channels in the data are correlated. Thus, the generated random noise needs to be correlated as well. This is achieved as described in Appendix~A of paper~I.

We are interested primarily in gas co-spatial with the dust in order to constrain the possibility of gas-dust interactions. Thus, the \citet{Boley_etal_2012} model of the dust spatial distribution is taken as a description of the CO (3--2) `emission density field' (energy emitted per volume). This density field is then projected onto the sky: figure \ref{Fomalhaut_belt}. This map needs to be multiplied by the primary beam\footnote{The primary beam describes the relative sensitivity of the observations in the observed field of view.} before being fit to the observed visibilities: figure \ref{Fomalhaut_belt_pbapplied}.

In addition to the relative line intensity across the map, we also need to know the radial velocity in each point of the sky to correctly account for red and blue shifts of the CO line. Thus, I calculated the radial velocity as a function of sky coordinates assuming Keplerian rotation of the Fomalhaut belt, taking also the systematic velocity of Fomalhaut into account. The rotation sense of the belt is assumed to follow the observed motion of Dagon. However, since it is not known which part of the belt is residing inside the sky plane, we have to consider two separate cases.

I used \texttt{UVMULTIFIT} \citep{Marti-Vidal_etal_2014} to fit the model described
\begin{WrapText}{Confidence intervals}
Confidence intervals (or upper/lower limits) are commonly encountered in science. Usually, the result of a measurement is given with error bars that denote a confidence interval at a certain confidence level. However, a confidence interval at, say, 95\% confidence level, does \emph{not} mean that the true value of the measured parameter lies within the confidence interval with 95\% probability. Consider the following example \citep{Barlow_1989}. A weight has a $1\sigma$ precision of 0.7\,g. One measures the weight of a bowl as $10\pm0.7$\,g. One then measures the weight of the bowl containing a sample as $11\pm0.7$\,g. Thus, the weight of the sample is $1\pm1$\,g. Naively, one might now say that there is a 32\% probability that the true weight of the sample is outside the error bars. Thus, there would be a 16\% chance that the weight of the sample is negative, which is obviously not possible. Either the confidence interval contains the true value, or it does not, but it is not a question of probability. The right interpretation is to say that if one repeats the measurement many times, in 68\% of the case will the confidence interval contain the true value. One can also say that if the true value is equal or exceeds the upper boundary of the confidence interval, then there is only a 16\% chance of getting the measured value. Note that for Bayesian statistics, the discussion is different.
\end{WrapText}
above to the measured visibilities. In practice, \texttt{UVMULTIFIT} considers the map of figure \ref{Fomalhaut_belt_pbapplied} as a collection of point sources. The emission of each point source is modelled with a Gaussian line profile with $\mathrm{FWHM}=2$\,km\,s$^{-1}$ and shifted according to the radial velocity at the position of the point source. The flux of each point source is scaled relative to the others according to the relative intensities of figure \ref{Fomalhaut_belt_pbapplied}. The only free parameter of the fit is a global scaling applied to all point sources simultaneously.

In order to test this procedure, a number of sanity checks were performed. As a zeroth order test, I fitted the absolute scaling of the model shown in figure \ref{Fomalhaut_belt_pbapplied} to the \emph{continuum} visibility data and compared the resulting total flux with the value published by \citet{Boley_etal_2012}. The agreement is very good (within a few percent). I also used the \texttt{CASA} task \texttt{simobserve} to generate simulated ALMA observations of a point source emitting a Gaussian line profile. I fitted the peak flux using \texttt{UVMULTIFIT} and estimated the error bars using the MC approach described above. These error bars can be compared to the errors provided directly by the \texttt{UVMULTIFIT}  software and to the noise estimated from a CLEANed image of the simulated data. These three different estimates of the error bars are within a factor of $\sim$2. Finally, to test a scenario closer to the non-detection of the CO line, I fitted the aforementioned point source to simulated data of a completely empty sky  (i.e. only noise). The derived upper limits on the flux of the point source are again within a factor of $\sim$2 for the three different methods, with the MC method giving the smallest error bars and the noise from the CLEANed image giving the largest.

Applying the MC procedure to the Fomalhaut ALMA data, I get an upper limit on the CO (3--2) emission from the belt. A problem is the distribution of the scaling parameter resulting from the repeated fitting to the new realisations of the data. Whereas in the case of the tests described above the distributions are of Gaussian shape, the distribution of the scaling parameter (corresponding to the total flux of the belt) fitted to the ALMA data is very asymmetric, making an estimation of the upper limit difficult (figure \ref{hist_outofskyplane}). Adopting a confidence level of 99\%, I get the following upper limits for the total CO (3--2) flux for the two possible orientations of the belt:
\begin{itemize}
\item Dagon moving into sky plane: $f_\textrm{CO}<1.0\times10^{-21}$\,W/m$^2$
\item Dagon moving out of sky plane: $f_\textrm{CO}<4.1\times10^{-21}$\,W/m$^2$
\end{itemize}
\figuremacroW{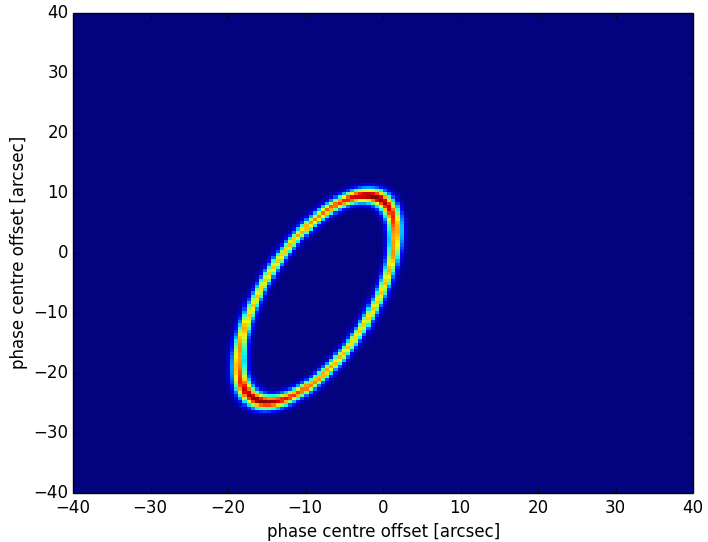}{CO (3--2) emission model}{Projection of the mm dust density distribution as presented by \citet{Boley_etal_2012}. This is taken as a description of the relative CO J=3-2 line intensity on the sky.}{0.7}
\figuremacroW{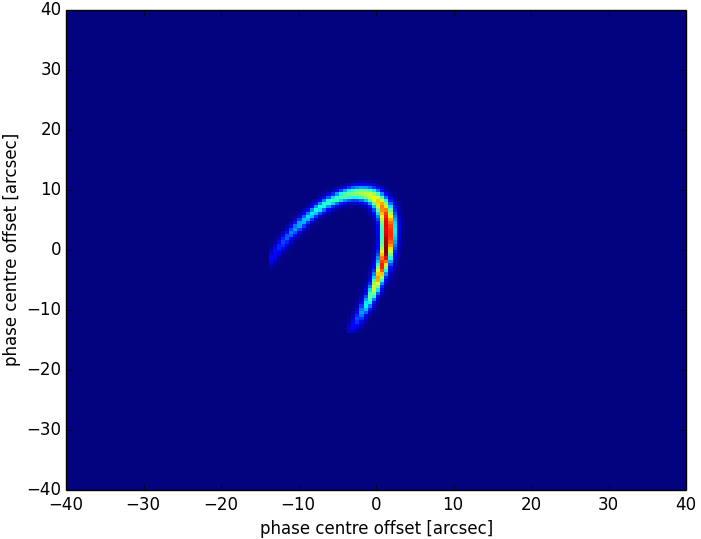}{Emission model with primary beam}{Same as figure \ref{Fomalhaut_belt}, but multiplied with the primary beam. This map serves as input for \textsc{UVMULTIFIT.}}{0.7}
\figuremacroW{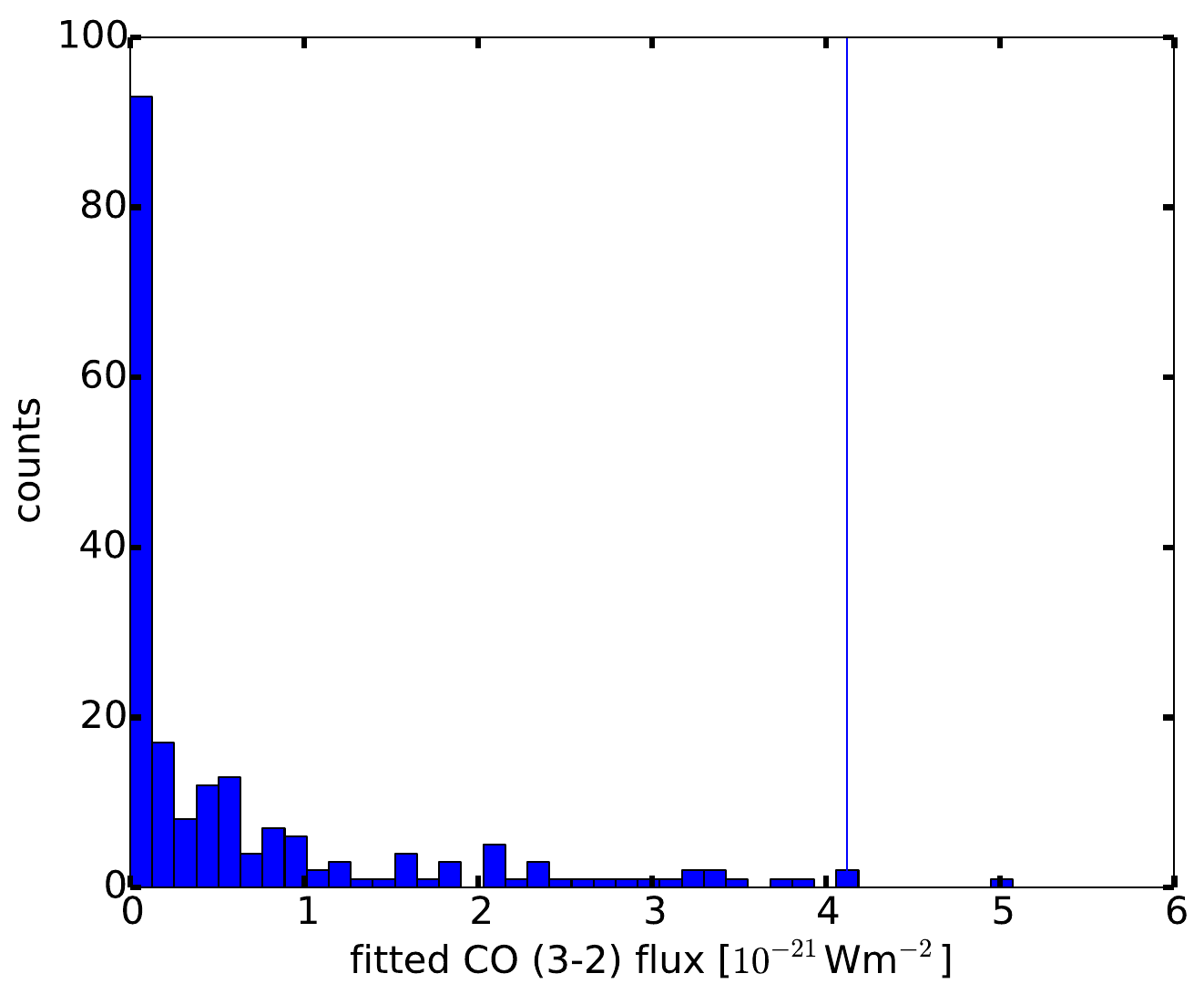}{Histogram of fitted fluxes}{Distribution of the flux fitted to new realisations of the CO data, assuming Dagon is rotating out of the sky plane. The asymmetric shape of the distribution makes a good estimation of the upper limit difficult. The vertical line indicates the upper limit at 99\% confidence level.}{0.7}

\subsubsection{Upper limit from CLEANed image}
An alternative method consists of using a CLEANed, primary beam corrected image of the data. I first produced an image of the continuum and derived a mask covering the regions of significant dust emission. Then, for each of the two possible orientations of the belt, I determined the range of radial velocity where CO emission is expected. I integrated the CLEANed image cube over this velocity range within the mask. No CO emission is detected. To get an upper limit on the integrated flux, I integrated the image cube over velocity ranges of the same width (and with the same mask), but for velocities where no CO emission is expected. The distribution of these integrated fluxes can be translated into an upper limit on the integrated CO flux. At a confidence level of 99\% (i.e.\ 2.33$\sigma$, with $\sigma$ the standard deviation of the sample of integrated fluxes), this gives the following upper limits:
\begin{itemize}
\item Dagon moving into sky plane: $f_\textrm{CO}<8.2\times10^{-21}$\,W/m$^2$
\item Dagon moving out of sky plane: $f_\textrm{CO}<8.0\times10^{-21}$\,W/m$^2$
\end{itemize}

\subsubsection{Discussion}
The derived upper limits may be compared to the upper limit by \citet{Matra_etal_2015}. They use a more evolved analysis of a CLEANed image. Their 3$\sigma$ upper limit of $1.8\times10^{-21}$\,W\,m$^{-2}$ corresponds to $1.4\times10^{-21}$\,W\,m$^{-2}$ at 99\% confidence. We see that the MC approach achieves a similar sensitivity, while our upper limit from the CLEANed image is less constraining.

\citet{Matra_etal_2015} converted their upper limit on the flux into an upper limit on the CO mass. However, since CO is very quickly photodissociated (on timescales of the order of $\sim$120\,yr), the absence of CO does not necessarily tell us a lot about the total gas content of the Fomalhaut belt. Constraints on the amount of atomic gas are more useful. Therefore, in paper~III, we analyse non-detections of C\,II and O\,I emission by \textit{Herschel}/PACS. Figures \ref{continuum_cii}--\ref{lineemission_oi} show the non-detections of the emission lines as well as the continuum where the dust belt is clearly detected (figure \ref{lineemission_cii} is also shown in paper~III). With reasonable assumptions about the abundances of other elements, the non-detections allowed us to put an upper limit on the total gas content of the Fomalhaut belt. The study showed that gas-dust interactions such as proposed by \citet{Lyra_Kuchner_2013} are not likely at the origin of the belt's eccentricity and sharp edges.
\figuremacroW{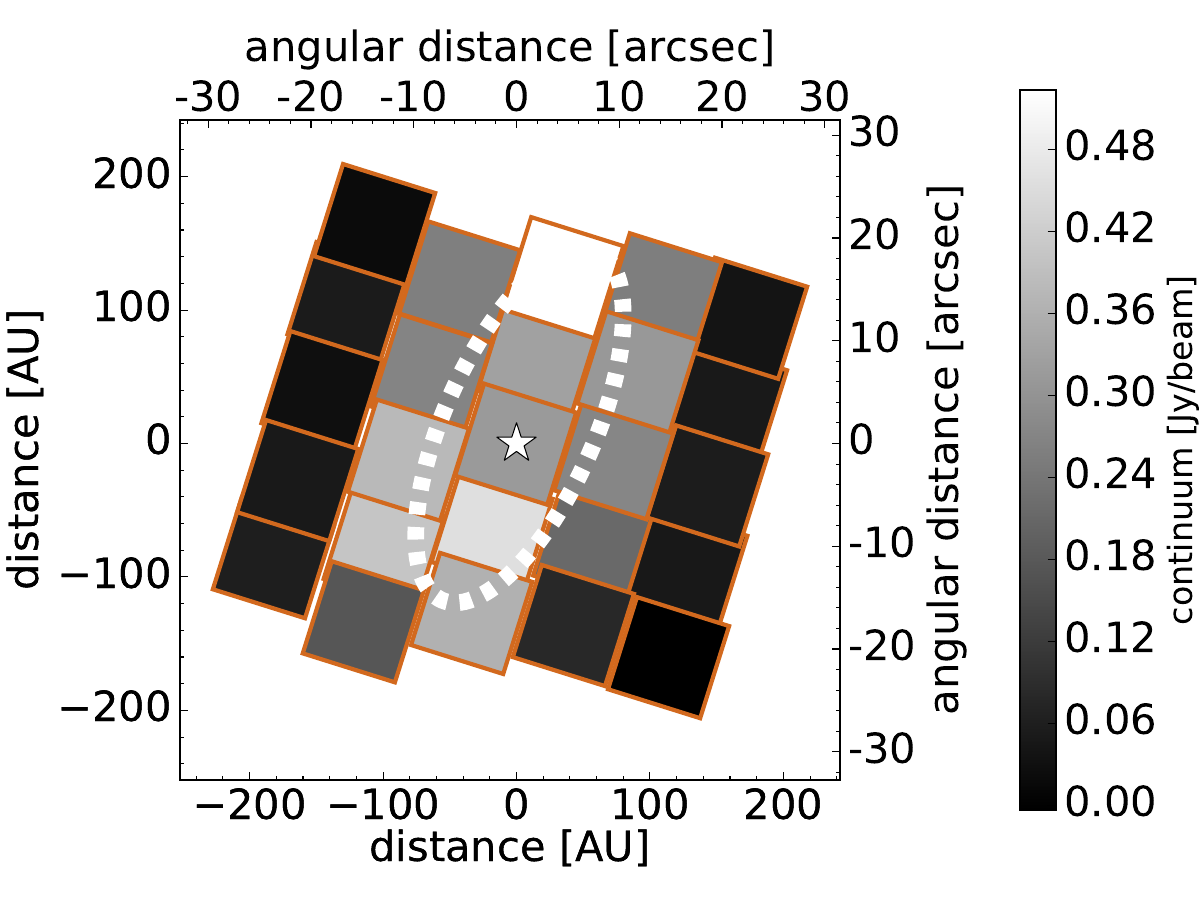}{Continuum at 158\,$\mu$m}{\textit{Herschel}/PACS observation of the continuum emission at 158\,$\mu$m from the Fomalhaut system. PACS is an integral field unit with 25 spaxels, each covering 9.4''$\times$9.4''. The dashed white line shows the position of the dust belt as inferred from ALMA observations. The white star denotes the position of Fomalhaut. Emission from the dust is clearly detected.}{0.8}
\figuremacroW{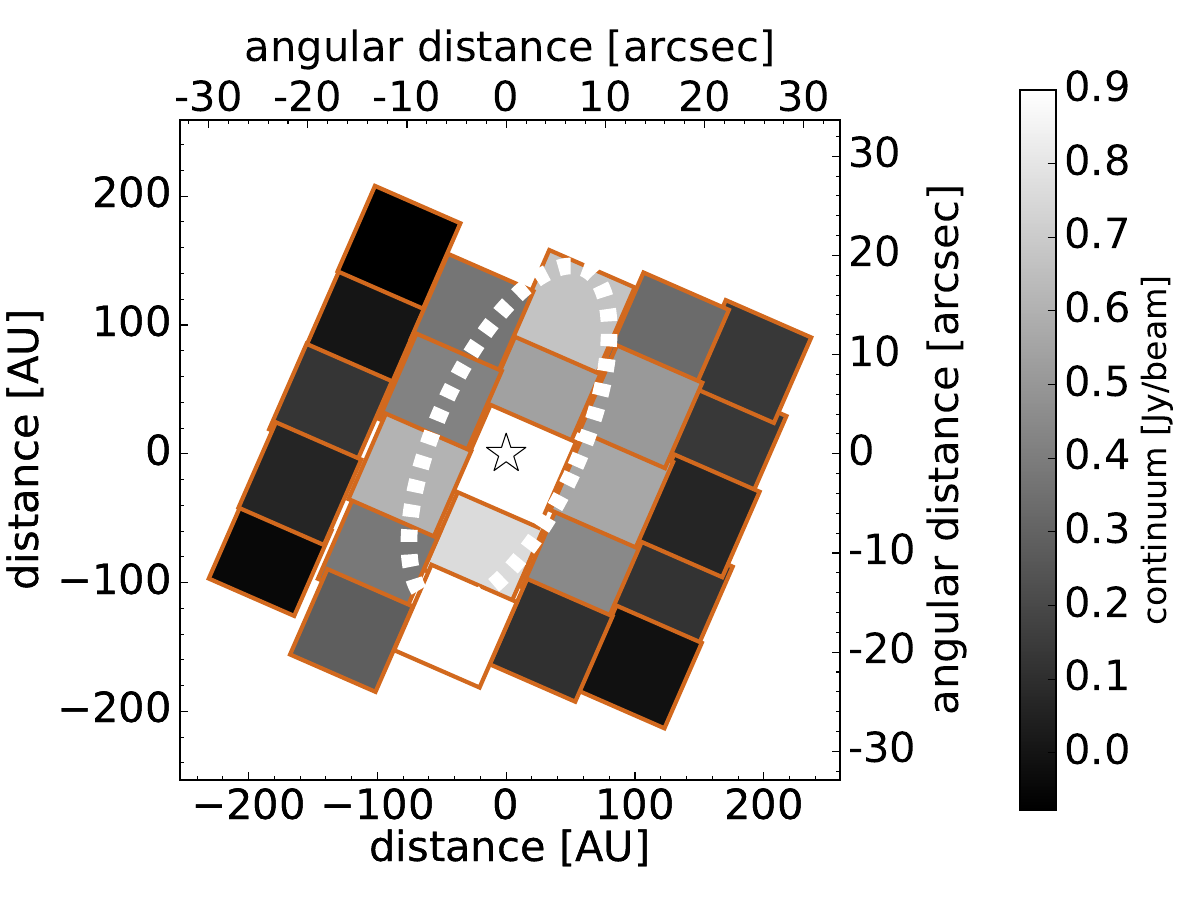}{Continuum at 63\,$\mu$m}{Same as figure \ref{continuum_cii}, but at a wavelength of 63\,$\mu$m. Also in this case, emission from the dust belt is detected.}{0.8}
\figuremacroW{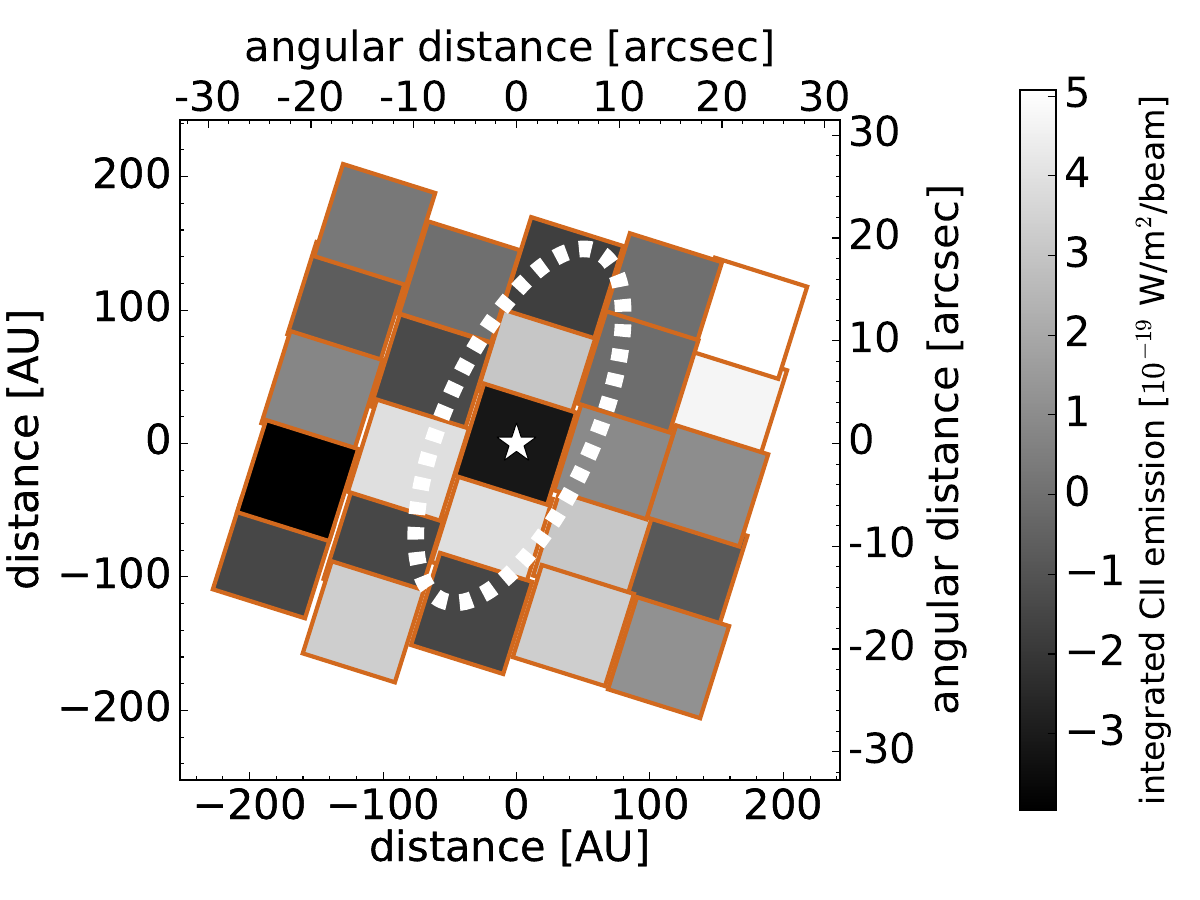}{C\,II non-detection at 158\,$\mu$m}{PACS non-detection of C\,II emission at 158\,$\mu$m.}{0.8}
\figuremacroW{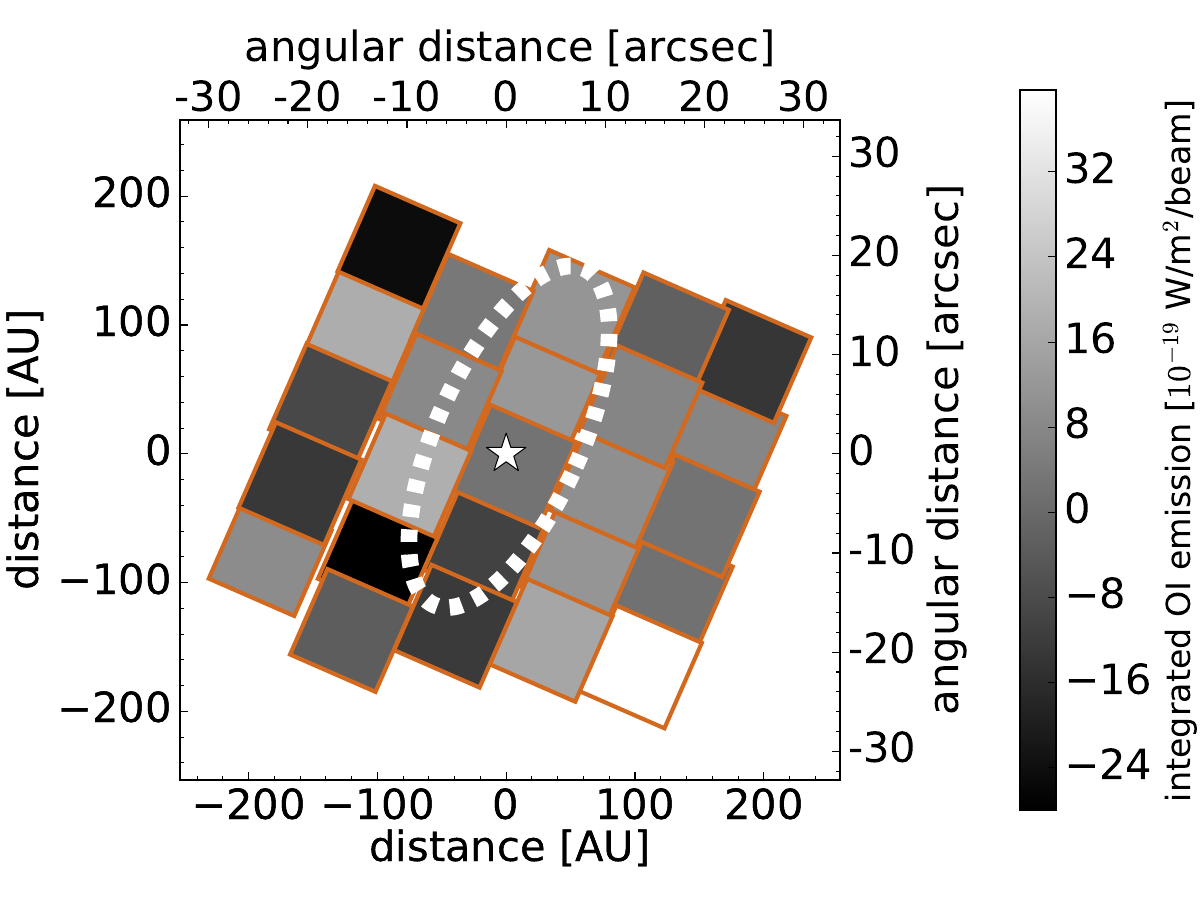}{O\,I non-detection at 63\,$\mu$m}{PACS non-detection of O\,II emission at 63\,$\mu$m. Together with the non-detection of C\,II (figure \ref{lineemission_cii}), these data allowed us to put upper limits on the gas co-spatial with the dust in the Fomalhaut system (paper~III), implying that gas-dust interactions are not efficiently operating.}{0.8}

\section{Gas in other debris disks}
The question arises whether gas is commonly found in debris disks. At the moment, the number of debris disks with detected gas is much smaller than the total number of known debris disks. I will try here to give an overview of the known gaseous debris disks. This is essentially an updated version of the inventory given in \citet{Cataldi_2013}. There is no claim of completeness, but I do not expect a lot of missed systems. \citet{Kospal_Moor_2015} identified only eight gaseous debris disks (including the controversial case of AU~Mic) in their inventory. Similarly, \citet{Riviere-Marichalar_etal_2014} list eight gaseous debris disks in their inventory, only six of which are the same as in the inventory by \citet{Kospal_Moor_2015}. The overview in table \ref{tab:gas_disks_overview} shows 14 disks instead. This is due to the fact that \citet{Kospal_Moor_2015} and \citet{Riviere-Marichalar_etal_2014} applied stricter criteria when constructing their inventory. \citet{Riviere-Marichalar_etal_2014} mention that they do not include objects with unclear evolutionary stage.

With the exception of AU~Mic, all gaseous debris disks are found around A-stars or late B-type stars. This might be connected to the production mechanism, for example, if the gas is produced by UV photodesorption. However, one should not forget that some searches for gaseous debris disks explicitly target A-type stars. 

There is a considerable number of additional objects not included in table \ref{tab:gas_disks_overview} showing both IR excess and circumstellar Ca absorption \citep{Montgomery_Welsh_2012,Welsh_Montgomery_2013,Welsh_Montgomery_2015}. Variability in these absorption lines is indicative of FEB (exocomet) activity. According to \citet{Welsh_Montgomery_2015}, the total number of stars showing evidence for FEB activity is 16.

In summary, whether gaseous debris disks are rare or whether gas is present in all debris disks at a certain level remains to be determined.

\begin{table}
\begin{center}
\hspace*{-1.5cm}
\vspace{-0.7cm}
\begin{tabular}{|l|c|l|c|l|l|}
\hline
object & spec. & \multicolumn{1}{c|}{detected} &age& \multicolumn{1}{c|}{references} & \multicolumn{1}{c|}{notes}\\
 & type & \multicolumn{1}{c|}{species}& [Myr] &  & \\
\hline
$\beta$~Pic & A6V [1] & CO, C, O, Na,&$23\pm3$ [2]& e.g.\ [3], [4]& KM15, R14\\
 &&Ca,\dots &&&FEBs\\
49~Ceti & A1V [5] & CO, C, O, Fe,\dots &$\sim$40 [6]& e.g.\ [7]&KM15, R14,\\
 &&&&&FEBs\\
$\sigma$~Her$^*$ & B9V$^\dagger$ [8]& C, N &$\sim$140 [8]& [8]& R14\\
HD~32297 &A0V [9]& C, Na &$\sim$30 [10]& [11], [12] &KM15, R14\\
HD~172555 & A7V [1] & O, SiO\,(?) & $23\pm3$ [2]& [13] & KM15, R14\\
HD~21997 & A3IV/V [5] & CO &$\sim$30 [15]& [14], [15] &KM15, R14\\
51~Oph & B9.5IV/V & CO, CO$_2$, H$_2$O, &$0.7^{+0.4}_{-0.4}$ [17]& e.g.\ [16], [18]& R14, FEBs\\
 &[16]&C, O, N\dots &&&\\
AU~Mic & M1V [9] & H$_2$ (?)& $23\pm3$ [2]& [19] &KM15\\
HD~181296&A0 [20]& C & $23\pm3$ [2] &[20]&KM15, R14\\
HD~131835&A2IV [21]&CO&$\sim$16 [22]&[22]&KM15\\
HD~158352 & A8V [23] & H, K, Ti &$750\pm150$ [24]& [25], [26], [27] &\\
HD~118232 & A4V [23] & Ti &?& [25], [27] &\\
HD~21620$^*$ & A0V$^\dagger$ [27]& Ca &80 [27]& [27], [28]&FEBs\\
HD~142926 & B9pV [27] & Fe &78 [29]& [27], [30]&\\
\hline
\end{tabular}
\end{center}
\captionsetup{width=14.9cm}
\caption[caption]{Inventory of gaseous debris disks.\\
$^*$ binary system; (?) tentative detection; $^\dagger$ primary; KM15 appears in the \citet{Kospal_Moor_2015} inventory; R14 appears in the \citet{Riviere-Marichalar_etal_2014} inventory.\\
references: [1] \citet{Gray_etal_2006}, [2] \citet{Mamajek_Bell_2014}, [3] \citet{Roberge_etal_2006}, [4] \citet{Brandeker_etal_2004}, [5] \citet{Houk_Smith-Moore_1988}, [6] \citet{Zuckerman_Song_2012}, [7] \citet{Roberge_etal_2014}, [8] \citet{Chen_Jura_2003}, [9] \citet{Torres_etal_2006}, [10] \citet{Kalas_2005}, [11] \citet{Redfield_2007}, [12] \citet{Donaldson_etal_2013}, [13] \citet{Riviere-Marichalar_etal_2012}, [14] \citet{Moor_etal_2011}, [15] \citet{Kospal_etal_2013}, [16] \citet{Thi_etal_2013}, [17] \citet{Montesinos_etal_2009}, [18] \citet{Roberge_etal_2002}, [19] \citet{France_etal_2007}, [20] \citet{Lowrance_etal_2000}, [20] \citet{Riviere-Marichalar_etal_2014}, [21] \citet{David_etal_2013}, [22] \citet{Moor_etal_2015}, [23] \citet{Mora_etal_2001}, [24] \citet{Moor_etal_2006}, [25] \citet{Abt_Moyd_1973}, [26] \citet{Jaschek_etal_1988}, [27] \citet{Roberge_Weinberger_2008}, [28] \citet{Welsh_Montgomery_2013}, [29] \citet{Zorec_etal_2005}, [30] \citet{Slettebak_1982}
}
\label{tab:gas_disks_overview}
\end{table}






\chapter{Astrobiology: a brief overview}
One of the main challenges in exoplanetary science today is to answer the question whether there exist inhabited worlds other than the Earth. In order to progress, an interdisciplinary approach involving various fields such as biology, geology, geochemistry or astrophysics is needed. In this section, I give a very brief overview of some aspects of astrobiology. This serves as a general context for paper~IV.

\section{Life beyond Earth---an old debate}
The question about life beyond Earth is very old, presumably one of the oldest in science and philosophy. The greek philosopher Epicurus (341--270 BC), a member of the atomism school, was convinced of the existence of other worlds \citep{Ollivier_2007}:
\begin{quote}
It is not only the number of atoms, it is also the number of worlds which is infinite in the Universe. There is an infinite number of worlds, similar to ours and an infinite number of different worlds [\dots] One must agree that in all these worlds, without any exception, there are animals, plants and all the living beings we observe.
\end{quote}
This is opposed to ideas by Aristotle (384--322 BC), who was convinced of the uniqueness of our world\footnote{When reading these kind of quotes, one should not forget that for the ancient greeks, the word `world' might have had a different meaning than for us today.} \citep{Bennett_Shostak_2007}:
\begin{quote}
The world must be unique [\dots] There cannot be several worlds.
\end{quote}
In the middle ages, Giordano Bruno (1548--1600) wrote in his book \textit{On the Infinite Universe and Worlds} \citep{Cockell_2015}:
\begin{quote}
In space there are countless constellations, suns and planets; we see only the suns because they give light; the planets remain invisible, for they are small and dark. There are also numberless earths circling around their suns, no worse and no less than this globe of ours. For no reasonable mind can assume that heavenly bodies that may be far more magnificent than ours would not bear upon them creatures similar or even superior to those upon our human earth.
\end{quote}
These views were in strong conflict to the Church's world view and were part of the reasons why Bruno was finally executed. In the late 19th and early 20th century, a controversy arose around the alleged presence of canals on Mars. Straight-line features were first reported by the Italian astronomer Giovanni Schiaparelli in 1877. Some people interpreted these features as irrigation canals built by a Martian civilisation. In particular, American astronomer Percival Lowell spent a large fraction of his time to observe Mars and put forward the picture of an advanced Martian civilisation that desperately tries to transfer water from the polar caps to the equatorial region, fighting the slowly increasing dryness on the planet \citep[e.g.][]{Kerrod_2000}. Other observers did not see the canals and disputed their existence. \citet{Evans_Maunder_1903} conducted experiments with school boys to demonstrate that the illusion of a canal network can arise for unbiassed observers even if no such network exists in reality. The school boys were asked to sketch a circular disk placed in front of them. The disk contained larger and smaller black dots. The experiment showed that the dots were often involuntarily connected to `canals'. Eventually the consensus arose that the canals were due to optical illusions. Still, the believe that Mars is inhabited was apparently quite widespread. For example, the \citet{AcademieDesSciences_1900} announced the Prix Pierre Guzman (Pierre Guzman Prize) with the following text\footnote{Translated from French by the present author.}:
\begin{quote}
[\dots] a sum of one hundred thousand francs [\dots] will be awarded to the one that has found a way to communicate with a celestial body other than the planet Mars.
\end{quote}
Obviously, to establish communication with the Martians was considered too easy to be worth the price.

The development of modern science and astronomy led to a much more complete picture of Earth's place in the universe. Since the Copernican revolution, we know that Earth is not at the centre of the universe with the Sun orbiting it, but rather in an orbit around the Sun. Today it is clear that the solar system is part of the Milky Way galaxy that contains roughly $10^{11}$ stars, and that there are countless other galaxies in the universe. Since the discovery of the first exoplanet around Helvetios\footnote{A.k.a.\ 51~Pegasi.} \citep{Mayor_Queloz_1995}, thousands of exoplanets have been detected and a fascinating diversity of planetary systems has been revealed. Still, we do not know whether the phenomenon of life is widespread in the universe or unique to Earth. From our current data, we cannot tell whether the probability of a `suitable' exoplanet to develop life is one or essentially zero. However, from a technological point of view, it might become possible to detect traces of life during the coming decades, should life be present on nearby exoplanets\footnote{On the other hand, it will practically be impossible to prove that life does \emph{not} exist on any exoplanet.}.

\section{What is life?}
Astrobiology can be defined as the study of life in the universe (including the Earth). In particular, astrobiologists are interested in the evolution and distribution of life. However, the question arises how to define life in the first place. This is important because otherwise, it is not clear what we are actually looking for. To define life turns out to be a tricky task, although in our daily life, it is usually easy to tell whether a certain quantity is alive or not. Actually, there is at the moment no unequivocal definition of life, although various definitions have been proposed. The difficulty can be illustrated by considering a number of examples. One might define live by the capability of feeding, growing and reproducing. With such a definition, one would seriously need to consider whether a fire or a crystal are alive. On the other hand, a mule would need to be considered non-life since it is not capable of reproducing. A more evolved definition of life might request a living entity to be capable of evolving and adapting to its environment. New generations of computer viruses autonomously adapt their source code. Are they alive? For biological viruses, it is also difficult to decide whether they are alive or not. Indeed, viruses possess genetic material. They reproduce and evolve, but they do not metabolise and they need a host cell to produce new viruses.

A commonly adapted `working definition' by \citet{Joyce_1994} reads as follows:
\begin{quote}
Life is a self-sustained chemical system capable of undergoing Darwinian evolution.
\end{quote}
In other words, living entities sustain themselves by collecting energy and matter from their environment \citep{Hazen_2005}. Also, life adapts to its environment by natural selection. This definition restricts life to chemical system, i.e.\ robots or computer viruses are excluded. As any definition of life, it has its own advantages and problems. For example. the aforementioned mule is not capable of undergoing Darwinian evolution since it is sterile. In the light of the considerable difficulty to define life, \citet{Cleland_Chyba_2002} argue that it is not possible to find an unequivocal definition of life before a biological theory allowing a deeper understanding of living systems is developed. In this sense, we are in the same position as someone that is asked to define water without having any knowledge about atoms or molecules. It would merely be possible to define water by describing its properties. Such a definition would face various problems. For example, substances with similar properties might erroneously be classified as water. With a molecular theory, the simple statement `water is H$_2$O' removes all ambiguity \citep{Cleland_Chyba_2002}. However, for the time being, we have to content ourselves with definitions as the one proposed by \citet{Joyce_1994}. It is evident that the detection and characterisation of extraterrestrial life would help us understanding what life actually is.

\section{Habitability}
When searching for extraterrestrial life, an important question is \emph{where} to look. Which planets are the most promising sites for life to develop? To answer this question, we need to know in what environments life is able to exist. So far, the only possibility to find out is to study in detail life on Earth. In particular, astrobiologists are interested in so-called extremophiles, organisms thriving in `extreme' environments. Extremophiles are often archaea. These organisms can be used to define the borders of the physical parameter space where life exists on Earth. For example, they live in environments of extreme temperatures (cold or hot), extreme dryness or extreme salinity. \citet{Takai_etal_2008} showed that the hyperthermophilic methanogen \textit{Methanopyrus kandleri} strain 116, which lives in deep sea hydrothermal vents, can grow at temperatures as high as 122$^\circ$\,C (for a pressure of 20\,MPa). A famous extremophile is \textit{Deinococcus radiodurans}\footnote{A.k.a.\ Conan the Bacterium.} (figure \ref{Deinococcus_radiodurans}) that, as its name suggests, tolerates large amounts of radiation. \textit{D.\ radiodurans} can survive a dose of 5000\,Gy without loss of viability\footnote{One \emph{gray} (symbol: Gy) corresponds to the absorption of one Joule of energy in the form of ionising radiation per kilogram of matter.} \citep{Moseley_Mattingly_1971}. As a comparison, the median lethal dose\footnote{The median lethal dose $\mathrm{LD}_{50}$ is the dose that kills half of the tested population.} for humans is only about $\sim$5\,Gy \citep{Goans_2013}. \textit{D.\ radiodurans} achieves its extraordinary radioresistance with a very efficient DNA repair mechanism, the details of which are not yet well understood \citep{Blasius_etal_2008}. From an evolutionary point of view, one might ask why a specie should develop such a strong radioresistance, given that natural radiation levels are low. It has been suggested that the radioresistance is actually a side effect of a mechanism to survive desiccation, which also relies on efficient repair of damaged DNA \citep[e.g.][]{Mattimore_Battista_1996}.
\figuremacroW{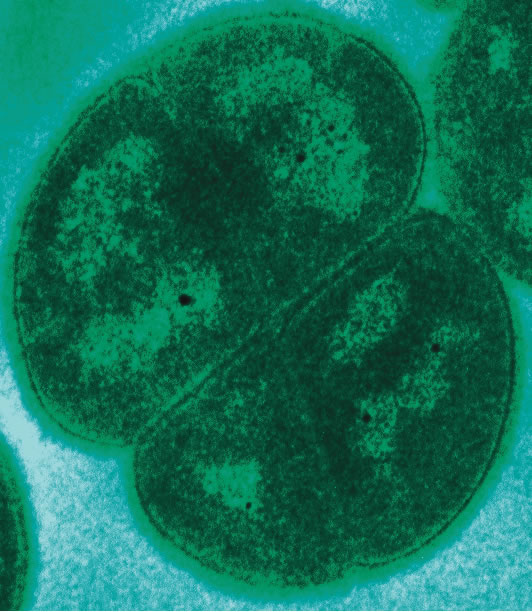}{TEM of \textit{D.\ radiodurans}}{This transmission electron microscopy (TEM) image shows \textit{Deinococcus radiodurans}, an extremophile tolerating large amounts of radiation. It possesses the ability to very efficiently repair damaged DNA, which also allows it to survive extreme desiccation. Image credit: Laboratory of Michael Daly, Uniformed Services University, Bethesda, MD, USA. Public domain, via Wikimedia Commons.}{0.5}

Whether the so-called extremophiles should really be considered `extreme' is a matter of debate and depends on one's point of view. During Earth's history, `extreme' environments have been much more common than conditions perceived as comfortable by humans. For example, Earth's atmosphere may have contained enough oxygen for humans to survive only for the last 10\% of its history \citep{Bennett_Shostak_2007}.

Extremophiles teach us that life is possible in a broad range of conditions. However, no matter how extreme the environment, all life on Earth is dependent on liquid water to survive. Thus, it seems that a prerequisite for a planet to be habitable is the presence of liquid water. This leads to the definition of the habitable zone \citep{Bennett_Shostak_2007}:
\begin{quote}
At any particular time, a star's habitable zone is the range of distances around it at which a planet could potentially have surface temperatures that would allow for abundant liquid water.
\end{quote}
Note that being in the habitable zone is a necessary, but not a sufficient condition for a planet to have liquid surface water. On the other hand, liquid water can also be present outside the habitable zone, as is demonstrated by the subsurface ocean of the Jovian moon Europa in the solar system. Where around a star the habitable zone is located depends first of all on the spectral type of the star: for cold stars, the habitable zone naturally lies closer in than for hot stars. For a solar-type star, the `conservative habitable zone' as calculated with the Habitable Zone Calculator\footnote{\url{http://depts.washington.edu/naivpl/content/hz-calculator}} \citep{Kopparapu_etal_2013,Kopparapu_etal_2014} for an Earth-mass planet extends from 0.95 to 1.68\,AU. For the A6 star $\beta$~Pic, we get a habitable zone between 2.5 and 4.4\,AU, while for the M3 star Gliese~581, which is hosting a multiple planet system, the habitable zone lies between 0.12 and 0.23\,AU. Consequently, with current planet detection techniques (in particular transit observations), it is much easier to find planets in the habitable zone of M-type stars than early type stars. Indeed, from the currently 32 potentially habitable exoplanets listed in the Habitable Exoplanets Catalog\footnote{\url{http://phl.upr.edu/projects/habitable-exoplanets-catalog}, accessed 21.02.2016.}, 19 are orbiting M stars, 10 K stars and only 3 G-type stars. For various reasons, there is a debate whether M stars are actually suitable to host inhabited planets. On the one hand, M stars are very common in the Galaxy: about 75\% of all main-sequence stars belong to this spectral class \citep[e.g.][]{Bennett_Shostak_2007}. They also have very extended main-sequence lifetimes (longer than the age of the universe, typically hundreds of billions of years), giving life plenty of time to evolve. On the other hand, as we have seen with the example of Gliese~581, the habitable zone of an M stars lies very close in and is narrow. A planet orbiting an M star in its habitable zone can become tidally locked, i.e.\ the orbital period of the planet equals the time it takes to rotate around its own axis\footnote{This is analogue to the Earth-Moon system.}: one side of the planet constantly faces the star while the other side is perpetually dark. The day side is thus expected to be at high temperature while the night side is constantly cold, potentially causing atmospheric volatiles to freeze out. This was considered a serious problem for the habitability of such planets. However, modelling efforts show that a relatively thin atmosphere can efficiently transport heat from the day to the night side, thus preventing the collapse of the atmosphere \citep[e.g.][and references therein]{Tarter_etal_2007}. Another potential problem arises from the fact that M stars are magnetically very active with corresponding flares, rising the flux of UV photons hitting the planet to high levels. This danger might be mitigated by the production of a protective layer of ozone \citep{Segura_etal_2005}. Finally, another caveat is the decline in luminosity of an M star during its long-lasting pre-main-sequence phase (see section \ref{YSO}). This means that the habitable zone moves inwards during the pre-main-sequence phase. As a consequence, a planet in the habitable zone of a main-sequence M star was located inwards of the habitable zone during the pre-main-sequence phase. Simulations by \citet{Tian_Ida_2015} show that this leads to a bimodal water content distribution of planets in the habitable zone. Planets that start with a high water content (`ocean planets' without continents) can retain their water during the pre-main-sequence phase. On the other hand, planets that do not start with sufficient water completely dry out (`dune planets') due to the higher luminosity of the pre-main-sequence star. These simulations thus suggest that the formation of Earth twins (with both oceans and continents) in the habitable zone of M stars is difficult.

In general, if the luminosity of a star is not constant, the habitable zone is not fixed either. The luminosity of the Sun is in fact steadily increasing and the habitable zone is consequently moving outwards. At present, the Sun is 30\% more luminous than when the solar system formed \citep{Bennett_Shostak_2007}. As a result, Earth will become uninhabitable within the next 1.75--3.25\,Gyr \citep{Rushby_etal_2013}. With temperatures on Earth continuously rising, the biosphere is expected to become dominated by unicellular organisms again \citep{OMalley-James_etal_2013,OMalley-James_etal_2014}. Lifeforms adapted to high-temperature, high-salinity environments might survive in high-altitude or high-latitude niches for up to 2.8\,Gyr from present according to the models by \citet{OMalley-James_etal_2013}. Note that for M stars, a planet in the habitable zone can remain there during 100\,Gyr \citep{Tarter_etal_2007}.

There might be factors other than the location within the habitable zone that are important for the habitability of a planet. Looking at Earth, plate tectonics is essential to stabilise our long-term climate by means of the carbon dioxide cycle, which acts as a thermostat. Also, in contrast to Venus or Mars, the Earth possesses a global magnetic field that protects the surface and the atmosphere from the energetic particles of the solar wind. It is believed that without the magnetic field, Earth's atmosphere would have been eroded away \citep{Bennett_Shostak_2007}. Furthermore, the Earth is the only terrestrial planet in the solar system with a large Moon, which has been suggested to stabilise the Earth's axis tilt and thus contributing to a stable climate. However, the validity of this argument is unclear. Indeed, it is possible that if the Moon never formed, the Earth would rotate faster and thus be self-stabilising \citep[e.g.][and references therein]{Chyba_Hand_2005}. Concerning the overall architecture of the solar system, it was long thought that Jupiter plays the role of a `cosmic shield'. In this picture, Jupiter would lower the impact rate of minor bodies on Earth thanks to its large mass. Since large impacts can have catastrophic consequences for a biosphere, the presence of Jupiter was seen as an important factor for life to evolve on Earth. However, more recent work strongly challenges this view, suggesting instead that Jupiter can actually increase rather than decrease the impact rate on Earth \citep[e.g.][and references therein]{Horner_Jones_2010,Grazier_2016}. In fact, Jupiter might be important for the development of life for another reason, namely by \emph{increasing} the flux of planetesimals into the inner solar system, delivering large amounts of volatiles needed for life to develop \citep[e.g.][]{Grazier_2016}. Finally, some authors suggested that not the whole Galaxy is suitable for life to develop and thrive. This idea leads to the (debated) concept of a galactic habitable zone, based on parameters such as metallicity or the rate of potentially detrimental supernovae \citep{Bennett_Shostak_2007}.

\section{Searching for life in the solar system}
After having discussed the conditions for life to thrive on Earth (in particular the presence of liquid water) in the previous section, we turn our attention to bodies in the solar system that potentially could harbour a biosphere. The proximity of the solar system's planets allows to search for life \emph{in situ}, in contrast to extrasolar systems.

One reason why the discovery of extraterrestrial life would have a very large impact on our understanding of life is that it would inform us about the probability for life to arise on a planet. This probability is at the moment completely unconstrained, apart from the fact that we know it is larger than zero. However, in the case of the solar system, one would need to make sure that the detected life really arose independently. Indeed, various studies showed that interplanetary transfer of life is in principle possible \citep[e.g.][]{Mileikowsky_etal_2000,Worth_etal_2013}.

Considering merely the planet sizes and their radial distances from the Sun, the most promising guess for an inhabited planet is probably Venus, sometimes also called Earth's `sister' or `twin'. Indeed, Venus is a rocky body orbiting at 0.72\,AU and its diameter is only 5\% smaller than the diameter of Earth. However, we know today that the climate and surface conditions on Earth and Venus differ radically. Venus' atmosphere is very thick and generates a surface pressure $\sim$90 times that on Earth. In addition, the atmosphere consists mostly (by more than 96\%) of carbon dioxide, leading to a very strong greenhouse effect with surface temperature of $\sim$470$^\circ$\,C \citep{Bennett_Shostak_2007}. It is possible that Venus had had oceans in the distant past \citep[e.g.][]{Hashimoto_etal_2008}, but if so, the water was lost long time ago, probably through photodissociation and subsequent thermal escape of hydrogen. This might have happened as a consequence of a runaway greenhouse effect: the increasing luminosity of the Sun would have caused increased water evaporation. The additional water enhanced the greenhouse effect, leading to even more evaporation \citep{Bennett_Shostak_2007}. Today, Venus is almost completely dry and an unattractive place to search for life. There are however two bodies in the solar system that are considered reasonable candidates for hosting life: Mars and Jupiter's moon Europa, described in the next sections.

\subsection{Mars}
Within the astrobiology community, Mars is considered one of the prime targets to search for past or present life. Today, liquid water is not stable at Mars' surface because of the low temperature and atmospheric pressure. Water is still present in the form of water ice (at the polar caps and in the subsurface) and as gas in the atmosphere \citep{Forget_2007}. Recent studies also observed small quantities of transient liquid, briny water\footnote{This is possible because certain salts lower the freezing point of the brine.} flowing on the surface of Mars \citep{Martin-Torres_etal_2015,Ojha_etal_2015}. However, it is now well established that liquid water was once present on Mars' surface in large quantities, perhaps even in the form of oceans. The evidence comes from various geological features such as dried-up river beds, valleys or deltas. In addition, minerals and rocks that typically form in water have been detected on the surface. This suggests that Mars' atmosphere was once much warmer and thicker than today. However, because Mars is relatively small, its core eventually solidified and consequently Mars lost its global magnetic field. The atmosphere was then no longer protected from the solar wind and steadily removed. This led to a climate change with a suppressed greenhouse effect and lower temperatures---Mars became the essentially frozen planet we observe today \citep[e.g.][]{Bennett_Shostak_2007}.

If Mars had abundant liquid water on its surface early in its history \citep[prior to $\sim$3.7\,Gyr ago, e.g.][]{Forget_2007}, it is not unthinkable that life arose as well. Indeed, on Earth, evidence suggests that life arose quickly---it is widely accepted that life was present on Earth prior to $\sim$3.5\,Gyr ago \citep[e.g.][]{Linder_etal_2000,Bennett_Shostak_2007}. This may or may not suggest that life develops easily in general under the right conditions, but it certainly does not argue against the possibility that life was present on early Mars. It is even conceivable that life arose first on Mars and was than transferred to Earth in rocks ejected from the Martina surface during impact events \citep[e.g.][]{Mileikowsky_etal_2000}. Traces of Martian life, if it existed, might still be present on Mars today. Perhaps life adapted to the climate change and still exists on Mars today, for example in the subsurface. Actually, the Viking landers conducted a number of biological in situ experiments in 1976. Some experiments suggested that biological metabolism was occurring in the Martina soil, but most scientists agree that the results can be explained by chemical processes \citep[e.g.][]{Bennett_Shostak_2007}. Still, Mars will continue to be an object of intense study in astrobiology. A key question remains whether water was flowing on Mars over geologically long periods of time. Indeed, since Mars is relatively far away from the Sun, which was in addition fainter at early times, Mars might have been too cold for liquid water in general. Liquid water might have been present only during geologically short periods, for example during intensiv volcanic outgassing with corresponding stronger greenhouse effect (M.\ Way 2016, private communication). Thus, early Venus might be a better candidate for a second genesis in the solar system \citep{Way_etal_2015}, but it might be difficult to find traces of such life on modern Venus.

\subsection{Europa}
Another solar system body considered a candidate site for extraterrestrial life is Jupiter's moon Europa. Once again, it is the presence of liquid water that makes the body interesting. However, in Europa's case, the water is not floating on the surface, but exists as a subsurface ocean beneath a layer of ice. In total, it is believed that Europa has a 100--150\,km thick layer of water overlying a rocky mantel \citep[e.g.][]{Greenberg_2005,Prockter_Pappalardo_2014}. Measurements of the magnetic field by the Galileo spacecraft indicate that most of this water is in the liquid state and forms a briny ocean. It is not entirely clear how thick the ice shell above the ocean is. Recent modelling by \citet{Park_etal_2015} suggests an ice shell of $\sim$10\,km.

Where does the energy that maintains Europa's ocean in the liquid state come from? The answer is \emph{tidal heating}. Because of an orbital resonance with two other Jovian moons, Io and Ganymede, the eccentricity of Europa's orbit around Jupiter is non-zero. This causes tidal forces to heat Europa's interior. Together with radiogenic heating, this energy input can keep Europa's ocean liquid \citep[e.g.][]{Prockter_Pappalardo_2014}.

Europa's surface shows only a limited number of craters, suggesting it is young and pointing towards geological activity. Figure \ref{Europa} shows a part of Europa's surface known as `chaos region'. Blocks of surface material have been disrupted and moved and rotated with respect to each other. The image resembles pack-ice on Earth. Chaos might form by a local melt-through of ocean water to the surface or the rising of `warm' ice that mobilised the surface \citep{Prockter_Pappalardo_2014}.
\figuremacroW{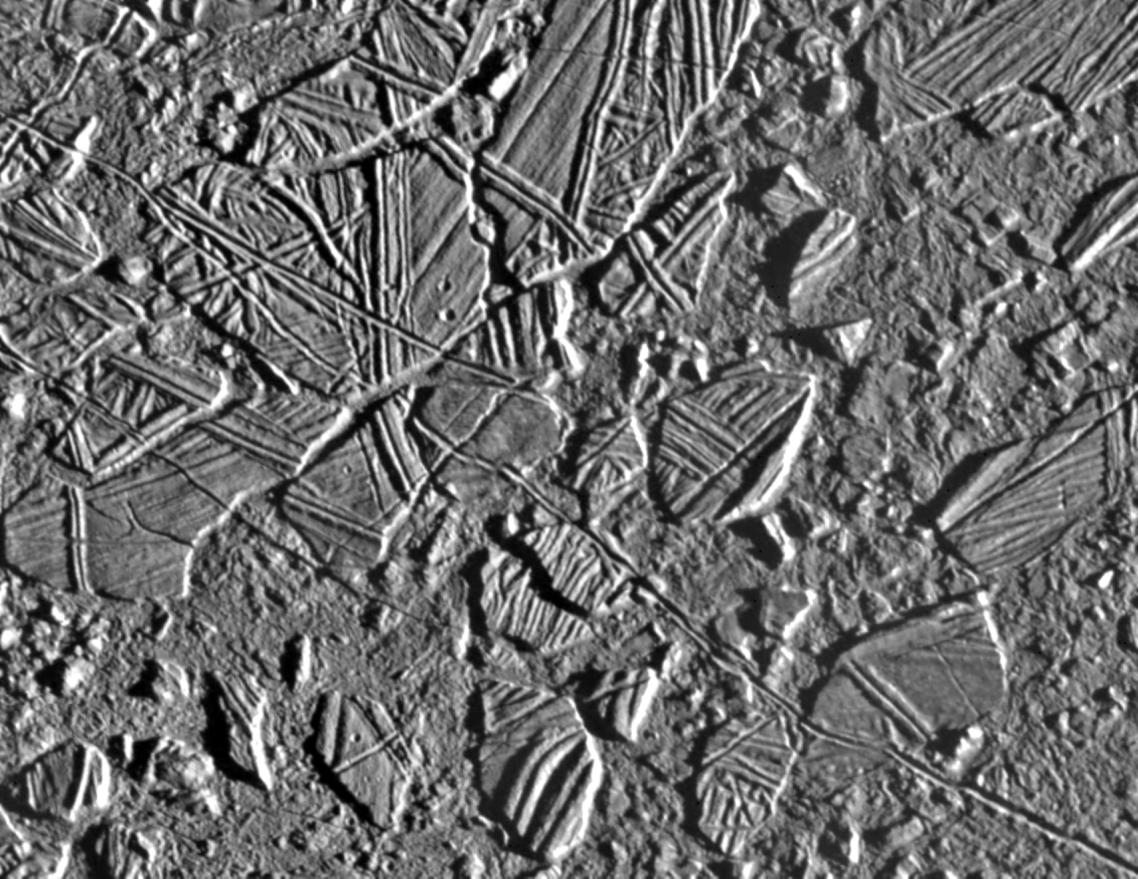}{Chaotic terrain on Europa}{This image taken by the Galileo spacecraft shows a 34\,km by 42\,km patch of Europa's surface that is part of the `chaos region'. It consists of blocks of surface material broken apart, shifted and rotated with respect to each other, similar to pack-ice on Earth. The image resembles a jigsaw. Chaos regions are believed to indicate spots where Europa's heat flow from the interior has been enhanced and allowed the mobilisation of the surface \citep[e.g.][]{Prockter_Pappalardo_2014}. Image credit: NASA/JPL/ASU.}{0.85}

Besides its ocean of liquid water, Europa has other properties that make it an interesting object for astrobiologists. For example, there is direct contact between the ocean and Europa's rocky interior, potentially enabling chemical reactions important for life. In addition, tidal heating might create hydrothermal vents at the bottom of the ocean. In the case of Earth, hydrothermal vents have been proposed as site for the origin of life \citep[e.g.][]{Bennett_Shostak_2007}.

\subsection{Other bodies of astrobiological interest in the solar system}
There are a number of other bodies in the solar system that are considered potential habitats for extraterrestrial life.

The two Jovian moons Ganymede and Callisto are thought to harbour subsurface oceans of liquid water. In contrast to Europa, these oceans are sandwiched between different phases of ice \citep{Collins_Johnson_2014}.

Enceladus is an icy satellite of Saturn and famous for its erupting jets of water vapour, ice grains and organic compounds. The latter is especially promising from an astrobiological viewpoint. Enceladus is thought to harbour a subsurface ocean in the south polar region that is in contact with a silicate core \citep{Nimmo_Porco_2014}. Since the aforementioned geysers are probably directly connected to the ocean, they offer straight access to this potentially habitable environment, making it a very attractive target for future space missions \citep{Nimmo_Porco_2014}.

Another of Saturn's moons, Titan, has also attracted the interest of astrobiologists. It is the only satellite in the solar system known to have a dense atmosphere, consisting mostly of N$_2$ and with a complex organic chemistry. Compared to Earth, the atmospheric surface pressure is enhanced by 45\%. Like the moons discussed previously, Titan is thought to harbour a subsurface ocean of liquid water \citep[e.g.][]{Coustenis_2014}. Interestingly, Titan's surface is covered by numerous lakes. However, these lakes are not made of water, but rather contain liquid hydrocarbons \citep[e.g.][]{Coustenis_2014}. Indeed, the surface temperature is only 94\,K. Thus, any water would be in the form of solid ice. It has been discussed whether the liquid hydrocarbons could replace water as a solvent in a biological system. However, because of the low temperature, chemical reactions would proceed very slowly \citep{Bennett_Shostak_2007}.

\section{Life beyond the solar system}
The past few years have seen a flood of newly discovered exoplanets. Currently, almost 2000 confirmed exoplanets are known\footnote{\url{http://exoplanetarchive.ipac.caltech.edu}, accessed 15.01.2016.}. Exoplanetary systems show a wide range of architectures that can vary radically from our solar system. For example, the so-called hot Jupiters are Jupiter-mass exoplanets orbiting very close to their host star (often closer than Mercury orbits the Sun). Super-Earths are exoplanets more massive than Earth, but significantly less massive than the ice giants Uranus and Neptune. While absent from the solar system, Super-Earths are common around other stars \citep{Fressin_etal_2013,Petigura_etal_2013}. Another class of planets absent in the solar system are massive planets in wide orbits, as in the HR~8799 system that hosts four massive, directly imaged planets with semi-major axis\footnote{\url{http://exoplanet.eu/catalog/?f='HR\%208799'+in+name}, accessed 17.01.2016} between 14.5 and 68\,AU.

In general, current data suggest that stars hosting planets are the rule rather than the exception. From statistical analysis of microlensing data, \citet{Cassan_etal_2012} find that each star in the Galaxy should have more than one planet on average (in the orbital range 0.5--10\,AU). From an astrobiological viewpoint, a relevant quantity is the fraction of stars with planets in the habitable zone. \citet{Petigura_etal_2013} estimate that 8.6\% of the Sun-like stars host an Earth-like planet (1--2\,R$_\oplus$) in the habitable zone as defined by \citet{Kopparapu_etal_2013}. Assuming that the planet formation efficiency is the same throughout the Milky Way, this results in a huge number of potentially habitable planets. If life exists on one of these planets, can we detect it? In contrast to the solar system, the only possibility to do so is by using remote observations. This is the topic of the next chapter, which will set paper~IV into context.



\chapter{Remote detection of life}
In the solar system, it is possible to search for traces of past or present life \emph{in situ}. However, for exoplanets, we rely on remote observations. In this chapter, I give a brief overview of ideas to remotely detect life. In paper~IV, we explore the possibility of detecting biosignatures in the debris created during an impact onto an inhabited exoplanet.

In general, false positives will inevitably be an issue for remote sensing. One way to decrease the risk of a false positive is to interpret a positive signal in context, i.e.\ to characterise the exoplanetary system in question (including the host star) as accurately as possible.

\section{Planetary atmospheres}
The most common idea to detect a biosphere on an exoplanet is to observe the composition of the exoplanet's atmosphere. This is based on the fact that Earth's atmosphere is heavily influenced by the presence of life. Indeed, the oxygen we breath is almost entirely of biological origin. Actually, Earth's early atmosphere was dominated by N$_2$, CO$_2$, water vapour and possibly sulphur---oxygen was only present in trace amounts for the first billion years after the formation of Earth \citep[e.g.][]{Gilmour_2004}. Approximately 2.3\,Gyr ago, oxygen started building up in the atmosphere, presumably due to photosynthesis by cyanobacteria \citep[e.g.][]{Bennett_Shostak_2007}. This is known as the Great Oxygenation Event (GOE). At least 2\,Gyr were then required to build up atmospheric oxygen to present levels. The rise of oxygen had important consequences for life on Earth. Prior to the GOE, life on Earth was anaerobic. For this kind of organisms, oxygen can actually be poisonous. On the other hand, oxygen allows for more efficient energy production and thus fostered the evolution of more complex organisms with new, energy-intensive capabilities.

Because life has such an important influence on the composition of Earth's atmosphere, the question arises whether it is possible to detect a biosphere on an exoplanet from atmospheric observations. The atmospheres of gas giants can already be studied with current instrumentation. For example, \citet{Fraine_etal_2014} reported the detection of water vapour in the atmosphere of an exoplanet as small as Neptune, observed in transmission. However, it is not yet
\begin{WrapText}{Why do we not fly there?}
A much more detailed characterisation of exoplanets and search for signs of life could be conducted by space probes sent to exoplanetary systems. However, the vast distances between individual stars prevents space probes with current propulsion technology to reach extrasolar systems within reasonable timescales. For example, Voyager~1, launched in 1977, is currently the farthest spacecraft from Earth at a distance\footnote{\url{http://voyager.jpl.nasa.gov/where/}, accessed 21.01.2016.} of 134\,AU, which corresponds to 0.05\% of the distance between the Sun and the nearest star, Proxima~Centauri. In the future, it might become possible to accelerate spacecrafts to speeds of a few percent of $c$, for example using solar sails \citep{Bennett_Shostak_2007}. A journey to the closest stars would still take of the order of a century. If it became possible to accelerate spacecrafts to speeds close to $c$, relativistic effects would appear. Interestingly, a crew aboard a ship with a constant acceleration of 1\,g would actually be able to travel to any point in the Galaxy within a human lifetime because of relativistic time dilation \citep{Sagan_1963}, but for the observers on Earth, the travel time of such a ship would still be approximately the light travel time to the target.
\end{WrapText}
possible to observe the atmospheres of small rocky planets in the habitable zone. Such observations would require advanced instruments, for example a space nulling interferometer \citep[e.g.][]{Cockell_etal_2009}. In this concept, the beams of the telescopes are combined in a way that destructive interference occurs, thus `nulling' the star and allowing the imaging and spectral characterisation of faint companions.

While the oxygen on Earth is due to biological activity, oxygen can also be produced abiotically \citep[e.g.][]{Wordsworth_Pierrehumbert_2014} and the detection of oxygen alone is thus not necessarily a robust biosignature. Rather than the detection of a single species, it has been suggested to search for species out of chemical equilibrium due to biological forcing. For instance, methane is expected to be rapidly destroyed in an oxygen-rich atmosphere by oxidation to CO$_2$ and H$_2$O. However, due to biological sources, methane is present in Earth's atmosphere at levels exceeding the equilibrium value by many orders of magnitude \citep{Sagan_etal_1993}. This was an important observation in the control experiment by \citet{Sagan_etal_1993} that attempted to detect life on Earth with the Galileo spacecraft.

The idea of using species out of equilibrium as an indication for a biosphere is actually quite old \citep{Lederberg_1965,Lovelock_1965}. The basic idea is still the same today, and using atmospheric biosignatures is the most popular approach to find an exoplanetary biosphere. With new instruments and the knowledge acquired from the observation of giant planet atmospheres, the characterisation of the atmospheres of Earth-like exoplanets might become possible within the next decades \citep[e.g.][]{Seager_2014}.

\section{Direct detection of living matter: reflectance spectroscopy}
Information about the surface of an exoplanet can also be used to find hints for a biosphere \citep[e.g.][]{DesMarais_etal_2002}. Vegetation on Earth shows a strong increase in reflectance between the red part of the visible spectrum and the near-IR. This is known as the \emph{red edge} and serves plants as a cooling mechanism. \citet{Sagan_etal_1993} detected this feature with the Galileo spacecraft. As another example, \citet{Knacke_2003} discussed the prospects to detect algae on the surface of exoplanets in reflected light.

Interestingly, microorganisms make up most of the biomass on Earth \citep{Madigan_2012}. In contrast to complex multicellular organisms, they have been present over a large fraction of Earth's history and are also expected to dominate Earth's future biosphere \citep{OMalley-James_etal_2013,OMalley-James_etal_2014}. Thus, microorganisms might provide attractive biosignatures. It turns out that microorganisms possess distinctive spectral reflectance signatures in the near-IR. \citet{Dalton_etal_2003} measured the reflectance spectra of three different microbial species and compared their data to spectra of Europa's surface acquired by the Galileo spacecraft. Recently, \citet{Hegde_etal_2015} extended this work substantially by measuring the reflectance spectra of 137 different microorganisms. Further data was provided by \citet{Schwieterman_etal_2015} who measured reflectance spectra of non-photosynthetic microorganisms. In paper~IV, the possibility of detecting microorganisms ejected from a planet during an impact event is explored.

Polarisation measurements have also been discussed as a potential way to detect life in reflected light. \citet{Berdyugina_etal_2016} investigated linear polarisation signatures associated with biological pigments used for photosynthesis or protection. Based on laboratory measurements, they argue that linearly polarised spectra might be a powerful tool to detect photosynthetic pigments remotely. In addition, circular polarisation spectra are proposed as a tool to indicate homochirality\footnote{The fact that all life on Earth only uses left-handed L-amino acids in proteins and only right-handed D-sugars in nucleic acids is known as homochirality, the origin of which is not clearly understood.}, which is regarded a universal biosignature \citep[e.g.][and references therein]{Sparks_etal_2012}.

In summary, light reflected from the surface of an exoplanet has the potential to hint towards the presence of a biosphere and can be an interesting complement to atmospheric studies.

\section{Geosphere-biosphere interactions}
Another possibility to detect the presence of extraterrestrial life remotely (and also in situ) is to make use of the influence the biosphere has on the geosphere. Compared to Venus or Mars, Earth possesses a much larger diversity of minerals. This is largely a consequence of the presence of life \citep{Hazen_etal_2008}. The biosphere can influence Earth's mineralogy indirectly, for example by changing the composition of the atmosphere, which in turn allows new minerals to form. It has also been suggested that photosynthesis can change geochemical cycles and is responsible for the presence of stable granitic continents on Earth \citep{Rosing_etal_2006}. Direct production of minerals by life (biomineralisation, e.g.\ skeletons) is also possible. Minerals as a tool for remote sensing of life are discussed in more detail in paper~IV.

\section{SETI}
Arguably the most direct evidence for the existence of life beyond Earth would be the reception of a message from an extraterrestrial intelligence\footnote{It is actually not easy to define intelligence. A convenient definition is to declare a life form intelligent if it is able to communicate over interstellar distances \citep[e.g.][]{Conway_2004}.}. The search for extraterrestrial intelligence (SETI) is looking for such messages, may they be intentional or not.

For technological reasons, SETI is primarily carried out at radio frequencies. Different questions need to be considered when conducting SETI, for example at which frequencies to search. Note also that our technological advances have been very substantial over the last decades and centuries, i.e.\ extremely short timescales compared to the lifetime of the solar system or the Galaxy. One may thus argue that it is extremely unlikely that an extraterrestrial civilisation, if it exists, would be at the same technological stage as we are today and use radio waves for interstellar communication.

Besides the detection of communication signals, there are other ways to infer the existence of an advanced civilisation, for example from observations suggesting astro-engineering. The Kardashev scale attempts to quantify the advancement of a civilisation in the following way: a type I civilisation uses the energy of its home planet, a type II civilisation collects all the energy available from its host star, while a type III civilisation harvests the energy from an entire galaxy \citep[e.g.][]{Bennett_Shostak_2007}. A proposed way to collect the energy radiated from an entire star is to astro-engineer a sphere covered with `solar cells' around the star. This is known as a Dyson sphere. Such a sphere would radiate away waste heat, typically in the infrared. Thus, the SED of the star is altered in a potentially observable way: the star appears with reduced optical luminosity, but increased infrared luminosity. Similarly, a type III civilisation constructing a Dyson sphere around every star of its galaxy would cause observable changes in the galaxy's SED \citep[e.g.][]{Zackrisson_etal_2015}. Astro-engineered megastructures might also be observable in transit. For example, the star KIC~8462852 shows unpredictable dips in flux that are difficult to explain \citep{Boyajian_etal_2016}. The dips might be due to the break-up of massive exocomets, but it has also been speculated that we are seeing the transits of artificial objects \citep{Wright_etal_2016}. So far,  radio SETI observations towards the source showed no evidence for a technology-related signal \citep{Harp_etal_2015}.

\chapter{Summary of papers}
Below follows a brief summary of the papers included in this thesis.
\section{Paper~I}
As discussed in section \ref{beta_Pic_gas}, the gas observed in the circumstellar debris disk around $\beta$~Pictoris is thought to be of secondary origin, i.e.\ produced from the dust. The gas is also known to be strongly overabundant in C (and O) compared to solar abundances of metallic elements such as Na and Fe. The overabundance of C is important since it stabilises the gas disk against radiative blowout. However, it is unclear how the overabundance arises in the first place. In paper~I, we use \textit{Herschel}/HIFI observations of C\,II emission at 158\,$\mu$m to constrain the C spatial distribution in the disk and learn something about the origin of the gas. HIFI can \emph{spectrally} resolve the emission line. Since the disk is seen edge-on, the line profile depends on the spatial distribution of the C$^{+}$ gas. We fit models of the spatial gas profile to the data in the following way: for a given gas density profile, we use the \texttt{ONTARIO} code to compute the gas temperature and level populations. Then, the disk model is projected onto the sky by implementing a standard ray-tracing method to solve the radiative transfer equation, taking into account the effect of the rotation of the disk. Finally, a model line profile is derived that can be directly compared to the observations. By fitting a density profile derived from previous observations of Fe\,I, we show that the observed profile is consistent with the hypothesis of a well-mixed gas (constant C/Fe ratio throughout the disk). We then fit a spatial profile consisting of a number of concentric rings to the data. These fits showed that most of the C$^{+}$ is located beyond 30\,AU, at around 100\,AU or beyond, implying that falling evaporating bodies (`exocomets') are not producing the bulk of the gas. We also find tentative evidence for an asymmetry in the disk with more gas in the SW side. Interestingly, recent ALMA observations showed a massive CO clump in the SW \citep{Dent_etal_2014}. Finally, we find that a simple accretion disk profile has problems to satisfactorily fit the data. Thus, the data suggest that the C (and O) overabundance might be due to preferential depletion\footnote{The gas is produced at `normal' abundances (e.g.\ via collisional evaporation of dust grains), but elements such as Na and Fe are preferentially removed by radiation pressure.} rather than preferential production\footnote{In other words, C and O are produced at a higher rate than other elements, for example via UV photodesorption of CO from icy grains.}. In the latter case, one would indeed expect all the gas to accrete onto the star \citep{Xie_etal_2013}.

As an outlook, we also present simulations of C\,I by ALMA, assuming the C spatial distribution derived from the HIFI data. These simulations were the basis for an ALMA proposal that was granted time to observe the $\beta$~Pic disk in C\,I. An upcoming analysis of these data will gives us a much more detailed picture of the C distribution in the disk.

\section{Paper~II}
This paper presents observations of C\,II 158\,$\mu$m and O\,I 63\,$\mu$m emission with \textit{Herschel}/PACS, a 5 by 5 integral field unit. The oxygen emission turns out to be much stronger than expected. As in paper~I, we model the data by using \texttt{ONTARIO} and a simple ray-tracing code to solve the radiative transfer. For a given density profile, a map of line emission on the sky is derived. This can be used to produce synthetic PACS observations that can be compared to the data. We find it difficult to reproduce the O\,I and C\,II line strength simultaneously. Basically, one needs a lot of electrons to excite oxygen sufficiently to reproduce the observed O\,I flux. However, since the main electron donor is carbon, the observed C\,II line emission puts a limit on the amount of carbon that can be put in the model. Models assuming that the C and O are well-mixed with metallic species such as Fe, and thus follow a power law density profile, turn out to be incompatible with the data. When assuming instead that the C and O is produced from the CO clump at 85\,AU seen by ALMA and located in a torus at the same radial distance, it is also difficult to explain the strong oxygen emission. We conclude that a region of relatively high density, possibly analogous to the CO clump, is needed to produce the observed O\,I emission. 

\section{Paper~III}
In this paper, we consider the Fomalhaut debris belt, which is characterised by its eccentricity and its sharp edges. Such a morphology can be attributed to a perturbing planet (or a pair of shepherding planets). However, the only planetary candidate so far, Dagon, is found to follow an extremely eccentric orbit. Thus, Dagon cannot be at the origin of the observed morphology. Alternatively, the morphology might be due to gas-dust interactions. We aim to test this possibility by analysing non-detections of C\,II and O\,I emission by \textit{Herschel}/PACS (figures \ref{lineemission_cii} and \ref{lineemission_oi}). Since we want to assess the possibility of gas-dust interactions, we assume a model where the gas exactly follows the known dust distribution. We then calculate how much flux would be registered in each individual spaxel of PACS for a given total C\,II or O\,I emission. This allows us to put an upper limit on the C\,II and O\,I emission from gas co-spatial with the dust. The second step is to convert this flux upper limits into upper limits on the gas mass. To do this, we treat the kinetic temperature as a free parameter and use \texttt{RADEX} to model non-LTE excitation by electrons and hydrogen as well as the radiative transfer. We assume either solar abundances or $\beta$~Pic-like abundances (i.e.\ gas dominated by C and O). We find that for both types of abundances, the gas mass remains below the dust mass over a wide range of kinetic temperatures. This implies that gas-dust interactions cannot work efficiently in the Fomalhaut belt. Thus, the belt's morphology is probably due to a yet unseen planet or stellar encounters.

\section{Paper~IV}
In this paper, we assess the possibility to detect biosignatures in the debris created during an impact event on an exoplanet. The standard approach to find a biosphere on an exoplanet is to look at the planet's atmosphere, because we know that on Earth, life is heavily influencing the atmosphere's composition. One might also look at light reflected from the planetary surface. Observing biosignatures in impact-generated debris (for example, certain minerals indicating the presence of life or ejected microorganisms) instead would be complementary to these methods in a number of ways. For example, the largest number of debris is produced from impacts on small planets, the atmospheres of which are hardest to observe. Also, ejected debris might carry information about an otherwise invisible subsurface biosphere.

For given parameters of the impact event such as the size of the impactor (we take 20\,km, i.e.\ comparable to the Chicxulub impactor), the impact velocity or the size of the target planet, we calculate the total mass that can escape the planet using scaling laws determined from laboratory measurements. We distinguish between material escaping in the solid state (spalled debris) and material that melts or vaporises during the impact event (recondensed debris). This is important if one is interested in the detection of material that is destroyed at high temperature. The second step is to calculate the collisional evolution of the newly created circumstellar `debris disk'. We use a simple analytical model to infer the dust production and thus the fractional luminosity of the debris over time. We find that the spalled population of debris evolves on timescales of millions of years, while the recondensed population is much brighter initially, but faints away on much shorter timescales. We then investigate whether current or future instruments could detect impact-generated debris. We find that the amount of dust produced during the considered impact is potentially sufficient to be detected with current instrumentation. However, to study the dust \emph{composition}, one has to wait for future instruments. Our calculations also suggest that the direct detection of living matter in the debris is near impossible.


\backmatterSU


\selectlanguage{swedish}

\chapter{Svensk sammanfattning}
Fragmentskivor \"ar cirkumstell\"ara skivor av rymdstoft. De motsvarar asteroidb\"altet eller Kuiperb\"altet i solsystemet. Stoftet i fragmentskivor produceras fr\r{a}n kometer och asteroider som kolliderar. Fragmentskivor \"ar ett resultat av planetbildning. Genom att observera fragmentskivor kan vi d\"arf\"or testa teorier av planetbildning och l\"ara oss n\r{a}gonting om solsystemets uppkomst. Dessutom kan planeter f\"or\"andra en fragmentskivas utseende och till exempel f\"ororsaka gap. Det betyder att man i vissa fall kan uppt\"acka planeter genom att titta p\r{a} fragmentskivors struktur. F\"or att sammanfatta kan man s\"aga att studier av fragmentskivor \"ar en viktig del av exoplanetarisk forskning. Den h\"ar avhandlingen handlar om n\r{a}gra aspekter av fragmentskivor.

N\r{a}gra fragmentskivor best\r{a}r inte bara av rymdstoft, men har \"aven synliga m\"angder av gas. Ett exempel \"ar den unga A-stj\"arnan $\beta$~Pictoris. Troligtvis \"ar gasen runt $\beta$~Pic inte bara en rest av den protoplanet\"ara skivan. Man tror att den kontinuerligt produceras fr\r{a}n stoftet sj\"alv. Gasen kan till exempel produceras n\"ar stoftpartiklar kolliderar med h\"og hastighet eller n\"ar UV fotoner fr\r{a}n stj\"arnan sl\r{a}r ut atomer eller molekyler fr\r{a}n stoftpartiklar. \"Aven kolliderande kometer kan producera gas. Det betyder att man kan l\"ara sig n\r{a}gonting om stoftet (och d\"arf\"or planeters byggstenar) genom att observera gasen, till exempel stoftets sammans\"attning. I fallet av gasen runt $\beta$~Pic \"ar f\"orekomst av kol och syre mycket h\"ogt j\"amf\"ort med f\"orekomsten i solsystemet. Det \"ar dock oklart varf\"or det finns s\r{a} mycket kol och syre. I artikel I presenterar vi observationer av C\,II emission med rymdteleskopet \textit{Herschel}/HIFI. Vi anv\"ander observationer f\"or att l\"ara oss hur kol-gasen \"ar f\"ordelad inom skivan. Det kan hj\"alpa att f\"orst\r{a} hur gasen produceras av stoft och varf\"or det finns s\r{a} mycket kol, \"aven om det finns fortfarande m\r{a}nga ol\"osta fr\r{a}gor. Nya observationer av C\,I med ALMA kommer hj\"alpa att kartl\"agga kolens f\"ordelning inom skivan med mycket h\"ogre noggrannhet. I artikel II analyserar vi observationer av C\,II och O\,I med \textit{Herschel}/PACS. Emissionen fr\r{a}n syre \"ar mycket h\"ogre \"an f\"orv\"antat. Det visar sig att man m\r{a}ste anta att syre befinner sig i en region av h\"og t\"athet, m\"ojligtvis en klump, f\"or att f\"orklara det. Kanske finns det en koppling till CO klumpen som observerats med ALMA. \"Aven h\"ar kommer v\r{a}ra ALMA observationer av C\,I att vara till hj\"alp med att f\"orst\r{a} varf\"or vi observerar s\r{a} mycket syre-emission.

En annan v\"alk\"and fragmentskiva finns runt A-stj\"arnan Fomalhaut. Den beskrivs faktiskt b\"attre som ett excentriskt fragment\emph{b\"alte} med skarpa kanter, en morfologi som antyder att det finns en planet. Den enda planetkandidaten vi k\"anner till har dock en omloppsbana som visar att den inte kan vara ansvarig f\"or morfologin av fragmentb\"altet. \"Aven interaktioner mellan stoft och gas kan g\"ora att ett fragmentb\"alte blir excentriskt och har skarpa kanter. S\r{a} kanske beh\"over man ingen planet f\"or att f\"orklara b\"altets morfologi? I artikel III testar vi det genom att analysera icke-detektioner av C\,II och O\,I emission fr\r{a}n Fomalhaut med \textit{Herschel}/PACS. Vi visar att det inte finns tillr\"ackligt mycket gas f\"or effektiva gas-stoft interaktioner. Det betyder att morfologin av stoftb\"altet f\"orklaras b\"ast med en \"annu osedd planet.

En av de st\"orsta fr\r{a}gorna inom exoplanetforskning \"ar om det finns andra planeter med liv. Eftersom vi uppt\"acker fler och fler jordlika planeter inom den beboeliga zonen kan vi kanske svara p\r{a} den fr\r{a}gan inom de n\"armaste decennierna. Det finns olika metoder att uppt\"acka liv p\r{a} exoplaneter. Man kan till exempel titta p\r{a} planetens atmosf\"ar eftersom vi vet att p\r{a} jorden s\r{a} influerar liv atmosf\"arens sammans\"attning kraftigt. I artikel IV studerar vi om det \"ar ist\"allet m\"ojligt att hitta sp\r{a}r av liv (till exempel vissa mineraler eller mikroorganismer) inom en fragmentskiva som best\r{a}r av stoft som kommer fr\r{a}n planetens yta. En s\r{a}dan fragmentskiva kan bildas om en asteroid eller komet kolliderar med planeten. Vi r\"aknar ut hur mycket stoft som bildas f\"or ett nedslag i samma storleksordning som Chicxulub-nedslaget. Vi unders\"oker ocks\r{a} hur ljusstyrkan av stoft f\"or\"andras med tiden efter nedslaget. Det visar sig att ett nedslag kan producera stoft med ett totalt tv\"arsnitt som \"ar mycket st\"orre en planetens tv\"arrsnitt. V\r{a}r slutsats \"ar att det i princip \"ar m\"ojligt att hitta stj\"arnor med stoft fr\r{a}n ett nedslag med nu existerande teleskop. F\"or att studera sammans\"attningen av stoft och hitta eventuella sp\r{a}r av liv beh\"ovs dock nya, k\"ansligare teleskop.

\selectlanguage{english} 

\chapter{\ECFAugie Acknowledgements}
\vspace{-1cm}
It would have been impossible to complete my PhD studies without the help of various people. They are all part of this thesis in one or the other way. I want to thank them all here, and I apologise for anyone I forget.

First, Alexis, my supervisor. Thank you for guiding me through this four and a half year long journey. Thank you for always being available for questions and discussions and for your endless patience. Perhaps the most important lesson I have learned from you is that science can be great fun. And thanks for being a good entertainer (`pajas') at lunch! G\"oran, my second supervisor, has been our life insurance. Whenever you are really stuck, ask G\"oran! He will have some ideas\dots Besides that, G\"oran is really en trevlig kille, always ready for a small chat. Tack G\"oran! I also want to thank my mentor Peter for taking care of me all the time and bringing the two aforementioned guys back down to Earth in numerous thesis group meetings.

I would like to thank all the members of the star and planet formation group, G\"osta ($y=\ln\left(\frac{x}{m}-sa \right)/r^2$), Aage, H-G, Bengt and Markus, for a nice working environment. Thanks Markus for establishing a lunch tradition for our group.

Being part of the Astrobiology Centre has been an enormous pleasure to me. Special thanks go to Wolf for organising the best summer schools ever and to Axel for all his enthusiasm for science. I also enjoyed the company of the other astrobiology students Pantea, Fredrik, Tao and especially Engy, who was a great collaborator for the biosignature project.

I am grateful to Sandra and Rocio for making our daily life at the department as smooth as possible. And thanks Sergio for helping to solve my computer issues. I generated quite some help tickets over these years\dots

\begin{CJK}{UTF8}{gbsn}
The department of astronomy is a great working place and I want to thank all people for contributing. Special thanks go to Garrelt (for taking care of me during the first weeks of my studies), Magnus (for support with the telescope at numerous occasions), Margareta (for always talking Swedish to me), Emily (for organising seminar bingos; I won some absolutely awesome prizes),  Hiva (for being the only one playing bingo with me), Jens (for recognising that exoplanets make great student exercises), Andrii (the knife is yours alone now, use it responsibly), Francesco (for always being ready to talk football), Emanuel (for sharing a passion for good beer), Hannes (for some funny Munchkin sessions), Martina (for helping me a lot to get started here), Utte (for nice board game nights), Vasco (for taking me to Dalarnas Schweiz), Serena, Ben and Simona (for being very pleasant office mates), Florent (for a tasty raclette in the VLR), Ricky (my precursor---thanks for giving me some useful tips), Andreas (for telling me everything about Swedish culture), Saghar (for believing that I actually speak Persian), Anders (for always greeting in Sk\r{a}nska), Illa (for keeping up the social life of the PhD students), J\"orn and Johannes (for speaking German to me), Veronica (for speaking Bolivian German to me), Tine (for organising parties, camping, kayaking,\dots), Emir (for bringing the American spirit), Katia and Th\o ger (for always being in a good mood), Rub\'en (for the foundation of the astrogang), Matteo (for leaving some fruit to the people of the Eastern part of the corridor), Esha (for nice chats during fika), Johan (for organising FIFA sessions as well as an amazing cabin trip to Kopparberg), Maria (for interpreting my T-shirts), Carolina (for amazing fikas), Mattias (for helping me in a desperate situation during a telescope show), Jayant (for taking care of my plants---they survived), and Kai Yan (for numerous lunches at 朝, for being a friend, and for trying to take care of my plants---they did not survive it).
\end{CJK}

My second passion besides science is football. Jag vill tacka alla grabbar fr\r{a}n BK Tr\"asket f\"or att jag kunde vara en del av denna h\"arliga lag! Vi \"ar kanske inte Sveriges mest framg\r{a}ngsrika lag, men det spelar ingen roll---det var alltid grymt att lira med er! Ich danke auch der 3.\ Mannschaft des FC Rebstein daf\"ur dass ich immer ein Teil dieser Mannschaft bleiben durfte obwohl ich nicht mehr in der Schweiz wohne. Danke den Trainern Pius und Erwin dass ihr mich immer ein bisschen spielen lasst wenn ich wieder mal zu Besuch bin! Speziellen Dank an meine Freunde Hugi, Flo, Adi, Nici, Michi, Meier, Mike, Dago, J\"ungling, T\"ongi f\"ur die zwei besten Ferien aller Zeiten! Danke Hugi dass du noch in Stockholm vorbeigeschaut hast!

Meine tiefste Dankbarkeit gilt meiner Familie: meinen Eltern Giovanni und Angelina und meinen Br\"udern Fabiano und Filippo. Ihr habt mich immer unterst\"utzt, selbst wenn es vielleicht nicht immer klar war was ich eigentlich mache. Ich komme immer gerne nach Hause, denn mein Zuhause seid ihr.

Finally, I want to thank you, Sarah, for all the love you give me every day. Thanks that you are in my life and take care of me, you are the best! And yes, if I ever discover extraterrestrial life, it will be named after you. \RL{mers_A sArA!}  



\def\bibfont{\scriptsize}

\bibliographystyle{aa}

\renewcommand{\bibname}{References} 

\bibliography{references} 

\begin{thebibliography}{214}
\expandafter\ifx\csname natexlab\endcsname\relax\def\natexlab#1{#1}\fi

\bibitem[{{Abt} \& {Moyd}(1973)}]{Abt_Moyd_1973}
{Abt}, H.~A. \& {Moyd}, K.~I. 1973, \apj, 182, 809

\bibitem[{{Acad\'emie des sciences}(1900)}]{AcademieDesSciences_1900}
{Acad\'emie des sciences}. 1900, {Comptes rendus hebdomadaires des s\'eances de
  l'Acad\'emie des sciences} (Bachelier (Paris)), 1147

\bibitem[{{Acke} {et~al.}(2012){Acke}, {Min}, {Dominik}, {Vandenbussche},
  {Sibthorpe}, {Waelkens}, {Olofsson}, {Degroote}, {Smolders}, {Pantin},
  {Barlow}, {Blommaert}, {Brandeker}, {De Meester}, {Dent}, {Exter}, {Di
  Francesco}, {Fridlund}, {Gear}, {Glauser}, {Greaves}, {Harvey}, {Henning},
  {Hogerheijde}, {Holland}, {Huygen}, {Ivison}, {Jean}, {Liseau}, {Naylor},
  {Pilbratt}, {Polehampton}, {Regibo}, {Royer}, {Sicilia-Aguilar}, \&
  {Swinyard}}]{Acke_etal_2012}
{Acke}, B., {Min}, M., {Dominik}, C., {et~al.} 2012, \aap, 540, A125

\bibitem[{{ALMA Partnership} {et~al.}(2015){ALMA Partnership}, {Brogan},
  {P{\'e}rez}, {Hunter}, {Dent}, {Hales}, {Hills}, {Corder}, {Fomalont},
  {Vlahakis}, {Asaki}, {Barkats}, {Hirota}, {Hodge}, {Impellizzeri}, {Kneissl},
  {Liuzzo}, {Lucas}, {Marcelino}, {Matsushita}, {Nakanishi}, {Phillips},
  {Richards}, {Toledo}, {Aladro}, {Broguiere}, {Cortes}, {Cortes}, {Espada},
  {Galarza}, {Garcia-Appadoo}, {Guzman-Ramirez}, {Humphreys}, {Jung}, {Kameno},
  {Laing}, {Leon}, {Marconi}, {Mignano}, {Nikolic}, {Nyman}, {Radiszcz},
  {Remijan}, {Rod{\'o}n}, {Sawada}, {Takahashi}, {Tilanus}, {Vila Vilaro},
  {Watson}, {Wiklind}, {Akiyama}, {Chapillon}, {de Gregorio-Monsalvo}, {Di
  Francesco}, {Gueth}, {Kawamura}, {Lee}, {Nguyen Luong}, {Mangum}, {Pietu},
  {Sanhueza}, {Saigo}, {Takakuwa}, {Ubach}, {van Kempen}, {Wootten},
  {Castro-Carrizo}, {Francke}, {Gallardo}, {Garcia}, {Gonzalez}, {Hill},
  {Kaminski}, {Kurono}, {Liu}, {Lopez}, {Morales}, {Plarre}, {Schieven},
  {Testi}, {Videla}, {Villard}, {Andreani}, {Hibbard}, \&
  {Tatematsu}}]{ALMA_etal_2015}
{ALMA Partnership}, {Brogan}, C.~L., {P{\'e}rez}, L.~M., {et~al.} 2015, \apjl,
  808, L3

\bibitem[{Apai \& Lauretta(2010)}]{Apai_Lauretta_2010}
Apai, D. \& Lauretta, D.~S. 2010, in Protoplanetary Dust, ed. D.~Apai \& D.~S.
  Lauretta (Cambridge University Press), 1--26

\bibitem[{Armitage(2009)}]{Armitage_2009}
Armitage, P.~J. 2009, Astrophysics of Planet Formation (Cambridge University
  Press)

\bibitem[{{Augereau} {et~al.}(2001){Augereau}, {Nelson}, {Lagrange},
  {Papaloizou}, \& {Mouillet}}]{Augereau_etal_2001}
{Augereau}, J.~C., {Nelson}, R.~P., {Lagrange}, A.~M., {Papaloizou}, J.~C.~B.,
  \& {Mouillet}, D. 2001, \aap, 370, 447

\bibitem[{{Backman} \& {Paresce}(1993)}]{Backman_Paresce_1993}
{Backman}, D.~E. \& {Paresce}, F. 1993, in Protostars and Planets III, ed.
  E.~H. {Levy} \& J.~I. {Lunine}, 1253--1304

\bibitem[{Balbus(2009)}]{Balbus_2009}
Balbus, S.~A. 2009, Scholarpedia, 4, 2409, {revision \#91455}

\bibitem[{Barlow(1989)}]{Barlow_1989}
Barlow, R. 1989, Statistics: A Guide to the Use of Statistical Methods in the
  Physical Sciences, Manchester Physics Series (Wiley)

\bibitem[{{Beichman} {et~al.}(2011){Beichman}, {Lisse}, {Tanner}, {Bryden},
  {Akeson}, {Ciardi}, {Boden}, {Dodson-Robinson}, {Salyk}, \&
  {Wyatt}}]{Beichman_etal_2011}
{Beichman}, C.~A., {Lisse}, C.~M., {Tanner}, A.~M., {et~al.} 2011, \apj, 743,
  85

\bibitem[{Bennett \& Shostak(2007)}]{Bennett_Shostak_2007}
Bennett, J. \& Shostak, G. 2007, Life in the Universe (Pearson Addison Wesley)

\bibitem[{{Benz} \& {Asphaug}(1999)}]{Benz_Asphaug_1999}
{Benz}, W. \& {Asphaug}, E. 1999, \icarus, 142, 5

\bibitem[{{Berdyugina} {et~al.}(2016){Berdyugina}, {Kuhn}, {Harrington}, {{\v
  S}antl-Temkiv}, \& {Messersmith}}]{Berdyugina_etal_2016}
{Berdyugina}, S.~V., {Kuhn}, J.~R., {Harrington}, D.~M., {{\v S}antl-Temkiv},
  T., \& {Messersmith}, E.~J. 2016, International Journal of Astrobiology, 15,
  45

\bibitem[{{Besla} \& {Wu}(2007)}]{Besla_Wu_2007}
{Besla}, G. \& {Wu}, Y. 2007, \apj, 655, 528

\bibitem[{{Beust}(2014)}]{Beust_2014}
{Beust}, H. 2014, in Thirty years of Beta Pic and Debris Disk Studies, ed.
  A.-M. {Lagrange} \& A.~{Boccaletti}

\bibitem[{{Beust} {et~al.}(2014){Beust}, {Augereau}, {Bonsor}, {Graham},
  {Kalas}, {Lebreton}, {Lagrange}, {Ertel}, {Faramaz}, \&
  {Th{\'e}bault}}]{Beust_etal_2014}
{Beust}, H., {Augereau}, J.-C., {Bonsor}, A., {et~al.} 2014, \aap, 561, A43

\bibitem[{{Beust} \& {Valiron}(2007)}]{Beust_Valiron_2007}
{Beust}, H. \& {Valiron}, P. 2007, \aap, 466, 201

\bibitem[{Blasius {et~al.}(2008)Blasius, H\"ubscher, \&
  Sommer}]{Blasius_etal_2008}
Blasius, M., H\"ubscher, U., \& Sommer, S. 2008, Critical Reviews in
  Biochemistry and Molecular Biology, 43, 221

\bibitem[{{Boch} \& {Fernique}(2014)}]{Boch_Fernique_2014}
{Boch}, T. \& {Fernique}, P. 2014, in Astronomical Society of the Pacific
  Conference Series, Vol. 485, Astronomical Data Analysis Software and Systems
  XXIII, ed. N.~{Manset} \& P.~{Forshay}, 277

\bibitem[{Bohren \& Huffman(1998)}]{Bohren_Huffman_1998_AppendixA}
Bohren, C.~F. \& Huffman, D.~R. 1998, Appendix A: Homogeneous Sphere (Wiley
  Science Paperback Series), 477--482

\bibitem[{{Boley}(2009)}]{Boley_2009}
{Boley}, A.~C. 2009, \apjl, 695, L53

\bibitem[{{Boley} {et~al.}(2012){Boley}, {Payne}, {Corder}, {Dent}, {Ford}, \&
  {Shabram}}]{Boley_etal_2012}
{Boley}, A.~C., {Payne}, M.~J., {Corder}, S., {et~al.} 2012, \apjl, 750, L21

\bibitem[{{Bonnarel} {et~al.}(2000){Bonnarel}, {Fernique}, {Bienaym{\'e}},
  {Egret}, {Genova}, {Louys}, {Ochsenbein}, {Wenger}, \&
  {Bartlett}}]{Bonnarel_etal_2000}
{Bonnarel}, F., {Fernique}, P., {Bienaym{\'e}}, O., {et~al.} 2000, \aaps, 143,
  33

\bibitem[{{Booth} {et~al.}(2013){Booth}, {Kennedy}, {Sibthorpe}, {Matthews},
  {Wyatt}, {Duch{\^e}ne}, {Kavelaars}, {Rodriguez}, {Greaves}, {Koning},
  {Vican}, {Rieke}, {Su}, {Moro-Mart{\'{\i}}n}, \& {Kalas}}]{Booth_etal_2013}
{Booth}, M., {Kennedy}, G., {Sibthorpe}, B., {et~al.} 2013, \mnras, 428, 1263

\bibitem[{{Booth} {et~al.}(2009){Booth}, {Wyatt}, {Morbidelli},
  {Moro-Mart{\'{\i}}n}, \& {Levison}}]{Booth_etal_2009}
{Booth}, M., {Wyatt}, M.~C., {Morbidelli}, A., {Moro-Mart{\'{\i}}n}, A., \&
  {Levison}, H.~F. 2009, \mnras, 399, 385

\bibitem[{{Boss}(2000)}]{Boss_2000}
{Boss}, A.~P. 2000, \apjl, 536, L101

\bibitem[{{Boulanger} {et~al.}(2000){Boulanger}, {Cox}, \&
  {Jones}}]{Boulanger_etal_2000}
{Boulanger}, F., {Cox}, P., \& {Jones}, A.~P. 2000, in Infrared Space
  Astronomy, Today and Tomorrow, ed. F.~{Casoli}, J.~{Lequeux}, \& F.~{David},
  251

\bibitem[{{Boyajian} {et~al.}(2016){Boyajian}, {LaCourse}, {Rappaport},
  {Fabrycky}, {Fischer}, {Gandolfi}, {Kennedy}, {Korhonen}, {Liu}, {Moor},
  {Olah}, {Vida}, {Wyatt}, {Best}, {Brewer}, {Ciesla}, {Cs{\'a}k}, {Deeg},
  {Dupuy}, {Handler}, {Heng}, {Howell}, {Ishikawa}, {Kov{\'a}cs}, {Kozakis},
  {Kriskovics}, {Lehtinen}, {Lintott}, {Lynn}, {Nespral}, {Nikbakhsh},
  {Schawinski}, {Schmitt}, {Smith}, {Szabo}, {Szabo}, {Viuho}, {Wang},
  {Weiksnar}, {Bosch}, {Connors}, {Goodman}, {Green}, {Hoekstra}, {Jebson},
  {Jek}, {Omohundro}, {Schwengeler}, \& {Szewczyk}}]{Boyajian_etal_2016}
{Boyajian}, T.~S., {LaCourse}, D.~M., {Rappaport}, S.~A., {et~al.} 2016, \mnras

\bibitem[{{Brandeker}(2011)}]{Brandeker_2011}
{Brandeker}, A. 2011, \apj, 729, 122

\bibitem[{{Brandeker} {et~al.}(2004){Brandeker}, {Liseau}, {Olofsson}, \&
  {Fridlund}}]{Brandeker_etal_2004}
{Brandeker}, A., {Liseau}, R., {Olofsson}, G., \& {Fridlund}, M. 2004, \aap,
  413, 681

\bibitem[{{Burns} {et~al.}(1979){Burns}, {Lamy}, \& {Soter}}]{Burns_etal_1979}
{Burns}, J.~A., {Lamy}, P.~L., \& {Soter}, S. 1979, \icarus, 40, 1

\bibitem[{{Cassan} {et~al.}(2012){Cassan}, {Kubas}, {Beaulieu}, {Dominik},
  {Horne}, {Greenhill}, {Wambsganss}, {Menzies}, {Williams}, {J{\o}rgensen},
  {Udalski}, {Bennett}, {Albrow}, {Batista}, {Brillant}, {Caldwell}, {Cole},
  {Coutures}, {Cook}, {Dieters}, {Prester}, {Donatowicz}, {Fouqu{\'e}}, {Hill},
  {Kains}, {Kane}, {Marquette}, {Martin}, {Pollard}, {Sahu}, {Vinter},
  {Warren}, {Watson}, {Zub}, {Sumi}, {Szyma{\'n}ski}, {Kubiak}, {Poleski},
  {Soszynski}, {Ulaczyk}, {Pietrzy{\'n}ski}, \&
  {Wyrzykowski}}]{Cassan_etal_2012}
{Cassan}, A., {Kubas}, D., {Beaulieu}, J.-P., {et~al.} 2012, \nat, 481, 167

\bibitem[{Cataldi(2013)}]{Cataldi_2013}
Cataldi, G. 2013, Licentiate thesis, Debris disks from an astronomical and an
  astrobiological viewpoint, Department of Astronomy, Stockholm University

\bibitem[{{Chen} \& {Jura}(2003)}]{Chen_Jura_2003}
{Chen}, C.~H. \& {Jura}, M. 2003, \apj, 582, 443

\bibitem[{{Chen} {et~al.}(2007){Chen}, {Li}, {Bohac}, {Kim}, {Watson}, {van
  Cleve}, {Houck}, {Stapelfeldt}, {Werner}, {Rieke}, {Su}, {Marengo},
  {Backman}, {Beichman}, \& {Fazio}}]{Chen_etal_2007}
{Chen}, C.~H., {Li}, A., {Bohac}, C., {et~al.} 2007, \apj, 666, 466

\bibitem[{{Chiang} {et~al.}(2007){Chiang}, {Lithwick}, {Murray-Clay}, {Buie},
  {Grundy}, \& {Holman}}]{Chiang_etal_2007}
{Chiang}, E., {Lithwick}, Y., {Murray-Clay}, R., {et~al.} 2007, Protostars and
  Planets V, 895

\bibitem[{{Chyba} \& {Hand}(2005)}]{Chyba_Hand_2005}
{Chyba}, C.~F. \& {Hand}, K.~P. 2005, \araa, 43, 31

\bibitem[{{Cleland} \& {Chyba}(2002)}]{Cleland_Chyba_2002}
{Cleland}, C.~E. \& {Chyba}, C.~F. 2002, Origins of Life and Evolution of the
  Biosphere, 32, 387

\bibitem[{Cockell(2015)}]{Cockell_2015}
Cockell, C. 2015, Astrobiology: Understanding Life in the Universe (Wiley)

\bibitem[{{Cockell} {et~al.}(2009){Cockell}, {L{\'e}ger}, {Fridlund}, {Herbst},
  {Kaltenegger}, {Absil}, {Beichman}, {Benz}, {Blanc}, {Brack}, {Chelli},
  {Colangeli}, {Cottin}, {Coud{\'e} du Foresto}, {Danchi}, {Defr{\`e}re}, {den
  Herder}, {Eiroa}, {Greaves}, {Henning}, {Johnston}, {Jones}, {Labadie},
  {Lammer}, {Launhardt}, {Lawson}, {Lay}, {LeDuigou}, {Liseau}, {Malbet},
  {Martin}, {Mawet}, {Mourard}, {Moutou}, {Mugnier}, {Ollivier}, {Paresce},
  {Quirrenbach}, {Rabbia}, {Raven}, {Rottgering}, {Rouan}, {Santos}, {Selsis},
  {Serabyn}, {Shibai}, {Tamura}, {Thi{\'e}baut}, {Westall}, \&
  {White}}]{Cockell_etal_2009}
{Cockell}, C.~S., {L{\'e}ger}, A., {Fridlund}, M., {et~al.} 2009, Astrobiology,
  9, 1

\bibitem[{Collins \& Johnson(2014)}]{Collins_Johnson_2014}
Collins, G. \& Johnson, T.~V. 2014, in Encyclopedia of the Solar System, 3rd
  edn., ed. T.~Spohn, D.~Breuer, \& T.~V. Johnson (Boston: Elsevier), 813 --
  829

\bibitem[{Conway(2004)}]{Conway_2004}
Conway, A. 2004, in An Introduction to Astrobiology, ed. I.~{Gilmour} \& M.~A.
  Sephton (Open University), 281--302

\bibitem[{Coustenis(2014)}]{Coustenis_2014}
Coustenis, A. 2014, in Encyclopedia of the Solar System, 3rd edn., ed.
  T.~Spohn, D.~Breuer, \& T.~V. Johnson (Boston: Elsevier), 831 -- 849

\bibitem[{{Currie} {et~al.}(2013){Currie}, {Burrows}, {Madhusudhan},
  {Fukagawa}, {Girard}, {Dawson}, {Murray-Clay}, {Kenyon}, {Kuchner},
  {Matsumura}, {Jayawardhana}, {Chambers}, \& {Bromley}}]{Currie_etal_2013}
{Currie}, T., {Burrows}, A., {Madhusudhan}, N., {et~al.} 2013, \apj, 776, 15

\bibitem[{{Czechowski} \& {Mann}(2007)}]{Czechowski_Mann_2007}
{Czechowski}, A. \& {Mann}, I. 2007, \apj, 660, 1541

\bibitem[{{Dalton} {et~al.}(2003){Dalton}, {Mogul}, {Kagawa}, {Chan}, \&
  {Jamieson}}]{Dalton_etal_2003}
{Dalton}, J.~B., {Mogul}, R., {Kagawa}, H.~K., {Chan}, S.~L., \& {Jamieson},
  C.~S. 2003, Astrobiology, 3, 505

\bibitem[{{D'Angelo} {et~al.}(2010){D'Angelo}, {Durisen}, \&
  {Lissauer}}]{DAngelo_etal_2010}
{D'Angelo}, G., {Durisen}, R.~H., \& {Lissauer}, J.~J. 2010, {Giant Planet
  Formation}, ed. S.~{Seager} (University of Arizona Press), 319--346

\bibitem[{{David} {et~al.}(2013){David}, {Hensberge}, \&
  {Nitschelm}}]{David_etal_2013}
{David}, M., {Hensberge}, H., \& {Nitschelm}, C. 2013, \aap, 557, A47

\bibitem[{{de Vries} {et~al.}(2012){de Vries}, {Acke}, {Blommaert}, {Waelkens},
  {Waters}, {Vandenbussche}, {Min}, {Olofsson}, {Dominik}, {Decin}, {Barlow},
  {Brandeker}, {di Francesco}, {Glauser}, {Greaves}, {Harvey}, {Holland},
  {Ivison}, {Liseau}, {Pantin}, {Pilbratt}, {Royer}, \&
  {Sibthorpe}}]{deVries_etal_2012}
{de Vries}, B.~L., {Acke}, B., {Blommaert}, J.~A.~D.~L., {et~al.} 2012, \nat,
  490, 74

\bibitem[{{Dent} {et~al.}(2014){Dent}, {Wyatt}, {Roberge}, {Augereau},
  {Casassus}, {Corder}, {Greaves}, {de Gregorio-Monsalvo}, {Hales}, {Jackson},
  {Hughes}, {Lagrange}, {Matthews}, \& {Wilner}}]{Dent_etal_2014}
{Dent}, W.~R.~F., {Wyatt}, M.~C., {Roberge}, A., {et~al.} 2014, Science, 343,
  1490

\bibitem[{{Des Marais} {et~al.}(2002){Des Marais}, {Harwit}, {Jucks},
  {Kasting}, {Lin}, {Lunine}, {Schneider}, {Seager}, {Traub}, \&
  {Woolf}}]{DesMarais_etal_2002}
{Des Marais}, D.~J., {Harwit}, M.~O., {Jucks}, K.~W., {et~al.} 2002,
  Astrobiology, 2, 153

\bibitem[{{Dohnanyi}(1969)}]{Dohnanyi_1969}
{Dohnanyi}, J.~S. 1969, \jgr, 74, 2531

\bibitem[{{Donaldson} {et~al.}(2013){Donaldson}, {Lebreton}, {Roberge},
  {Augereau}, \& {Krivov}}]{Donaldson_etal_2013}
{Donaldson}, J.~K., {Lebreton}, J., {Roberge}, A., {Augereau}, J.-C., \&
  {Krivov}, A.~V. 2013, \apj, 772, 17

\bibitem[{{Evans} \& {Maunder}(1903)}]{Evans_Maunder_1903}
{Evans}, J.~E. \& {Maunder}, E.~W. 1903, \mnras, 63, 488

\bibitem[{{Faramaz} {et~al.}(2015){Faramaz}, {Beust}, {Augereau}, {Kalas}, \&
  {Graham}}]{Faramaz_etal_2015}
{Faramaz}, V., {Beust}, H., {Augereau}, J.-C., {Kalas}, P., \& {Graham}, J.~R.
  2015, \aap, 573, A87

\bibitem[{{Fern{\'a}ndez} {et~al.}(2006){Fern{\'a}ndez}, {Brandeker}, \&
  {Wu}}]{Fernandez_etal_2006}
{Fern{\'a}ndez}, R., {Brandeker}, A., \& {Wu}, Y. 2006, \apj, 643, 509

\bibitem[{{Ferri{\`e}re}(2001)}]{Ferriere_2001}
{Ferri{\`e}re}, K.~M. 2001, Reviews of Modern Physics, 73, 1031

\bibitem[{Forget(2007)}]{Forget_2007}
Forget, F. 2007, in Lectures in Astrobiology, ed. M.~Gargaud, H.~Martin, \&
  P.~Claeys, Advances in Astrobiology and Biogeophysics (Springer Berlin
  Heidelberg), 103--122

\bibitem[{{Fraine} {et~al.}(2014){Fraine}, {Deming}, {Benneke}, {Knutson},
  {Jord{\'a}n}, {Espinoza}, {Madhusudhan}, {Wilkins}, \&
  {Todorov}}]{Fraine_etal_2014}
{Fraine}, J., {Deming}, D., {Benneke}, B., {et~al.} 2014, \nat, 513, 526

\bibitem[{{France} {et~al.}(2007){France}, {Roberge}, {Lupu}, {Redfield}, \&
  {Feldman}}]{France_etal_2007}
{France}, K., {Roberge}, A., {Lupu}, R.~E., {Redfield}, S., \& {Feldman}, P.~D.
  2007, \apj, 668, 1174

\bibitem[{{Fressin} {et~al.}(2013){Fressin}, {Torres}, {Charbonneau}, {Bryson},
  {Christiansen}, {Dressing}, {Jenkins}, {Walkowicz}, \&
  {Batalha}}]{Fressin_etal_2013}
{Fressin}, F., {Torres}, G., {Charbonneau}, D., {et~al.} 2013, \apj, 766, 81

\bibitem[{{Freudling} {et~al.}(1995){Freudling}, {Lagrange}, {Vidal-Madjar},
  {Ferlet}, \& {Forveille}}]{Freudling_etal_1995}
{Freudling}, W., {Lagrange}, A.-M., {Vidal-Madjar}, A., {Ferlet}, R., \&
  {Forveille}, T. 1995, \aap, 301, 231

\bibitem[{{Fujiwara} {et~al.}(2013){Fujiwara}, {Ishihara}, {Onaka}, {Takita},
  {Kataza}, {Yamashita}, {Fukagawa}, {Ootsubo}, {Hirao}, {Enya}, {Marshall},
  {White}, {Nakagawa}, \& {Murakami}}]{Fujiwara_etal_2013}
{Fujiwara}, H., {Ishihara}, D., {Onaka}, T., {et~al.} 2013, \aap, 550, A45

\bibitem[{Gilmour(2004)}]{Gilmour_2004}
Gilmour, I. 2004, in An Introduction to Astrobiology, ed. I.~{Gilmour} \& M.~A.
  Sephton (Open University), 43--84

\bibitem[{{Goans}(2013)}]{Goans_2013}
{Goans}, R.~E. 2013, Medical management of radiological causalities, 4th edn.
  (Armed Forces Radiobiology Research Institute)

\bibitem[{{Gomes} {et~al.}(2005){Gomes}, {Levison}, {Tsiganis}, \&
  {Morbidelli}}]{Gomes_etal_2005}
{Gomes}, R., {Levison}, H.~F., {Tsiganis}, K., \& {Morbidelli}, A. 2005, \nat,
  435, 466

\bibitem[{{Gorti} {et~al.}(2015){Gorti}, {Liseau}, {Sandor}, \&
  {Clarke}}]{Gorti_etal_2015}
{Gorti}, U., {Liseau}, R., {Sandor}, Z., \& {Clarke}, C. 2015, ArXiv e-prints

\bibitem[{{Gray} {et~al.}(2006){Gray}, {Corbally}, {Garrison}, {McFadden},
  {Bubar}, {McGahee}, {O'Donoghue}, \& {Knox}}]{Gray_etal_2006}
{Gray}, R.~O., {Corbally}, C.~J., {Garrison}, R.~F., {et~al.} 2006, \aj, 132,
  161

\bibitem[{{Grazier}(2016)}]{Grazier_2016}
{Grazier}, K.~R. 2016, Astrobiology, 16, 23

\bibitem[{{Greenberg}(2005)}]{Greenberg_2005}
{Greenberg}, R. 2005, {Europa - the Ocean Moon} (Springer Berlin Heidelberg)

\bibitem[{{Grigorieva} {et~al.}(2007){Grigorieva}, {Th{\'e}bault},
  {Artymowicz}, \& {Brandeker}}]{Grigorieva_etal_2007}
{Grigorieva}, A., {Th{\'e}bault}, P., {Artymowicz}, P., \& {Brandeker}, A.
  2007, \aap, 475, 755

\bibitem[{{Harp} {et~al.}(2015){Harp}, {Richards}, {Shostak}, {Tarter},
  {Vakoch}, \& {Munson}}]{Harp_etal_2015}
{Harp}, G.~R., {Richards}, J., {Shostak}, S., {et~al.} 2015, ArXiv e-prints

\bibitem[{{Hartmann}(1998)}]{Hartmann_1998}
{Hartmann}, L. 1998, {Accretion Processes in Star Formation} (Cambridge
  University Press)

\bibitem[{{Hashimoto} {et~al.}(2008){Hashimoto}, {Roos-Serote}, {Sugita},
  {Gilmore}, {Kamp}, {Carlson}, \& {Baines}}]{Hashimoto_etal_2008}
{Hashimoto}, G.~L., {Roos-Serote}, M., {Sugita}, S., {et~al.} 2008, Journal of
  Geophysical Research (Planets), 113, E00B24

\bibitem[{Hazen(2005)}]{Hazen_2005}
Hazen, R. 2005, Genesis: The Scientific Quest for Life's Origins (National
  Academies Press)

\bibitem[{Hazen {et~al.}(2008)Hazen, Papineau, Bleeker, Downs, Ferry, McCoy,
  Sverjensky, \& Yang}]{Hazen_etal_2008}
Hazen, R.~M., Papineau, D., Bleeker, W., {et~al.} 2008, American Mineralogist,
  93, 1693

\bibitem[{{Heap} {et~al.}(2000){Heap}, {Lindler}, {Lanz}, {Cornett}, {Hubeny},
  {Maran}, \& {Woodgate}}]{Heap_etal_2000}
{Heap}, S.~R., {Lindler}, D.~J., {Lanz}, T.~M., {et~al.} 2000, \apj, 539, 435

\bibitem[{Hegde {et~al.}(2015)Hegde, Paulino-Lima, Kent, Kaltenegger, \&
  Rothschild}]{Hegde_etal_2015}
Hegde, S., Paulino-Lima, I.~G., Kent, R., Kaltenegger, L., \& Rothschild, L.
  2015, Proceedings of the National Academy of Sciences, 112, 3886

\bibitem[{{Heng} \& {Keeton}(2009)}]{Heng_Keeton_2009}
{Heng}, K. \& {Keeton}, C.~R. 2009, \apj, 707, 621

\bibitem[{{Holland} {et~al.}(2009){Holland}, {Carpenter}, {Goldsmith},
  {Greaves}, \& {Dowell}}]{Holland_etal_2009}
{Holland}, W., {Carpenter}, J., {Goldsmith}, P., {Greaves}, J., \& {Dowell}, D.
  2009, in ArXiv Astrophysics e-prints, Vol. 2010, astro2010: The Astronomy and
  Astrophysics Decadal Survey

\bibitem[{{Horner} \& {Jones}(2010)}]{Horner_Jones_2010}
{Horner}, J. \& {Jones}, B.~W. 2010, Astronomy and Geophysics, 51, 6.16

\bibitem[{{Houk} \& {Smith-Moore}(1988)}]{Houk_Smith-Moore_1988}
{Houk}, N. \& {Smith-Moore}, M. 1988, {Michigan Catalogue of Two-dimensional
  Spectral Types for the HD Stars. Volume 4, Declinations -26$^{\circ}$.0 to
  -12$^{\circ}$.0.}

\bibitem[{{Hundertmark} {et~al.}(2009){Hundertmark}, {Hessman}, \&
  {Dreizler}}]{Hundertmark_etal_2009}
{Hundertmark}, M., {Hessman}, F.~V., \& {Dreizler}, S. 2009, \aap, 500, 929

\bibitem[{{Jackson} \& {Wyatt}(2012)}]{Jackson_Wyatt_2012}
{Jackson}, A.~P. \& {Wyatt}, M.~C. 2012, \mnras, 425, 657

\bibitem[{{Janson} {et~al.}(2012){Janson}, {Carson}, {Lafreni{\`e}re},
  {Spiegel}, {Bent}, \& {Wong}}]{Janson_etal_2012}
{Janson}, M., {Carson}, J.~C., {Lafreni{\`e}re}, D., {et~al.} 2012, \apj, 747,
  116

\bibitem[{{Jaschek} {et~al.}(1988){Jaschek}, {Jaschek}, \&
  {Andrillat}}]{Jaschek_etal_1988}
{Jaschek}, M., {Jaschek}, C., \& {Andrillat}, Y. 1988, \aaps, 72, 505

\bibitem[{Jimenez(2009)}]{Jimenez_2009}
Jimenez, J. 2009, Vacuum, 84, 2

\bibitem[{{Johnson} {et~al.}(2012){Johnson}, {Lisse}, {Chen}, {Melosh},
  {Wyatt}, {Thebault}, {Henning}, {Gaidos}, {Elkins-Tanton}, {Bridges}, \&
  {Morlok}}]{Johnson_etal_2012}
{Johnson}, B.~C., {Lisse}, C.~M., {Chen}, C.~H., {et~al.} 2012, \apj, 761, 45

\bibitem[{{Joyce}(1994)}]{Joyce_1994}
{Joyce}, G.~F. 1994, {Foreword}, ed. W.~D. {Deamer} \& G.~R. {Fleischaker}
  (Jones \& Bartlett, Boston), xi--xii

\bibitem[{{Kalas}(2005)}]{Kalas_2005}
{Kalas}, P. 2005, \apjl, 635, L169

\bibitem[{{Kalas} {et~al.}(2008){Kalas}, {Graham}, {Chiang}, {Fitzgerald},
  {Clampin}, {Kite}, {Stapelfeldt}, {Marois}, \& {Krist}}]{Kalas_etal_2008}
{Kalas}, P., {Graham}, J.~R., {Chiang}, E., {et~al.} 2008, Science, 322, 1345

\bibitem[{{Kalas} {et~al.}(2005){Kalas}, {Graham}, \&
  {Clampin}}]{Kalas_etal_2005}
{Kalas}, P., {Graham}, J.~R., \& {Clampin}, M. 2005, \nat, 435, 1067

\bibitem[{{Kalas} {et~al.}(2013){Kalas}, {Graham}, {Fitzgerald}, \&
  {Clampin}}]{Kalas_etal_2013}
{Kalas}, P., {Graham}, J.~R., {Fitzgerald}, M.~P., \& {Clampin}, M. 2013, \apj,
  775, 56

\bibitem[{{Kennedy} \& {Wyatt}(2011)}]{Kennedy_Wyatt_2011}
{Kennedy}, G.~M. \& {Wyatt}, M.~C. 2011, \mnras, 412, 2137

\bibitem[{{Kenyon} \& {Bromley}(2006)}]{Kenyon_Bromley_2006}
{Kenyon}, S.~J. \& {Bromley}, B.~C. 2006, \aj, 131, 1837

\bibitem[{Kerrod(2000)}]{Kerrod_2000}
Kerrod, R. 2000, Mars, Planet Library (Lerner Publications)

\bibitem[{{Kiefer} {et~al.}(2014){Kiefer}, {Lecavelier des Etangs}, {Boissier},
  {Vidal-Madjar}, {Beust}, {Lagrange}, {H{\'e}brard}, \&
  {Ferlet}}]{Kiefer_etal_2014}
{Kiefer}, F., {Lecavelier des Etangs}, A., {Boissier}, J., {et~al.} 2014, \nat,
  514, 462

\bibitem[{{Klahr} \& {Lin}(2005)}]{Klahr_Lin_2005}
{Klahr}, H. \& {Lin}, D.~N.~C. 2005, \apj, 632, 1113

\bibitem[{{Knacke}(2003)}]{Knacke_2003}
{Knacke}, R.~F. 2003, Astrobiology, 3, 531

\bibitem[{{Kopparapu} {et~al.}(2013){Kopparapu}, {Ramirez}, {Kasting}, {Eymet},
  {Robinson}, {Mahadevan}, {Terrien}, {Domagal-Goldman}, {Meadows}, \&
  {Deshpande}}]{Kopparapu_etal_2013}
{Kopparapu}, R.~K., {Ramirez}, R., {Kasting}, J.~F., {et~al.} 2013, \apj, 765,
  131

\bibitem[{{Kopparapu} {et~al.}(2014){Kopparapu}, {Ramirez}, {SchottelKotte},
  {Kasting}, {Domagal-Goldman}, \& {Eymet}}]{Kopparapu_etal_2014}
{Kopparapu}, R.~K., {Ramirez}, R.~M., {SchottelKotte}, J., {et~al.} 2014,
  \apjl, 787, L29

\bibitem[{{K{\'o}sp{\'a}l} \& {Mo{\'o}r}(2015)}]{Kospal_Moor_2015}
{K{\'o}sp{\'a}l}, {\'A}. \& {Mo{\'o}r}, A. 2015, in IAU Symposium, Vol. 314,
  Young Stars \& Planets Near the Sun, ed. J.~H. {Kastner}, B.~{Stelzer}, \&
  S.~A. {Metchev}, 183--188

\bibitem[{{K{\'o}sp{\'a}l} {et~al.}(2013){K{\'o}sp{\'a}l}, {Mo{\'o}r},
  {Juh{\'a}sz}, {{\'A}brah{\'a}m}, {Apai}, {Csengeri}, {Grady}, {Henning},
  {Hughes}, {Kiss}, {Pascucci}, \& {Schmalzl}}]{Kospal_etal_2013}
{K{\'o}sp{\'a}l}, {\'A}., {Mo{\'o}r}, A., {Juh{\'a}sz}, A., {et~al.} 2013,
  \apj, 776, 77

\bibitem[{{Kral} {et~al.}(2013){Kral}, {Th{\'e}bault}, \&
  {Charnoz}}]{Kral_etal_2013}
{Kral}, Q., {Th{\'e}bault}, P., \& {Charnoz}, S. 2013, \aap, 558, A121

\bibitem[{{Kral} {et~al.}(2015){Kral}, {Wyatt}, \& {Pringle}}]{Kral_etal_2015}
{Kral}, Q., {Wyatt}, M., \& {Pringle}, J. 2015, in AAS/Division for Extreme
  Solar Systems Abstracts, Vol.~3, 120.08

\bibitem[{{Lagrange} {et~al.}(2010){Lagrange}, {Bonnefoy}, {Chauvin}, {Apai},
  {Ehrenreich}, {Boccaletti}, {Gratadour}, {Rouan}, {Mouillet}, {Lacour}, \&
  {Kasper}}]{Lagrange_etal_2010}
{Lagrange}, A.-M., {Bonnefoy}, M., {Chauvin}, G., {et~al.} 2010, Science, 329,
  57

\bibitem[{{Lagrange} {et~al.}(1995){Lagrange}, {Vidal-Madjar}, {Deleuil},
  {Emerich}, {Beust}, \& {Ferlet}}]{Lagrange_etal_1995}
{Lagrange}, A.~M., {Vidal-Madjar}, A., {Deleuil}, M., {et~al.} 1995, \aap, 296,
  499

\bibitem[{{Laor} \& {Draine}(1993)}]{Laor_Draine_1993}
{Laor}, A. \& {Draine}, B.~T. 1993, \apj, 402, 441

\bibitem[{{Lawler} {et~al.}(2015){Lawler}, {Greenstreet}, \&
  {Gladman}}]{Lawler_etal_2015}
{Lawler}, S.~M., {Greenstreet}, S., \& {Gladman}, B. 2015, \apjl, 802, L20

\bibitem[{{Lecavelier des Etangs} {et~al.}(2001){Lecavelier des Etangs},
  {Vidal-Madjar}, {Roberge}, {Feldman}, {Deleuil}, {Andr{\'e}}, {Blair},
  {Bouret}, {D{\'e}sert}, {Ferlet}, {Friedman}, {H{\'e}brard}, {Lemoine}, \&
  {Moos}}]{LecavelierdesEtangs_etal_2001}
{Lecavelier des Etangs}, A., {Vidal-Madjar}, A., {Roberge}, A., {et~al.} 2001,
  \nat, 412, 706

\bibitem[{{Lederberg}(1965)}]{Lederberg_1965}
{Lederberg}, J. 1965, \nat, 207, 9

\bibitem[{{Linder} {et~al.}(2000){Linder}, {Bayrhuber}, {Kull}, {B\"assler},
  {Hopmann}, \& {R\"udiger}}]{Linder_etal_2000}
{Linder}, H., {Bayrhuber}, H., {Kull}, U., {et~al.} 2000, {Linder Biologie:
  Lehrbuch f\"ur die Oberstufe: Gesamtband}, 21st edn. (Schroedel Verlag GmbH,
  Hannover)

\bibitem[{{Liseau}(1999)}]{Liseau_1999}
{Liseau}, R. 1999, \aap, 348, 133

\bibitem[{{Lisse} {et~al.}(2012){Lisse}, {Wyatt}, {Chen}, {Morlok}, {Watson},
  {Manoj}, {Sheehan}, {Currie}, {Thebault}, \& {Sitko}}]{Lisse_etal_2012}
{Lisse}, C.~M., {Wyatt}, M.~C., {Chen}, C.~H., {et~al.} 2012, \apj, 747, 93

\bibitem[{{Lovelock}(1965)}]{Lovelock_1965}
{Lovelock}, J.~E. 1965, \nat, 207, 568

\bibitem[{{Lowrance} {et~al.}(2000){Lowrance}, {Schneider}, {Kirkpatrick},
  {Becklin}, {Weinberger}, {Zuckerman}, {Plait}, {Malmuth}, {Heap}, {Schultz},
  {Smith}, {Terrile}, \& {Hines}}]{Lowrance_etal_2000}
{Lowrance}, P.~J., {Schneider}, G., {Kirkpatrick}, J.~D., {et~al.} 2000, \apj,
  541, 390

\bibitem[{{Lyra} \& {Kuchner}(2013)}]{Lyra_Kuchner_2013}
{Lyra}, W. \& {Kuchner}, M. 2013, \nat, 499, 184

\bibitem[{Madigan(2012)}]{Madigan_2012}
Madigan, M. 2012, Brock Biology of Microorganisms (Benjamin Cummings)

\bibitem[{Maeder(2009{\natexlab{a}})}]{Maeder_2009_20}
Maeder, A. 2009{\natexlab{a}}, in Physics, Formation and Evolution of Rotating
  Stars, Astronomy and Astrophysics Library (Springer Berlin Heidelberg),
  513--538

\bibitem[{Maeder(2009{\natexlab{b}})}]{Maeder_2009_19}
Maeder, A. 2009{\natexlab{b}}, in Physics, Formation and Evolution of Rotating
  Stars, Astronomy and Astrophysics Library (Springer Berlin Heidelberg),
  497--512

\bibitem[{{Mamajek}(2012)}]{Mamajek_2012}
{Mamajek}, E.~E. 2012, \apjl, 754, L20

\bibitem[{{Mamajek} \& {Bell}(2014)}]{Mamajek_Bell_2014}
{Mamajek}, E.~E. \& {Bell}, C.~P.~M. 2014, \mnras, 445, 2169

\bibitem[{{Mart{\'{\i}}-Vidal} {et~al.}(2014){Mart{\'{\i}}-Vidal}, {Vlemmings},
  {Muller}, \& {Casey}}]{Marti-Vidal_etal_2014}
{Mart{\'{\i}}-Vidal}, I., {Vlemmings}, W.~H.~T., {Muller}, S., \& {Casey}, S.
  2014, \aap, 563, A136

\bibitem[{{Mart{\'{\i}}n-Torres} {et~al.}(2015){Mart{\'{\i}}n-Torres},
  {Zorzano}, {Valent{\'{\i}}n-Serrano}, {Harri}, {Genzer}, {Kemppinen},
  {Rivera-Valentin}, {Jun}, {Wray}, {Bo Madsen}, {Goetz}, {McEwen},
  {Hardgrove}, {Renno}, {Chevrier}, {Mischna}, {Navarro-Gonz{\'a}lez},
  {Mart{\'{\i}}nez-Fr{\'{\i}}as}, {Conrad}, {McConnochie}, {Cockell}, {Berger},
  {R.~Vasavada}, {Sumner}, \& {Vaniman}}]{Martin-Torres_etal_2015}
{Mart{\'{\i}}n-Torres}, F.~J., {Zorzano}, M.-P., {Valent{\'{\i}}n-Serrano}, P.,
  {et~al.} 2015, Nature Geoscience, 8, 357

\bibitem[{{Matr{\`a}} {et~al.}(2015){Matr{\`a}}, {Pani{\'c}}, {Wyatt}, \&
  {Dent}}]{Matra_etal_2015}
{Matr{\`a}}, L., {Pani{\'c}}, O., {Wyatt}, M.~C., \& {Dent}, W.~R.~F. 2015,
  \mnras, 447, 3936

\bibitem[{{Matthews} {et~al.}(2014){Matthews}, {Krivov}, {Wyatt}, {Bryden}, \&
  {Eiroa}}]{Matthews_etal_2014}
{Matthews}, B.~C., {Krivov}, A.~V., {Wyatt}, M.~C., {Bryden}, G., \& {Eiroa},
  C. 2014, Protostars and Planets VI, 521

\bibitem[{Mattimore \& Battista(1996)}]{Mattimore_Battista_1996}
Mattimore, V. \& Battista, J.~R. 1996, Journal of Bacteriology, 178, 633

\bibitem[{{Mayor} \& {Queloz}(1995)}]{Mayor_Queloz_1995}
{Mayor}, M. \& {Queloz}, D. 1995, \nat, 378, 355

\bibitem[{{McMullin} {et~al.}(2007){McMullin}, {Waters}, {Schiebel}, {Young},
  \& {Golap}}]{McMullin_etal_2007}
{McMullin}, J.~P., {Waters}, B., {Schiebel}, D., {Young}, W., \& {Golap}, K.
  2007, in Astronomical Society of the Pacific Conference Series, Vol. 376,
  Astronomical Data Analysis Software and Systems XVI, ed. R.~A. {Shaw},
  F.~{Hill}, \& D.~J. {Bell}, 127

\bibitem[{{Mileikowsky} {et~al.}(2000){Mileikowsky}, {Cucinotta}, {Wilson},
  {Gladman}, {Horneck}, {Lindegren}, {Melosh}, {Rickman}, {Valtonen}, \&
  {Zheng}}]{Mileikowsky_etal_2000}
{Mileikowsky}, C., {Cucinotta}, F.~A., {Wilson}, J.~W., {et~al.} 2000, \icarus,
  145, 391

\bibitem[{{Millar-Blanchaer} {et~al.}(2015){Millar-Blanchaer}, {Graham},
  {Pueyo}, {Kalas}, {Dawson}, {Wang}, {Perrin}, {moon}, {Macintosh}, {Ammons},
  {Barman}, {Cardwell}, {Chen}, {Chiang}, {Chilcote}, {Cotten}, {De Rosa},
  {Draper}, {Dunn}, {Duch{\^e}ne}, {Esposito}, {Fitzgerald}, {Follette},
  {Goodsell}, {Greenbaum}, {Hartung}, {Hibon}, {Hinkley}, {Ingraham},
  {Jensen-Clem}, {Konopacky}, {Larkin}, {Long}, {Maire}, {Marchis}, {Marley},
  {Marois}, {Morzinski}, {Nielsen}, {Palmer}, {Oppenheimer}, {Poyneer},
  {Rajan}, {Rantakyr{\"o}}, {Ruffio}, {Sadakuni}, {Saddlemyer}, {Schneider},
  {Sivaramakrishnan}, {Soummer}, {Thomas}, {Vasisht}, {Vega}, {Wallace},
  {Ward-Duong}, {Wiktorowicz}, \& {Wolff}}]{Millar-Blanchaer_etal_2015}
{Millar-Blanchaer}, M.~A., {Graham}, J.~R., {Pueyo}, L., {et~al.} 2015, \apj,
  811, 18

\bibitem[{{Montesinos} {et~al.}(2009){Montesinos}, {Eiroa}, {Mora}, \&
  {Mer{\'{\i}}n}}]{Montesinos_etal_2009}
{Montesinos}, B., {Eiroa}, C., {Mora}, A., \& {Mer{\'{\i}}n}, B. 2009, \aap,
  495, 901

\bibitem[{{Montgomery} \& {Welsh}(2012)}]{Montgomery_Welsh_2012}
{Montgomery}, S.~L. \& {Welsh}, B.~Y. 2012, \pasp, 124, 1042

\bibitem[{{Mo{\'o}r} {et~al.}(2006){Mo{\'o}r}, {{\'A}brah{\'a}m}, {Derekas},
  {Kiss}, {Kiss}, {Apai}, {Grady}, \& {Henning}}]{Moor_etal_2006}
{Mo{\'o}r}, A., {{\'A}brah{\'a}m}, P., {Derekas}, A., {et~al.} 2006, \apj, 644,
  525

\bibitem[{{Mo{\'o}r} {et~al.}(2011){Mo{\'o}r}, {{\'A}brah{\'a}m}, {Juh{\'a}sz},
  {Kiss}, {Pascucci}, {K{\'o}sp{\'a}l}, {Apai}, {Henning}, {Csengeri}, \&
  {Grady}}]{Moor_etal_2011}
{Mo{\'o}r}, A., {{\'A}brah{\'a}m}, P., {Juh{\'a}sz}, A., {et~al.} 2011, \apjl,
  740, L7

\bibitem[{{Mo{\'o}r} {et~al.}(2015){Mo{\'o}r}, {Henning}, {Juh{\'a}sz},
  {{\'A}brah{\'a}m}, {Balog}, {K{\'o}sp{\'a}l}, {Pascucci}, {Szab{\'o}},
  {Vavrek}, {Cur{\'e}}, {Csengeri}, {Grady}, {G{\"u}sten}, \&
  {Kiss}}]{Moor_etal_2015}
{Mo{\'o}r}, A., {Henning}, T., {Juh{\'a}sz}, A., {et~al.} 2015, \apj, 814, 42

\bibitem[{{Mora} {et~al.}(2001){Mora}, {Mer{\'{\i}}n}, {Solano}, {Montesinos},
  {de Winter}, {Eiroa}, {Ferlet}, {Grady}, {Davies}, {Miranda}, {Oudmaijer},
  {Palacios}, {Quirrenbach}, {Harris}, {Rauer}, {Collier Cameron}, {Deeg},
  {Garz{\'o}n}, {Penny}, {Schneider}, {Tsapras}, \&
  {Wesselius}}]{Mora_etal_2001}
{Mora}, A., {Mer{\'{\i}}n}, B., {Solano}, E., {et~al.} 2001, \aap, 378, 116

\bibitem[{{Morzinski} {et~al.}(2015){Morzinski}, {Males}, {Skemer}, {Close},
  {Hinz}, {Rodigas}, {Puglisi}, {Esposito}, {Riccardi}, {Pinna}, {Xompero},
  {Briguglio}, {Bailey}, {Follette}, {Kopon}, {Weinberger}, \&
  {Wu}}]{Morzinski_etal_2015}
{Morzinski}, K.~M., {Males}, J.~R., {Skemer}, A.~J., {et~al.} 2015, \apj, 815,
  108

\bibitem[{{Moseley} \& {Mattingly}(1971)}]{Moseley_Mattingly_1971}
{Moseley}, B.~E.~B. \& {Mattingly}, A. 1971, Journal of Bacteriology, 105(3),
  976

\bibitem[{{Natta}(2000)}]{Natta_2000}
{Natta}, A. 2000, in Infrared Space Astronomy, Today and Tomorrow, ed.
  F.~{Casoli}, J.~{Lequeux}, \& F.~{David}, 193

\bibitem[{{Nesvold} {et~al.}(2013){Nesvold}, {Kuchner}, {Rein}, \&
  {Pan}}]{Nesvold_etal_2013}
{Nesvold}, E.~R., {Kuchner}, M.~J., {Rein}, H., \& {Pan}, M. 2013, \apj, 777,
  144

\bibitem[{{Nesvorn{\'y}} {et~al.}(2010){Nesvorn{\'y}}, {Jenniskens}, {Levison},
  {Bottke}, {Vokrouhlick{\'y}}, \& {Gounelle}}]{Nesvorny_etal_2010}
{Nesvorn{\'y}}, D., {Jenniskens}, P., {Levison}, H.~F., {et~al.} 2010, \apj,
  713, 816

\bibitem[{{Neuh{\"a}user} {et~al.}(2015){Neuh{\"a}user}, {Hohle}, {Ginski},
  {Schmidt}, {Hambaryan}, \& {Schmidt}}]{Neuhauser_etal_2015}
{Neuh{\"a}user}, R., {Hohle}, M.~M., {Ginski}, C., {et~al.} 2015, \mnras, 448,
  376

\bibitem[{{Nilsson} {et~al.}(2012){Nilsson}, {Brandeker}, {Olofsson}, {Fathi},
  {Th{\'e}bault}, \& {Liseau}}]{Nilsson_etal_2012}
{Nilsson}, R., {Brandeker}, A., {Olofsson}, G., {et~al.} 2012, \aap, 544, A134

\bibitem[{Nimmo \& Porco(2014)}]{Nimmo_Porco_2014}
Nimmo, F. \& Porco, C. 2014, in Encyclopedia of the Solar System, 3rd edn., ed.
  T.~Spohn, D.~Breuer, \& T.~V. Johnson (Boston: Elsevier), 851 -- 859

\bibitem[{{{\"O}berg} {et~al.}(2009){{\"O}berg}, {van Dishoeck}, \&
  {Linnartz}}]{Oberg_etal_2009}
{{\"O}berg}, K.~I., {van Dishoeck}, E.~F., \& {Linnartz}, H. 2009, \aap, 496,
  281

\bibitem[{{Ojha} {et~al.}(2015){Ojha}, {Wilhelm}, {Murchie}, {McEwen}, {Wray},
  {Hanley}, {Mass{\'e}}, \& {Chojnacki}}]{Ojha_etal_2015}
{Ojha}, L., {Wilhelm}, M.~B., {Murchie}, S.~L., {et~al.} 2015, Nature
  Geoscience, 8, 829

\bibitem[{Ollivier(2007)}]{Ollivier_2007}
Ollivier, M. 2007, in Lectures in Astrobiology, ed. M.~Gargaud, H.~Martin, \&
  P.~Claeys, Advances in Astrobiology and Biogeophysics (Springer Berlin
  Heidelberg), 157--198

\bibitem[{{Olofsson} {et~al.}(2001){Olofsson}, {Liseau}, \&
  {Brandeker}}]{Olofsson_etal_2001}
{Olofsson}, G., {Liseau}, R., \& {Brandeker}, A. 2001, \apjl, 563, L77

\bibitem[{{O'Malley-James} {et~al.}(2014){O'Malley-James}, {Cockell},
  {Greaves}, \& {Raven}}]{OMalley-James_etal_2014}
{O'Malley-James}, J.~T., {Cockell}, C.~S., {Greaves}, J.~S., \& {Raven}, J.~A.
  2014, International Journal of Astrobiology, 13, 229

\bibitem[{{O'Malley-James} {et~al.}(2013){O'Malley-James}, {Greaves}, {Raven},
  \& {Cockell}}]{OMalley-James_etal_2013}
{O'Malley-James}, J.~T., {Greaves}, J.~S., {Raven}, J.~A., \& {Cockell}, C.~S.
  2013, International Journal of Astrobiology, 12, 99

\bibitem[{{Park} {et~al.}(2015){Park}, {Bills}, {Buffington}, {Folkner},
  {Konopliv}, {Martin-Mur}, {Mastrodemos}, {McElrath}, {Riedel}, \&
  {Watkins}}]{Park_etal_2015}
{Park}, R.~S., {Bills}, B., {Buffington}, B.~B., {et~al.} 2015, \planss, 112,
  10

\bibitem[{{Petigura} {et~al.}(2013){Petigura}, {Howard}, \&
  {Marcy}}]{Petigura_etal_2013}
{Petigura}, E.~A., {Howard}, A.~W., \& {Marcy}, G.~W. 2013, Proceedings of the
  National Academy of Science, 110, 19273

\bibitem[{{Pilbratt} {et~al.}(2010){Pilbratt}, {Riedinger}, {Passvogel},
  {Crone}, {Doyle}, {Gageur}, {Heras}, {Jewell}, {Metcalfe}, {Ott}, \&
  {Schmidt}}]{Pilbratt_etal_2010}
{Pilbratt}, G.~L., {Riedinger}, J.~R., {Passvogel}, T., {et~al.} 2010, \aap,
  518, L1

\bibitem[{Prockter \& Pappalardo(2014)}]{Prockter_Pappalardo_2014}
Prockter, L.~M. \& Pappalardo, R.~T. 2014, in Encyclopedia of the Solar System,
  3rd edn., ed. T.~Spohn, D.~Breuer, \& T.~V. Johnson (Boston: Elsevier),
  793--811

\bibitem[{{Redfield}(2007)}]{Redfield_2007}
{Redfield}, S. 2007, \apjl, 656, L97

\bibitem[{{Ribas} {et~al.}(2015){Ribas}, {Bouy}, \&
  {Mer{\'{\i}}n}}]{Ribas_etal_2015}
{Ribas}, {\'A}., {Bouy}, H., \& {Mer{\'{\i}}n}, B. 2015, \aap, 576, A52

\bibitem[{{Riviere-Marichalar} {et~al.}(2012){Riviere-Marichalar}, {Barrado},
  {Augereau}, {Thi}, {Roberge}, {Eiroa}, {Montesinos}, {Meeus}, {Howard},
  {Sandell}, {Duch{\^e}ne}, {Dent}, {Lebreton}, {Mendigut{\'{\i}}a},
  {Hu{\'e}lamo}, {M{\'e}nard}, \& {Pinte}}]{Riviere-Marichalar_etal_2012}
{Riviere-Marichalar}, P., {Barrado}, D., {Augereau}, J.-C., {et~al.} 2012,
  \aap, 546, L8

\bibitem[{{Riviere-Marichalar} {et~al.}(2014){Riviere-Marichalar}, {Barrado},
  {Montesinos}, {Duch{\^e}ne}, {Bouy}, {Pinte}, {Menard}, {Donaldson}, {Eiroa},
  {Krivov}, {Kamp}, {Mendigut{\'{\i}}a}, {Dent}, \&
  {Lillo-Box}}]{Riviere-Marichalar_etal_2014}
{Riviere-Marichalar}, P., {Barrado}, D., {Montesinos}, B., {et~al.} 2014, \aap,
  565, A68

\bibitem[{{Roberge}(2014)}]{Roberge_2014}
{Roberge}, A. 2014, in Thirty years of Beta Pic and Debris Disk Studies, ed.
  A.-M. {Lagrange} \& A.~{Boccaletti}

\bibitem[{{Roberge} {et~al.}(2012){Roberge}, {Chen}, {Millan-Gabet},
  {Weinberger}, {Hinz}, {Stapelfeldt}, {Absil}, {Kuchner}, \&
  {Bryden}}]{Roberge_etal_2012}
{Roberge}, A., {Chen}, C.~H., {Millan-Gabet}, R., {et~al.} 2012, \pasp, 124,
  799

\bibitem[{{Roberge} {et~al.}(2002){Roberge}, {Feldman}, {Lecavelier des
  Etangs}, {Vidal-Madjar}, {Deleuil}, {Bouret}, {Ferlet}, \&
  {Moos}}]{Roberge_etal_2002}
{Roberge}, A., {Feldman}, P.~D., {Lecavelier des Etangs}, A., {et~al.} 2002,
  \apj, 568, 343

\bibitem[{{Roberge} {et~al.}(2006){Roberge}, {Feldman}, {Weinberger},
  {Deleuil}, \& {Bouret}}]{Roberge_etal_2006}
{Roberge}, A., {Feldman}, P.~D., {Weinberger}, A.~J., {Deleuil}, M., \&
  {Bouret}, J.-C. 2006, \nat, 441, 724

\bibitem[{{Roberge} \& {Weinberger}(2008)}]{Roberge_Weinberger_2008}
{Roberge}, A. \& {Weinberger}, A.~J. 2008, \apj, 676, 509

\bibitem[{{Roberge} {et~al.}(2014){Roberge}, {Welsh}, {Kamp}, {Weinberger}, \&
  {Grady}}]{Roberge_etal_2014}
{Roberge}, A., {Welsh}, B.~Y., {Kamp}, I., {Weinberger}, A.~J., \& {Grady},
  C.~A. 2014, \apjl, 796, L11

\bibitem[{{Rocchetto} {et~al.}(2015){Rocchetto}, {Farihi}, {G{\"a}nsicke}, \&
  {Bergfors}}]{Rocchetto_etal_2015}
{Rocchetto}, M., {Farihi}, J., {G{\"a}nsicke}, B.~T., \& {Bergfors}, C. 2015,
  \mnras, 449, 574

\bibitem[{Rosing {et~al.}(2006)Rosing, Bird, Sleep, Glassley, \&
  Albarede}]{Rosing_etal_2006}
Rosing, M.~T., Bird, D.~K., Sleep, N.~H., Glassley, W., \& Albarede, F. 2006,
  Palaeogeography, Palaeoclimatology, Palaeoecology, 232, 99

\bibitem[{{Rushby} {et~al.}(2013){Rushby}, {Claire}, {Osborn}, \&
  {Watson}}]{Rushby_etal_2013}
{Rushby}, A.~J., {Claire}, M.~W., {Osborn}, H., \& {Watson}, A.~J. 2013,
  Astrobiology, 13, 833

\bibitem[{Rybicki \& Lightman(2007)}]{Rybicki_Lightman_2007}
Rybicki, G.~B. \& Lightman, A.~P. 2007, Fundamentals of Radiative Transfer
  (Wiley-VCH Verlag GmbH), 1--50

\bibitem[{{Sagan}(1963)}]{Sagan_1963}
{Sagan}, C. 1963, \planss, 11, 485

\bibitem[{{Sagan} {et~al.}(1993){Sagan}, {Thompson}, {Carlson}, {Gurnett}, \&
  {Hord}}]{Sagan_etal_1993}
{Sagan}, C., {Thompson}, W.~R., {Carlson}, R., {Gurnett}, D., \& {Hord}, C.
  1993, \nat, 365, 715

\bibitem[{{Sajadian} \& {Rahvar}(2015)}]{Sajadian_Rahvar_2015}
{Sajadian}, S. \& {Rahvar}, S. 2015, \mnras, 454, 4429

\bibitem[{{Schwieterman} {et~al.}(2015){Schwieterman}, {Cockell}, \&
  {Meadows}}]{Schwieterman_etal_2015}
{Schwieterman}, E.~W., {Cockell}, C.~S., \& {Meadows}, V.~S. 2015,
  Astrobiology, 15, 341

\bibitem[{{Seager}(2014)}]{Seager_2014}
{Seager}, S. 2014, Proceedings of the National Academy of Science, 111, 12634

\bibitem[{{Segura} {et~al.}(2005){Segura}, {Kasting}, {Meadows}, {Cohen},
  {Scalo}, {Crisp}, {Butler}, \& {Tinetti}}]{Segura_etal_2005}
{Segura}, A., {Kasting}, J.~F., {Meadows}, V., {et~al.} 2005, Astrobiology, 5,
  706

\bibitem[{{Shannon} {et~al.}(2014){Shannon}, {Clarke}, \&
  {Wyatt}}]{Shannon_etal_2014}
{Shannon}, A., {Clarke}, C., \& {Wyatt}, M. 2014, \mnras, 442, 142

\bibitem[{{Slettebak}(1975)}]{Slettebak_1975}
{Slettebak}, A. 1975, \apj, 197, 137

\bibitem[{{Slettebak}(1982)}]{Slettebak_1982}
{Slettebak}, A. 1982, \apjs, 50, 55

\bibitem[{{Slettebak} \& {Carpenter}(1983)}]{Slettebak_Carpenter_1983}
{Slettebak}, A. \& {Carpenter}, K.~G. 1983, \apjs, 53, 869

\bibitem[{{Smith} \& {Terrile}(1984)}]{Smith_Terrile_1984}
{Smith}, B.~A. \& {Terrile}, R.~J. 1984, Science, 226, 1421

\bibitem[{{Snellen} {et~al.}(2014){Snellen}, {Brandl}, {de Kok}, {Brogi},
  {Birkby}, \& {Schwarz}}]{Snellen_etal_2014}
{Snellen}, I.~A.~G., {Brandl}, B.~R., {de Kok}, R.~J., {et~al.} 2014, \nat,
  509, 63

\bibitem[{{Sparks} {et~al.}(2012){Sparks}, {Hough}, {Germer}, {Robb}, \&
  {Kolokolova}}]{Sparks_etal_2012}
{Sparks}, W., {Hough}, J.~H., {Germer}, T.~A., {Robb}, F., \& {Kolokolova}, L.
  2012, \planss, 72, 111

\bibitem[{{Su} \& {Rieke}(2014)}]{Su_Rieke_2014}
{Su}, K.~Y.~L. \& {Rieke}, G.~H. 2014, in IAU Symposium, ed. M.~{Booth}, B.~C.
  {Matthews}, \& J.~R. {Graham}, Vol. 299, 318--321

\bibitem[{{Su} {et~al.}(2016){Su}, {Rieke}, {Defr{\'e}re}, {Wang}, {Lai},
  {Wilner}, {van Lieshout}, \& {Lee}}]{Su_etal_2016}
{Su}, K.~Y.~L., {Rieke}, G.~H., {Defr{\'e}re}, D., {et~al.} 2016, \apj, 818, 45

\bibitem[{Swartz(2016)}]{Swartz}
Swartz, N. 2016, Laws of Nature, The Internet Encyclopedia of Philosophy,
  {I}SSN 2161-0002, \url{http://www.iep.utm.edu/lawofnat/}, accessed 2016-01-01

\bibitem[{Takai {et~al.}(2008)Takai, Nakamura, Toki, Tsunogai, Miyazaki,
  Miyazaki, Hirayama, Nakagawa, Nunoura, \& Horikoshi}]{Takai_etal_2008}
Takai, K., Nakamura, K., Toki, T., {et~al.} 2008, Proceedings of the National
  Academy of Sciences, 105, 10949

\bibitem[{{Tamayo}(2014)}]{Tamayo_2014}
{Tamayo}, D. 2014, \mnras, 438, 3577

\bibitem[{{Tanaka} {et~al.}(1996){Tanaka}, {Inaba}, \&
  {Nakazawa}}]{Tanaka_etal_1996}
{Tanaka}, H., {Inaba}, S., \& {Nakazawa}, K. 1996, \icarus, 123, 450

\bibitem[{{Tarter} {et~al.}(2007){Tarter}, {Backus}, {Mancinelli}, {Aurnou},
  {Backman}, {Basri}, {Boss}, {Clarke}, {Deming}, {Doyle}, {Feigelson},
  {Freund}, {Grinspoon}, {Haberle}, {Hauck}, {Heath}, {Henry}, {Hollingsworth},
  {Joshi}, {Kilston}, {Liu}, {Meikle}, {Reid}, {Rothschild}, {Scalo}, {Segura},
  {Tang}, {Tiedje}, {Turnbull}, {Walkowicz}, {Weber}, \&
  {Young}}]{Tarter_etal_2007}
{Tarter}, J.~C., {Backus}, P.~R., {Mancinelli}, R.~L., {et~al.} 2007,
  Astrobiology, 7, 30

\bibitem[{{Th{\'e}bault} {et~al.}(2003){Th{\'e}bault}, {Augereau}, \&
  {Beust}}]{Thebault_etal_2003}
{Th{\'e}bault}, P., {Augereau}, J.~C., \& {Beust}, H. 2003, \aap, 408, 775

\bibitem[{{Thi} {et~al.}(2013){Thi}, {M{\'e}nard}, {Meeus}, {Carmona},
  {Riviere-Marichalar}, {Augereau}, {Kamp}, {Woitke}, {Pinte},
  {Mendigut{\'{\i}}a}, {Eiroa}, {Montesinos}, {Britain}, \&
  {Dent}}]{Thi_etal_2013}
{Thi}, W.~F., {M{\'e}nard}, F., {Meeus}, G., {et~al.} 2013, \aap, 557, A111

\bibitem[{{Tian} \& {Ida}(2015)}]{Tian_Ida_2015}
{Tian}, F. \& {Ida}, S. 2015, Nature Geoscience, 8, 177

\bibitem[{{Torres} {et~al.}(2006){Torres}, {Quast}, {da Silva}, {de La Reza},
  {Melo}, \& {Sterzik}}]{Torres_etal_2006}
{Torres}, C.~A.~O., {Quast}, G.~R., {da Silva}, L., {et~al.} 2006, \aap, 460,
  695

\bibitem[{{van der Tak} {et~al.}(2007){van der Tak}, {Black}, {Sch{\"o}ier},
  {Jansen}, \& {van Dishoeck}}]{vanderTak_etal_2007}
{van der Tak}, F.~F.~S., {Black}, J.~H., {Sch{\"o}ier}, F.~L., {Jansen}, D.~J.,
  \& {van Dishoeck}, E.~F. 2007, \aap, 468, 627

\bibitem[{{van Leeuwen}(2007)}]{vanLeeuwen_2007}
{van Leeuwen}, F. 2007, \aap, 474, 653

\bibitem[{{van Lieshout} {et~al.}(2014){van Lieshout}, {Dominik}, {Kama}, \&
  {Min}}]{vanLieshout_etal_2014}
{van Lieshout}, R., {Dominik}, C., {Kama}, M., \& {Min}, M. 2014, \aap, 571,
  A51

\bibitem[{Warren \& Brandt(2008)}]{Warren_Brandt_2008}
Warren, S.~G. \& Brandt, R.~E. 2008, Journal of Geophysical Research:
  Atmospheres, 113

\bibitem[{Way {et~al.}(2015)Way, Del~Genio, Kiang, Sohl, Clune, Aleinov, \&
  Kelley}]{Way_etal_2015}
Way, M.~J., Del~Genio, A., Kiang, N., {et~al.} 2015, Venus: The First Habitable
  World of Our Solar System?, NASA Technical Reports Server, Report Number
  GSFC-E-DAA-TN27812, Document ID 20150021298

\bibitem[{{Weinberger} {et~al.}(2015){Weinberger}, {Bryden}, {Kennedy},
  {Roberge}, {Defr{\`e}re}, {Hinz}, {Millan-Gabet}, {Rieke}, {Bailey},
  {Danchi}, {Haniff}, {Mennesson}, {Serabyn}, {Skemer}, {Stapelfeldt}, \&
  {Wyatt}}]{Weinberger_etal_2015}
{Weinberger}, A.~J., {Bryden}, G., {Kennedy}, G.~M., {et~al.} 2015, \apjs, 216,
  24

\bibitem[{{Welsh} \& {Montgomery}(2013)}]{Welsh_Montgomery_2013}
{Welsh}, B.~Y. \& {Montgomery}, S. 2013, \pasp, 125, 759

\bibitem[{{Welsh} \& {Montgomery}(2015)}]{Welsh_Montgomery_2015}
{Welsh}, B.~Y. \& {Montgomery}, S.~L. 2015, Advances in Astronomy, 2015, 980323

\bibitem[{{Wordsworth} \&
  {Pierrehumbert}(2014)}]{Wordsworth_Pierrehumbert_2014}
{Wordsworth}, R. \& {Pierrehumbert}, R. 2014, \apjl, 785, L20

\bibitem[{{Worth} {et~al.}(2013){Worth}, {Sigurdsson}, \&
  {House}}]{Worth_etal_2013}
{Worth}, R.~J., {Sigurdsson}, S., \& {House}, C.~H. 2013, Astrobiology, 13,
  1155

\bibitem[{{Wright} {et~al.}(2016){Wright}, {Cartier}, {Zhao}, {Jontof-Hutter},
  \& {Ford}}]{Wright_etal_2016}
{Wright}, J.~T., {Cartier}, K.~M.~S., {Zhao}, M., {Jontof-Hutter}, D., \&
  {Ford}, E.~B. 2016, \apj, 816, 17

\bibitem[{{Wyatt}(2008)}]{Wyatt_2008}
{Wyatt}, M.~C. 2008, \araa, 46, 339

\bibitem[{{Wyatt} \& {Dent}(2002)}]{Wyatt_Dent_2002}
{Wyatt}, M.~C. \& {Dent}, W.~R.~F. 2002, \mnras, 334, 589

\bibitem[{{Wyatt} {et~al.}(2007){Wyatt}, {Smith}, {Greaves}, {Beichman},
  {Bryden}, \& {Lisse}}]{Wyatt_etal_2007}
{Wyatt}, M.~C., {Smith}, R., {Greaves}, J.~S., {et~al.} 2007, \apj, 658, 569

\bibitem[{{Xie} {et~al.}(2013){Xie}, {Brandeker}, \& {Wu}}]{Xie_etal_2013}
{Xie}, J.-W., {Brandeker}, A., \& {Wu}, Y. 2013, \apj, 762, 114

\bibitem[{{Zackrisson} {et~al.}(2015){Zackrisson}, {Calissendorff}, {Asadi}, \&
  {Nyholm}}]{Zackrisson_etal_2015}
{Zackrisson}, E., {Calissendorff}, P., {Asadi}, S., \& {Nyholm}, A. 2015, \apj,
  810, 23

\bibitem[{{Zagorovsky} {et~al.}(2010){Zagorovsky}, {Brandeker}, \&
  {Wu}}]{Zagorovsky_etal_2010}
{Zagorovsky}, K., {Brandeker}, A., \& {Wu}, Y. 2010, \apj, 720, 923

\bibitem[{{Zheng} \& {M{\'e}nard}(2005)}]{Zheng_Menard_2005}
{Zheng}, Z. \& {M{\'e}nard}, B. 2005, \apj, 635, 599

\bibitem[{{Zorec} {et~al.}(2005){Zorec}, {Fr{\'e}mat}, \&
  {Cidale}}]{Zorec_etal_2005}
{Zorec}, J., {Fr{\'e}mat}, Y., \& {Cidale}, L. 2005, \aap, 441, 235

\bibitem[{{Zuckerman} \& {Song}(2012)}]{Zuckerman_Song_2012}
{Zuckerman}, B. \& {Song}, I. 2012, \apj, 758, 77

\end{thebibliography}






\end{document}